\documentclass{article}
\usepackage{arxiv}

\usepackage{booktabs} 
\usepackage[caption=false,font=small,labelfont=sf,textfont=sf]{subfig}
\usepackage{multirow}
\usepackage{ascmac}
\usepackage{comment}
\usepackage{moreverb,url}
\usepackage{listings}
\usepackage{mathtools}

\usepackage{amsmath}
\usepackage{amssymb}
\usepackage{amsfonts}

\usepackage{hyperref} 
\usepackage{natbib}

\newcommand\BibTeX{{\rmfamily B\kern-.05em \textsc{i\kern-.025em b}\kern-.08em
T\kern-.1667em\lower.7ex\hbox{E}\kern-.125emX}}

\def\vec#1{\mbox{\boldmath $#1$}}
\def\nnz{N_\textrm{nz}}
\def\ndiag{N_\textrm{diag}}
\def\nbl{n_\textrm{b}}
\def\bx#1{\underline{#1}}

\title{Accelerating the SpMV kernel on standard CPUs by exploiting the partially diagonal structures}

\author{
	\href{https://orcid.org/0000-0003-1217-6444}{\includegraphics[scale=0.06]{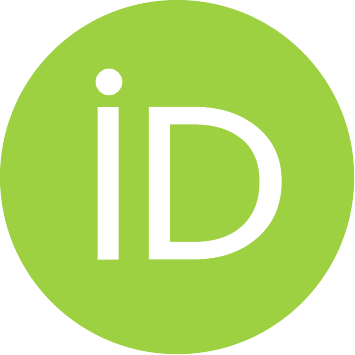}\hspace{1mm}Takeshi Fukaya}\\
	Information Initiative Center\\
	Hokkaido University\\
	JST PRESTO\\
	\texttt{fukaya@iic.hokudai.ac.jp}\\
	\And
	{\hspace{1mm}Koki Ishida}\\
	School of Engineering\\
	Hokkaido University\\
	\And
	{\hspace{1mm}Akie Miura}\\
	School of Engineering\\
	Hokkaido University\\
	\And
	\href{https://orcid.org/0000-0003-1938-1723}{\includegraphics[scale=0.06]{orcid.pdf}\hspace{1mm}Takeshi Iwashita}\\
	Information Initiative Center\\
	Hokkaido University\\
	\And
	{\hspace{1mm}Hiroshi Nakashima}\\
	Academic Center for Computing and Media Studies\\
	Kyoto University\\
}

\renewcommand{\shorttitle}{}

\begin{document}
\maketitle

\begin{abstract}
Sparse Matrix Vector multiplication (SpMV) is one of basic building blocks in scientific computing, 
and acceleration of SpMV has been continuously required. 
In this research, we aim for accelerating SpMV on recent CPUs for sparse matrices that have a specific sparsity structure, 
namely a diagonally structured sparsity pattern. 
We focus a hybrid storage format that combines the DIA and CSR formats, so-called the HDC format. 
First, we recall the importance of introducing cache blocking techniques into HDC-based SpMV kernels. 
Next, based on the observation of the cache blocked kernel, we present a modified version of the HDC formats, which we call the M-HDC format, 
in which partial diagonal structures are expected to be more efficiently picked up. 
For these SpMV kernels, we theoretically analyze the expected performance improvement based on performance models. 
Then, we conduct comprehensive experiments on state-of-the-art multi-core CPUs. 
By the experiments using typical matrices, we clarify the detailed performance characteristics of each SpMV kernel. 
We also evaluate the performance for matrices appearing in practical applications and demonstrate that our approach can accelerate SpMV for some of them. 
Through the present paper, we demonstrate the effectiveness of exploiting partial diagonal structures by the M-HDC format 
as a promising approach to accelerating SpMV on CPUs for a certain kind of practical sparse matrices.

\end{abstract}


\section{Introduction}
\label{sec:introduction}
Sparse Matrix Vector multiplication (SpMV), which computes the product of a sparse matrix and a dense vector, 
is one of basic building blocks in sparse linear algebra algorithms. 
Numerous algorithms such as those for solving linear systems of equations and eigenvalue problems~\citep{1994_tmplates, 2003_saad}, 
e.g. Krylov subspace methods, repeat the SpMV computations, 
and its execution time is usually dominant in their overall computational time. 
This fact has motivated various studies for accelerating the SpMV computation.  
\par
Compared with dense matrix vector multiplication, a characteristic technology in SpMV is storage formate, 
i.e. how to store the information of a sparse matrix in a computer. 
The sparsity of a matrix needs to be exploited to reduce both the memory footprint and arithmetic cost. 
A straightforward format is the coordinate (COO) format~\citep{2003_saad}, in which the row index, column index, and value of each nonzero element are stored. 
One of popular formats is the compressed sparse rows (CSR) format~\citep{1994_tmplates, 2003_saad}, 
in which the information of row index is compressed to reduce the memory footprint. 
It is worth noting that both the COO and CSR formats can store the information of an any sparse matrix without zero padding, 
which means the explicit store of zero elements.  
\par
The COO and CSR formats have nice versatility, however they are sometimes unsuitable for recent computer architectures, e.g. GPU. 
In order to exploit the advantages of recent computer architectures, e.g. SIMD, multi-thread, and hierarchical memory system, 
other storage formats are required. 
One of familiar formats that are designed for the suitability for the recent computer architectures is the ELLPACK (ELL) format~\citep{1989_kincaid}, 
and its variants, e.g. Sliced ELL~\citep{2010_monakov} and SELL-$C$-$\sigma$~\citep{2014_kreutzer}, are also known. 
As a result of improving the suitability, zero padding is basically needed in these formats. 
\par
A sparse matrix appearing in scientific computations often has a not random but structured sparsity pattern derived from its original problem or analytical method, 
e.g. discretization by FEM or FDM. 
Although the applicability is limited, assuming a specified sparsity pattern and designing a specialized storage format is a promising approach to accelerating SpMV. 
Because it is known that some specific sparsity patterns frequently appear in applications, 
it will be a feasible approach to adaptively select an appropriate format for a target matrix in practical situations~\citep{2018_elafrou}. 
\par
In the presented research, we focus on the diagonally structured sparsity pattern, 
which often appears in a sparse matrix derived by the FEM or FDM discretization based on a structured grid; 
a typical example is a heptadiagonal matrix obtained by the FDM discretization for 3-dimensional Poisson's equation. 
Although not fully diagonally structured, partial diagonal structures can be observed in a matrix in such analyses, 
and the SpMV computation for this kind of matrices has a strong demand in the field of computational science. 
\par
As a specialized storage format for matrices with such the sparsity pattern, the diagonal (DIA) format~\citep{2003_saad} has been already known. 
\citet{2010_yuan} considered optimizing SpMV using the DIA format and presented approaches to improving a SpMV kernel naively using the DIA format. 
They introduced a technique of splitting a matrix into the SpMV kernel in order to increase the reusability of the vector data. 
In addition, an idea of combining the DIA and CSR formats was also presented. 
A similar approach was reported by \citet{2014_yang}, where optimization of SpMV on GPU for quasi-diagonal matrices was considered. 
A hybrid format that combine the DIA and CSR formats is now familiar as the HDC (Hybrid DIA-CSR) format. 
\par
As far as the authors know, only the paper by \citet{2010_yuan} deals with DIA-based SpMV on standard CPUs. 
This paper indicated the potential of exploiting diagonal structures by the DIA format, 
however, the following points remain unclear: 
\begin{itemize}
\item All experiments were carried out in sequential computing on 4-core Intel Xeon and AMD CPU. 
However, a recent CPU has more cores, and parallel computing is now standard for it; 
for example, sequential computing often cannot fully use the memory bandwidth. 
\item There is flexibility in splitting a matrix and combining the DIA and CSR formats. 
However, their impact on the performance of SpMV has been not clear. 
\item Speedup over the SpMV kernel using the CSR format was reported for test matrices appearing in practical applications. 
However, both mechanisms for the speedup and the validity of the obtained speedup have not been fully discussed. 
\end{itemize}
To clarify these points is crucial for using DIA-based SpMV kernels in practical situations. 
\par
Our purpose in this research is to make the potential of DIA-based SpMV kernels on recent CPUs more clear. 
Taking the above unclear points into account, we conduct our research and provide the following contributions:
\begin{itemize}
\item We review the characteristics of DIA-based SpMV kernels and recall the importance of introducing cache blocking techniques into them. 
\item We present a modified version of the HDC format, which we call the M-HDC format, based on the observation of the cache blocked HDC-based SpMV kernel. 
We expect that the M-HDC format more efficiently picks up partial diagonal structures than the HDC format. 
\item We conduct theoretical analyses based on performance models, in which the expected performance improvement is theoretically investigated. 
\item We carry out detailed experiments on state-of-the-art multi-core CPUs using thread parallelized kernels. 
First, we investigate the performance behavior for typical matrices, so-called stencil matrices, and clarify the detailed characteristics of each SpMV kernel. 
Next, we report the performance for matrices appearing in practical applications, which are provided in the SuiteSparse Matrix Collection~\citep{2011_davis}, 
and establish that there are matrices whose SpMV computation on recent CPUs can be accelerated by employing the M-HDC format.
\item We show the validity of the experimental results by comparing with the theoretical analyses, 
which clarifies the mechanisms of obtaining the performance improvement by exploiting partial diagonal structures in SpMV. 
\end{itemize}
Through the present paper, we demonstrate that exploiting partial diagonal structures by the M-HDC format is a promising approach to accelerating SpMV on CPUs for a certain kind of practical sparse matrices. 

\par
The rest of the paper is organized as follows: 
in Section~\ref{sec:related_work}, we briefly summarize related work. 
In Section~\ref{sec:hdc}, we review the HDC format and the SpMV kernel based on it, together with the CSR and DIA formats. 
In Section~\ref{sec:blocking}, we consider introducing a cache blocking technique into the HDC-based SpMV kernel and propose the M-HDC format. 
In Section~\ref{sec:model}, we construct performance models and theoretically analyze the expected performance improvement by the SpMV kernels based on the HDC and M-HDC formats. 
In Section~\ref{sec:evaluation}, we present the results of numerical experiments and discuss them. 
Finally, we give the conclusion remarks.

\section{Related work}
\label{sec:related_work}
Enormous numbers of studies that aim for accelerating SpMV have been conducted. 
Basic optimization techniques in the days before multi-core CPUs and GPUs became common in scientific computing are summarized by \citet{2003_vuduc_phd}. 
With the appearance of the GPGPU computing, many attempts for the efficient SpMV computing on GPUs have been reported, 
such as an early study by \citet{2009_bell}, which spotlighted the suitability of ELLPACK for SpMV on GPU. 
Because we focus on the SpMV computation on standard CPUs in the present paper, 
we refer only the recent survey paper by \citet{2017_filippone}, in which a detailed list of storage formats for SpMV on GPUs is provided. 
\par
As discussed in the paper by \citet{2003_vuduc_phd}, one of basic optimization techniques for SpMV on CPU is blocking. 
It aims for increasing the reusability of data in memory~\citep{2004_im, 2004_mellor, 2007_nishtala} 
and reducing the amount of data transfer~\citep{1999_pinar, 2005_vuduc}. 
It was reported by \citet{2005_vuduc} that sparse matrices in practical applications often have block sub-structures, 
and thus these blocking techniques has been regarded as one of important and basic optimization techniques. 
\par
After multi-core CPUs and SIMD architectures became common, 
optimization techniques for parallelization, especially the SIMD vectorization, have attracted research interests. 
We can find studies for early hardware in \citet{2009_williams} and \citet{2013_liu}. 
For the efficient use of SIMD units, more suitable formats for SIMD than CSR were proposed, 
such as SELL-$C$-$\sigma$~\citep{2014_kreutzer}, CVR~\citep{2018_xie}, and CCF~\citep{2020_almasri},  
and improvement for the traditional CSR format to exploit SIMD units was also studied~\citep{2015_liu, 2020_bian}. 
Another important issue in SpMV on multi-core CPUs is load balancing as reported by \citet{2014_ohshima}, 
and a solution for this issue was studied in \citet{2016_merrill}. 
This issue generally becomes more difficult in symmetric SpMV, and some techniques were presented by \citet{2019_muro} and \citet{2019_elafrou}. 
\par
In this research, we focus the DIA format, which itself is one of well-known formats explained in textbooks such as \citet{2003_saad}. 
As mentioned in Section~\ref{sec:introduction}, the paper by \citet{2010_yuan} presented optimization techniques based on the DIA format, 
and several important points still remain unclear. 
\citet{2014_yang} also reported the similar approach for SpMV on GPUs. 
In the paper by \citet{2012_godwin}, an optimization technique for SpMV on GPUs by exploiting block diagonal structures was discussed. 
\par
Recently, a library for SpMV, named SparseX, was presented by \citet{2018_elafrou}, 
in which a storage format is adaptively optimized depending on an input matrix. 
The basic idea in this approach is appropriately selecting blocking techniques for each input matrix, 
which is based on their early studies~\citep{2009_karakasis, 2013_karakasis}. 
It is also worth noting that the Intel MKL library now provides so-called inspector-executor sparse BLAS routines~\citep{2020_fedorov}, 
in which the structure of an input matrix is first analyzed to optimize the SpMV kernel. 
These trends in the development of numerical libraries indicate that adaptive approaches become promising for accelerating SpMV, 
and the demand of preparing specialized optimization techniques for sparse matrices that have typical structures increases. 
Our research results contribute to the more efficient use of diagonal structures in these adaptive-type numerical libraries.

\section{The HDC format}
\label{sec:hdc}
In this section, we first give brief reviews on the CSR and DIA formats, together with standard implementations of the SpMV kernels using them. 
Then, we explain a hybrid storage format combining the DIA and CSR formats, namely the HDC format, and the SpMV kernel using it.

\subsection{Preliminary}
Let 
\begin{itemize}
\item $A$: an $n \times n$ general (nonsymmetric) real sparse matrix, 
\item $\vec{x}$: an $n$ dimensional real vector (input), 
\item $\vec{y}$: an $n$ dimensional real vector (output), 
\end{itemize}
and consider computing SpMV: 
\begin{equation*}
	\vec{y} = A \vec{x}.
\end{equation*}
\par
We denote the number of the nonzero elements in a matrix as $\nnz$. 
We use $i$ and $j$ as the indexes of the row and column in a matrix (and related loop indices in a pseudo-code), respectively. 
We use a sparse matrix in Figure~\ref{fig:sample_matrix} for demonstrating how a matrix is stored in each storage format, 
which we call \textit{Example matrix} throughout this paper.  
We present a pseudo-code of a SpMV kernel in the C language style with thread parallelization and SIMD vectorization using OpenMP directives. 
Corresponding to this, we employ the zero-based indexing for both arrays and elements in a matrix (i.e. row and column indexes). 

\begin{figure}[t]
\begin{center}
\[
\begin{pmatrix}
1 & 0 & 2 & 0 & 0 & 3 & 0 & 0 \\
0 & 4 & 0 & 5 & 0 & 0 & 6 & 0 \\
0 & 0 & 7 & 0 & 8 & 0 & 0 & 9 \\
0 & 0 & 0 & 10& 0 & 0 & 0 & 0 \\
11& 0 & 0 & 0 & 12& 0 & 13& 0 \\
0 & 0 & 0 & 0 & 0 & 14& 0 & 15\\
0 & 0 & 16& 0 & 0 & 0 & 17& 0 \\
18& 0 & 0 & 19& 0 & 0 & 0 & 20\\ 
\end{pmatrix}
\]
\caption{Example matrix: a matrix used for the demonstration of each storage format.}
\label{fig:sample_matrix}
\end{center}
\end{figure}

\subsection{The CSR format}
The CSR (Compressed Sparse Row) format~\citep{1994_tmplates, 2003_saad}, 
also known as the CRS (Compressed Row Storage) format, is one of standard storage formats in sparse matrix computations. 
The CSR format is applicable to a sparse matrix that has an any sparse pattern. 
The CSR format consists of three arrays: {\tt val[]}, {\tt col\_ind[]}, and {\tt row\_ptr[]}. 
From the first row ($i = 0$) to the last row ($i = n-1$), the value and column index of nonzero elements are continuously stored in {\tt val[]} and {\tt col\_ind[]}, respectively. 
The row partition information (the leading index in {\tt val[]} and {\tt col\_ind[]} for each row) is stored in {\tt row\_ptr[]} together with $\nnz$ stored in the last (i.e. $n$-th) position. 
Figure~\ref{fig:sample_csr} shows how Example matrix is stored in the CSR format. 
\par
Figure~\ref{fig:spmv_csr} presents a standard implementation of the SpMV kernel using the CSR format, 
which we call the CSR kernel hereafter. 
Here, in the innermost loop (i.e. the loop on {\tt k}), the vector $\vec{x}$ (the array {\tt x[]}) is accessed via the array {\tt col\_ind[]}, 
which is so-called \textit{indirect access}. 

\begin{figure*}[t]
\begin{center}
\begin{tabular}{|l||c|c|c|c|c|c|c|c|c|c|c|c|c|c|c|c|c|c|c|c|}
\hline
{\tt val[]}& 1 & 2 & 3 & 4 & 5 & 6 & 7 & 8 & 9 & 10 & 11 & 12 & 13 & 14 & 15 & 16 & 17 & 18 & 19 & 20\\
\hline
{\tt col\_ind[]} & 0 & 2 & 5 & 1 & 3 & 6 & 2 & 4 & 7 & 3 & 0 & 4 & 6 & 5 & 7 & 2 & 6 & 0 & 3 & 7 \\
\hline
\end{tabular}
\\
\smallskip
\begin{tabular}{|l||c|c|c|c|c|c|c|c|c|}
\hline
{\tt row\_ptr[]} & 0 & 3 & 6 & 9 & 10 & 13 & 15 & 17 & 20 \\
\hline
\end{tabular}
\caption{An example of the CSR format: Example matrix is stored in the CSR format.}
\label{fig:sample_csr}
\end{center}
\end{figure*}

\begin{figure}[t]
\begin{center}
\begin{lstlisting}
#pragma omp parallel for private(i, k, s)
for(i = 0; i < n; i++) {
	s = 0;
	#pragma omp simd reduction(+:s)
	for(k = row_ptr[i]; k < row_ptr[i+1]; k++) {
		s += val[k] * x[col_ind[k]];
	}
	y[i] = s;
}
\end{lstlisting}
\end{center}
\setlength\abovecaptionskip{-5pt}
\caption{The CSR kernel: a SpMV kernel using the CSR format.}
\label{fig:spmv_csr}
\end{figure}

\subsection{The DIA format}
The DIA (DIAgonal) format~\citep{2003_saad} is a storage format that focuses on diagonally structured sparse matrices. 
For each element in a matrix, let
\begin{equation*}
	\textrm{offset} \coloneqq i - j, 
\end{equation*}
and define a \textit{diagonal line} as a set of elements that have a same offset. 
We call a diagonal line \textit{nonzero} if it has at least one nonzero element. 
The DIA format consists of two arrays: {\tt val[][]} and {\tt offset[]}. 
All of the values including zero(s) in each nonzero diagonal line are continuously stored in {\tt val[][]}, and their offset are stored in {\tt offset[]}. 
Figure~\ref{fig:sample_dia} demonstrates how Example matrix is stored in the DIA format; 
Example matrix has five nonzero diagonal lines, and its $k$-th nonzero diagonal line's values and offset are stored in {\tt val[k][]} and {\tt col\_offset[k]}, respectively. 
\par
Figure~\ref{fig:spmv_dia} gives a pseudo-code of the SpMV kernel using the DIA format (the DIA kernel), where {\tt n\_diags} is the number of nonzero diagonal lines. 
Since the value of {\tt off} does not change in the innermost loop (i.e. the loop on {\tt i}), the array {\tt x[]} is sequentially accessed 
in the manner of \textit{direct access}. 
Compared with indirect access, this is a preferable feature in terms of the memory access.
In addition, if the number of zero elements explicitly stored in the nonzero diagonal lines is sufficiently small, the DIA format is superior to the CSR format in terms of the memory access cost; the memory access cost of {\tt offset[]} in the DIA format is much smaller than that of {\tt col\_ind[]} in the CSR format. 
However, if the nonzero diagonal lines contain a large number of zero elements, which means that a sparse matrix is far from diagonally structured, both the computation and memory access cost are increased by explicitly stored zero elements. 

\begin{figure}[t]
\begin{center}
\begin{tabular}{|l||c|c|c|c|c|c|c|c|}
\hline
{\tt val[0][]} & 0 & 0 & 0 & 0 & 0 & 0 & 0 & 17 \\
\hline
{\tt val[1][]} & 0 & 0 & 0 & 0 & 11& 0 & 16& 19 \\
\hline
{\tt val[2][]} & 1 & 4 & 7 & 10& 12& 14& 17& 20 \\
\hline
{\tt val[3][]} & 2 & 5 & 8 & 0 & 13& 15& 0 & 0  \\
\hline
{\tt val[4][]} & 3 & 6 & 9 & 0 & 0 & 0 & 0 & 0  \\
\hline
\end{tabular}
\\
\smallskip
\begin{tabular}{|l||c|c|c|c|c|}
\hline
{\tt offset[]} & -7 & -4 & 0 & 2 & 5 \\
\hline
\end{tabular}
\caption{An example of the DIA format: Example matrix is stored in the DIA format.}
\label{fig:sample_dia}
\end{center}
\end{figure}

\begin{figure}[t]
\begin{center}
\begin{lstlisting}
#pragma omp parallel private(i, k, off, is, ie)
{
	#pragma omp for simd
	for(i = 0; i < n; i++) {
		y[i] = 0;
	}
	
	for(k = 0; k < n_diags; k++) {
		off = offset[k];
		is = max(0, -off);
		ie = min(n, n-off);
		#pragma omp for simd
		for(i = is; i < ie; i++) {
			y[i] += val[k][i] * x[i+off];
		}
	}
}
\end{lstlisting}
\end{center}
\setlength\abovecaptionskip{-5pt}
\caption{The DIA kernel: a SpMV kernel using the DIA format.}
\label{fig:spmv_dia}
\end{figure}

\subsection{The HDC format}
In practical applications, we often deal with sparse matrices that are not fully diagonally structured. 
However, some of those matrices have partial diagonal structures, and combining the DIA and CSR format might provide better SpMV performance than simply using the CSR format. 
This is reason why we consider the HDC (Hybrid DIA-CSR) format~\citep{2014_yang}.
\par
The fundamental idea behind the HDC format is storing only nonzero diagonal lines that has sufficient number of nonzero elements in the DIA format and storing other nonzero elements in the CSR format. 
There are several rules of selecting nonzero diagonal lines to store in the DIA format, and in this paper, we employ the following rule: 
let $\nnz^{(d)}$ be the number of the nonzero elements in the diagonal line whose offset is $d \; (-n+1 \le d \le n-1)$, then we store a diagonal line in the DIA format if 
\begin{equation*}
	\frac{\nnz^{(d)}}{n} \ge \theta,
\end{equation*}
where $\theta \; (0 \le \theta \le 1)$ is a threshold that is given by a user.
\par
The HDC format consists of the arrays in both the DIA format ({\tt dia\_val[][]}, {\tt dia\_offset[]}) and the CSR format ({\tt csr\_val[]}, {\tt csr\_col\_ind[]}, {\tt csr\_row\_ptr[]}). 
Figure~\ref{fig:sample_matrix_in_hdc} shows an application of the HDC format to Example matrix, 
and Figure~\ref{fig:sample_hdc} illustrates how it is stored in the HDC format. 
In this example, we set $\theta = 0.6$, and two diagonal lines whose number of nonzero elements is larger than $4.8 (= 8 \times 0.6)$ are stored in the DIA format.
Compared with the example shown in Figure~\ref{fig:sample_dia}, the number of explicitly stored  zero elements is reduced in Figure~\ref{fig:sample_hdc}. 
\par
Figure~\ref{fig:spmv_hdc} outlines the SpMV kernel using the HDC format, which we call the HDC kernel. 
This is just a combination of the SpMV computation using the CSR format (the first part) and that using the DIA format (the second part). 

\begin{figure}[t]
\begin{center}
\[
\begin{pmatrix}
\bx{1} & 0 & \bx{2} & 0 & 0 & 3 & 0 & 0 \\
0 & \bx{4} & 0 & \bx{5} & 0 & 0 & 6 & 0 \\
0 & 0 & \bx{7} & 0 & \bx{8} & 0 & 0 & 9 \\
0 & 0 & 0 & \bx{10} & 0 & \bx{0} & 0 & 0 \\
11& 0 & 0 & 0 & \bx{12} & 0 & \bx{13} & 0 \\
0 & 0 & 0 & 0 & 0 & \bx{14} & 0 & \bx{15} \\
0 & 0 & 16& 0 & 0 & 0 & \bx{17} & 0 \\
18& 0 & 0 & 19& 0 & 0 & 0 & \bx{20} \\ 
\end{pmatrix}
\]
\caption{Selected diagonal lines in the application of the HDC format to Example matrix: underlined elements are stored in the DIA format.}
\label{fig:sample_matrix_in_hdc}
\end{center}
\end{figure}

\begin{figure}[t]
\begin{center}
\begin{tabular}{|l||c|c|c|c|c|c|c|c|}
\hline
{\tt dia\_val[0][]} & 1 & 4 & 7 & 10& 12& 14& 17& 20 \\
\hline
{\tt dia\_val[1][]} & 2 & 5 & 8 & 0 & 13& 15& 0 & 0  \\
\hline
\end{tabular}
\\
\smallskip
\begin{tabular}{|l||c|c|}
\hline
{\tt dia\_offset[]} & 0 & 2 \\
\hline
\end{tabular}
\\
\bigskip
\begin{tabular}{|l||c|c|c|c|c|c|c|}
\hline
{\tt csr\_val[]} & 3 & 6 & 9 & 11 & 16 & 18 & 19 \\
\hline
{\tt csr\_col\_ind[]} & 5 & 6 & 7 & 0 & 2 & 0 & 3 \\
\hline
\end{tabular}
\\
\smallskip
\begin{tabular}{|l||c|c|c|c|c|c|c|c|c|}
\hline
{\tt csr\_row\_ptr[]} & 0 & 1 & 2 & 3 & 3 & 4 & 4 & 5 & 7 \\
\hline
\end{tabular}
\caption{An example of the HDC format: Example matrix is stored in the HDC format.}
\label{fig:sample_hdc}
\end{center}
\end{figure}

\begin{figure}[t]
\begin{center}
\begin{lstlisting}
#pragma omp parallel private(i, k, s, off, is, ie)
{
	//SpMV using the CSR format
	#pragma omp for
	for(i = 0; i < n; i++) {
		s = 0;
		#pragma omp simd reduction(+:s)
		for(k = csr_row_ptr[i]; k < csr_row_ptr[i+1]; k++) {
			s += csr_val[k] * x[csr_col_ind[k]];
		}
		y[i] = s;
	}

	//SpMV using the DIA format
	for(k = 0; k < n_diags; k++) {
		off = dia_offset[k];
		is = max(0, -off);
		ie = min(n, n-off);
		#pragma omp for simd
		for(i = is; i < ie; i++) {
			y[i] += dia_val[k][i] * x[i+off];
		}
	}
}
\end{lstlisting}
\end{center}
\setlength\abovecaptionskip{-5pt}
\caption{The HDC kernel: a SpMV kernel using the HDC format.}
\label{fig:spmv_hdc}
\end{figure}

\section{Cache blocking for HDC-based SpMV}
\label{sec:blocking}
In this section, we introduce a cache blocking technique into the SpMV kernel using the HDC format. 
Next, we propose a modified version of the HDC format that is more suitable for cache blocking. 
The modified format is expected to more efficiently pick out partial diagonal structures than the HDC format.

\subsection{Motivation}
In the SpMV computation, the memory access cost for the input and output vectors is not negligible because $\nnz = c n \ll n^2$, where $c$ is a small number. 
It is well know that the data of the matrix cannot be reused in the matrix-vector multiplication (whichever a matrix is dense or sparse), 
but the data of the vectors can be reused. 
Therefore, we analyze the memory access patterns for the vectors in the SpMV kernel using each storage format. 
\par
\begin{figure}[t]
\begin{center}
\subfloat[SpMV using the CSR format]
{
	\includegraphics[scale=0.6]{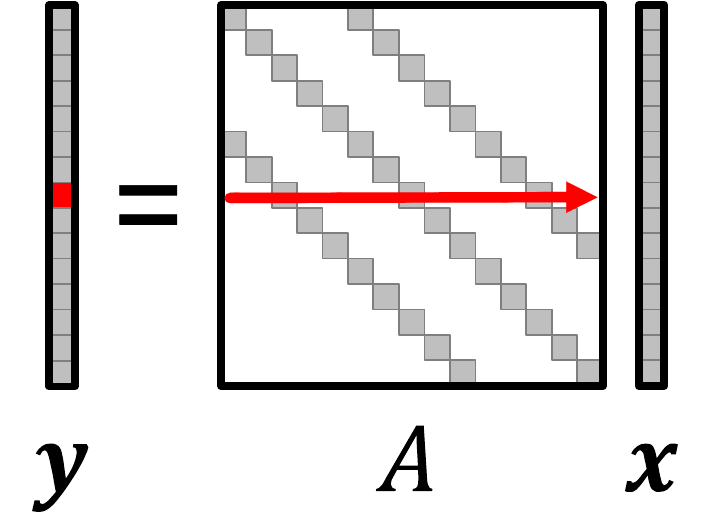}
	\label{fig:y_access_csr}
}
\subfloat[SpMV using the DIA format]
{
	\includegraphics[scale=0.6]{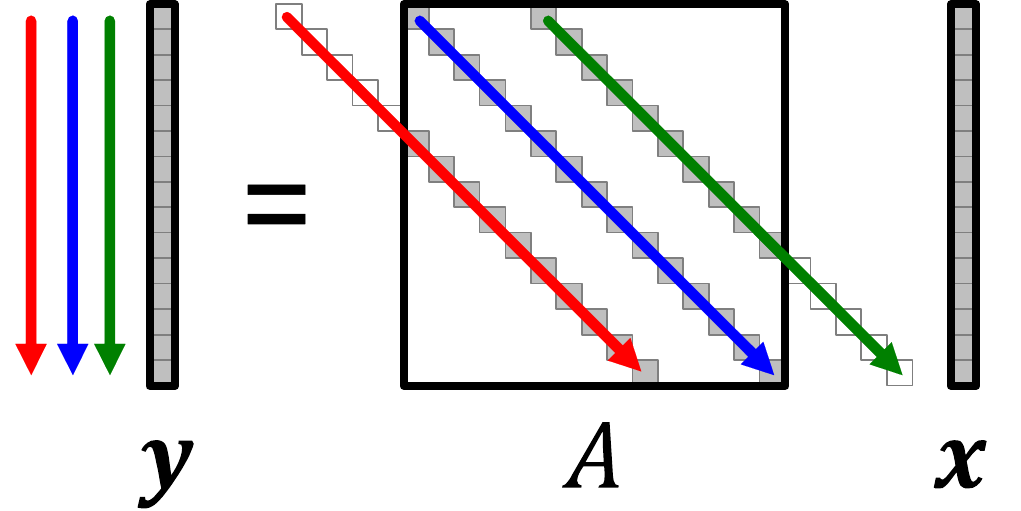}
	\label{fig:y_access_dia}
}

\end{center}
\caption{Memory access pattern for the output vector $\vec{y}$ in SpMV kernels.}
\label{fig:y_access}
\end{figure}

First, we consider the SpMV kernel using the CSR format. 
Figure~\ref{fig:spmv_csr} clearly shows that the memory access for the vector $\vec{y}$ has a good temporal locality (Figure~\ref{fig:y_access}~\subref{fig:y_access_csr}), and as the result, only one write is required for each element of $\vec{y}$ by using a register. 
On the other hand, the locality of the memory access for the vector $\vec{x}$ depends on the sparse pattern of the matrix $A$. 
\par
Next, we consider the SpMV kernel using the DIA format. 
From Figure~\ref{fig:spmv_dia}, we can find that the temporal locality of the memory access for both the vectors $\vec{x}$ and $\vec{y}$ is relatively low. 
If $n$ is sufficient large, it is difficult to efficiently use the cache memory; 
for each element of $\vec{y}$ (and $\vec{x}$), we need the access to the main memory ${\tt n\_diags}$ times (Figure~\ref{fig:y_access}~\subref{fig:y_access_dia}). 
\par
Compared with the SpMV kernel using the CSR format, the SpMV kernel using the DIA format has the following advantages:
\begin{itemize}
\item direct access to the the data of vector $\vec{x}$, 
\item less memory access cost, i.e. almost no access cost to the array that stores the index information such as {\tt col\_ind[]} in the CSR format.
\end{itemize}
However, the following issues are expected: 
\begin{itemize}
\item inefficient memory access to the vectors when $n$ is sufficiently large, 
\item additional cost (both in computation and memory access) for explicitly stored zero elements. 
\end{itemize}
In order to overcome the first issue, we introduce a cache blocking technique into the SpMV kernel using the DIA format.

\subsection{A cache blocking technique for SpMV using the HDC format}
We consider improving the the cache efficiency in the memory access for $\vec{y}$ in the SpMV kernel using the DIA format. 
Since the memory access pattern for $\vec{x}$ depends on the sparse pattern of $A$, we here discuss only the memory access for $\vec{y}$. 
\par
\begin{figure}[t]
\begin{center}
	\includegraphics[scale=0.6]{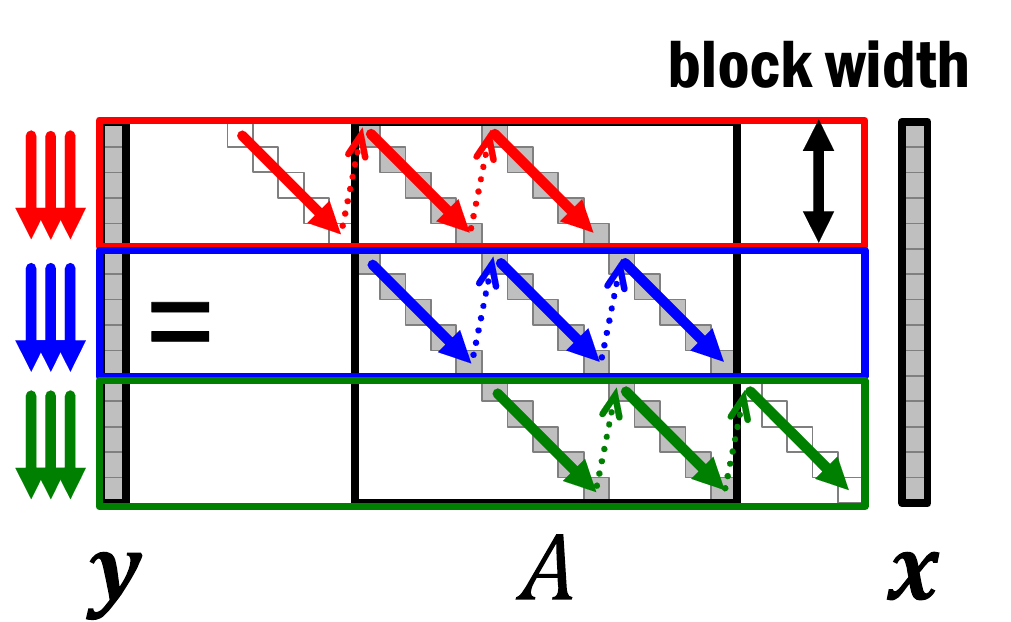}
\end{center}
\caption{A cache blocking technique for SpMV using the DIA format.}
\label{fig:cache_block_dia}
\end{figure}

In order to increase the temporal locality of the memory access for $\vec{y}$, we partition the rows into blocks and execute the SpMV computation block by block. 
Hereafter, we denote the block width as $bl$ and assume that $n$ can be divided by $bl$ for the simplicity. 
This technique is illustrated in Figure~\ref{fig:cache_block_dia}. 
By setting an appropriate $bl$, the data of $\vec{y}$ can remain on the cache memory through their related SpMV computation. 
On the other hand, this technique has a drawback that the memory access cost for {\tt offset[]} increases; 
roughly $n/bl$ times as many as the case without the cache blocking technique. 
The pseudo-code of the cache blocked SpMV kernel using the DIA format is shown in Figure~\ref{fig:spmv_dia_block} (the B-DIA kernel).  
\par
\begin{figure}[t]
\begin{center}
	\includegraphics[scale=0.6]{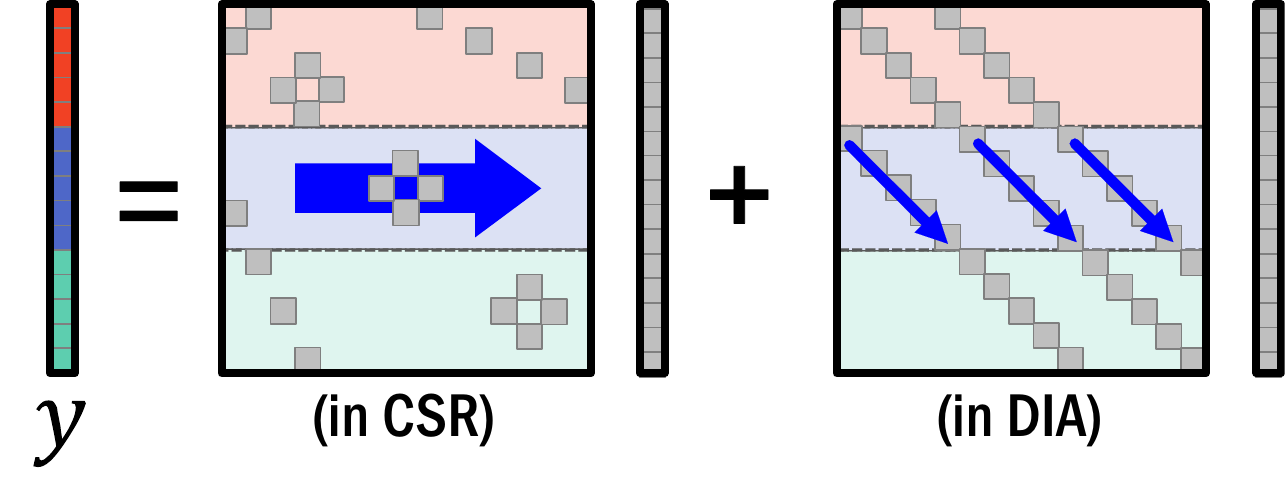}
\end{center}
\caption{A cache blocking technique for SpMV using the HDC format.}
\label{fig:cache_block_hdc}
\end{figure}

Then, we discuss cache blocking techniques for SpMV using the HDC format. 
Although a straightforward way is simply applying the above technique to the DIA-related part in SpMV using the HDC format, 
we employ an aggressive cache blocking technique, as illustrated in Figure~\ref{fig:cache_block_hdc}. 
In our technique, we combine the computation in SpVM using the CSR format and that in the DIA format so as to further improve the the temporal locality of the memory access for $\vec{y}$. 
For each block, we first execute SpMV using the CSR format and then do that using the DIA format. 
Figure~\ref{fig:spmv_hdc_block} gives the pseudo-code of the SpMV kernel using the HDC format with the cache blocking technique (the B-HDC kernel), 
where {\tt n\_blocks} is the number of blocks (i.e. ${\tt n\_blocks} = n/bl$).

\begin{figure}[t]
\begin{center}
\begin{lstlisting}
#pragma omp parallel for private(ib, i, k, off, is, ie)
for(ib = 0; ib < n_blocks; ib++) {
	#pragma omp simd
	for(i = ib*bl; i < (ib+1)*bl; i++) {
		y[i] = 0;
	}

	for(k = 0; k < n_diags; k++) {
		off = dia_offset[k];
		is = max(ib*bl, -off);
		ie = min((ib+1)*bl, n-off);
		#pragma omp simd
		for(i = is; i < ie; i++) {
			y[i] += dia_val[k][i] * x[i+off];
		}
	}	
}
\end{lstlisting}
\end{center}
\setlength\abovecaptionskip{-5pt}
\caption{The B-DIA kernel: a cache blocked SpMV kernel using the DIA format.}
\label{fig:spmv_dia_block}
\end{figure}

\begin{figure}[t]
\begin{center}
\begin{lstlisting}
#pragma omp parallel for private(ib, i, k, s, off, is, ie)
for(ib = 0; ib < n_blocks; ib++) {
	//SpMV using the CSR format
	for(i = ib*bl; i < (ib+1)*bl; i++) {
		s = 0;
		#pragma omp simd reduction(+:s)
		for(k = csr_row_ptr[i]; k < csr_row_ptr[i+1]; k++) {
			s += csr_val[k] * x[csr_col_ind[k]];
		}
		y[i] = s;
	}

	//SpMV using the DIA format
	for(k = 0; k < n_diags; k++) {
		off = dia_offset[k];
		is = max(ib*bl, -off);
		ie = min((ib+1)*bl, n-off);
		#pragma omp simd
		for(i = is; i < ie; i++) {
			y[i] += dia_val[k][i] * x[i+off];
		}
	}	
}
\end{lstlisting}
\end{center}
\setlength\abovecaptionskip{-5pt}
\caption{The B-HDC kernel: a cache blocked SpMV kernel using the HDC format.}
\label{fig:spmv_hdc_block}
\end{figure}

\subsection{The modified HDC format}
Taking the cache blocking technique described in the previous subsection into account, we modify the HDC format, which we call the M-HDC format. 
Since the computation is performed block by block in the cache blocked SpMV computation using the HDC format, 
it seems to be natural to select elements to store in the DIA format for each block rather than for a whole matrix. 
\par
We call a set of elements that belong to a same diagonal line and a same block \textit{partial diagonal line}, 
and define $\tilde{N}_\textrm{nz}^{(d, ib)}$ as the number of nonzero elements in the partial diagonal block whose offset is $d$ and that belongs to the $ib$-th block. 
Then, we store a partial diagonal line in the DIA format if 
\begin{equation*}
	\frac{\tilde{N}_\textrm{nz}^{(d, ib)}}{bl} \ge \theta. 
\end{equation*}
Figure~\ref{fig:sample_matrix_in_mhdc} gives an example of applying the M-HDC format to Example matrix, 
where $bl = 4$ and $\theta = 0.6$, which means partial diagonal lines that have more than $2.4 (= 4 \times 0.6)$ nonzero elements are stored in the DIA format. 
Figure~\ref{fig:sample_mhdc} shows how it is stored in the M-HDC format. 
Here, the additional array {\tt dia\_ptr[]} is introduced to store the partition information for {\tt dia\_val[][]} and {\tt dia\_col\_offset[]}. 
\par
Figure~\ref{fig:spmv_mhdc} presents the pseudo-code of the SpMV kernel using the M-HDC format, which we call the M-HDC kernel. 
Compared with the B-HDC kernel, the rage of {\tt k} for accessing {\tt dia\_val[k][]} slightly changes; 
the range is obtained via {\tt dia\_ptr[]} instead of 0 through {\tt n\_diags}. 
The memory access pattern for $\vec{y}$ in the M-HDC kernel is essentially the same as that in the the B-HDC kernel excepting the effect by the differences in the selection of (partial) diagonal lines, 
and the M-HDC format is expected to more efficiently exploit partial diagonal structures in a matrix than the original HDC format; 
it might store more nonzero elements and less zero elements in the DIA part. 
Therefore, for certain sparse matrices, the M-HDC kernel is expected to provide higher performance than the other SpMV kernels already mentioned in this paper. 

\begin{figure}[t]
\begin{center}
\[
\begin{pmatrix}
\bx{1} & 0 & \bx{2} & 0 & 0 & \bx{3} & 0 & 0 \\
0 & \bx{4} & 0 & \bx{5} & 0 & 0 & \bx{6} & 0 \\
0 & 0 & \bx{7} & 0 & \bx{8} & 0 & 0 & \bx{9} \\
0 & 0 & 0 & \bx{10} & 0 & \bx{0} & 0 & 0 \\
\hline
\bx{11} & 0 & 0 & 0 & \bx{12} & 0 & 13 & 0 \\
0 & \bx{0} & 0 & 0 & 0 & \bx{14} & 0 & 15 \\
0 & 0 & \bx{16} & 0 & 0 & 0 & \bx{17} & 0 \\
18& 0 & 0 & \bx{19} & 0 & 0 & 0 & \bx{20} \\ 
\end{pmatrix}
\]
\caption{Selected diagonal lines in the application of the M-HDC format to Example matrix ($bl = 4$): underlined elements are stored in the DIA format.}
\label{fig:sample_matrix_in_mhdc}
\end{center}
\end{figure}

\begin{figure}[t]
\begin{center}
\begin{tabular}{|l||c|c|c|c|}
\hline
{\tt dia\_val[0][]} & 1 & 4 & 7 & 10\\
\hline
{\tt dia\_val[1][]} & 2 & 5 & 8 & 0 \\
\hline
{\tt dia\_val[2][]} & 3 & 6 & 9 & 0 \\
\hline
{\tt dia\_val[3][]} & 11& 0 & 16& 19\\
\hline
{\tt dia\_val[4][]} & 12& 14& 17& 20\\
\hline
\end{tabular}
\\
\smallskip
\begin{tabular}{|l||c|c|c|c|c|}
\hline
{\tt dia\_col\_offset[]} & 0 & 2 & 5 & -4 & 0\\
\hline
\end{tabular}
\\
\smallskip
\begin{tabular}{|l||c|c|c|}
\hline
{\tt dia\_ptr[]} & 0 & 3 & 5\\
\hline
\end{tabular}
\\
\bigskip
\begin{tabular}{|l||c|c|c|}
\hline
{\tt csr\_val[]} & 13 & 15 & 18\\
\hline
{\tt csr\_col\_ind[]} & 6 & 7 & 0 \\
\hline
\end{tabular}
\\
\smallskip
\begin{tabular}{|l||c|c|c|c|c|c|c|c|c|}
\hline
{\tt csr\_row\_ptr[]} & 0 & 0 & 0 & 0 & 0 & 1 & 2 & 2 & 3 \\
\hline
\end{tabular}
\caption{An example of the M-HDC format ($bl = 4$): Example matrix is stored in the M-HDC format..}
\label{fig:sample_mhdc}
\end{center}
\end{figure}

\begin{figure}[t]
\begin{center}
\begin{lstlisting}
#pragma omp parallel for private(ib, i, k, s, off, is, ie)
for(ib = 0; ib < n_blocks; ib++) {
	//SpMV using the CSR format
	for(i = ib*bl; i < (ib+1)*bl; i++) {
		s = 0;
		#pragma omp simd reduction(+:s)
		for(k = csr_row_ptr[i]; k < csr_row_ptr[i+1]; k++) {
			s += csr_val[k] * x[csr_col_ind[k]];
		}
		y[i] = s;
	}

	//SpMV using the DIA format
	for(k = dia_ptr[ib]; k < dia_ptr[ib+1]; k++) {
		off = dia_offset[k];
		is = max(ib*bl, -off);
		ie = min((ib+1)*bl, n-off);
		#pragma omp simd
		for(i = is; i < ie; i++) {
			y[i] += dia_val[k][i] * x[i+off];
		}
	}	
}
\end{lstlisting}
\end{center}
\setlength\abovecaptionskip{-5pt}
\caption{The M-HDC kernel: a SpMV kernel using the M-HDC format.}
\label{fig:spmv_mhdc}
\end{figure}

\section{Analysis based on performance models}
\label{sec:model}
In this section, we provide a theoretical analysis based on performance models. 
Our objective here is to show the efficiency of the cache blocking technique introduced into the SpMV kernels using the DIA format. 
We first give a preliminary for performance modeling. 
Next, supposing that an input matrix is ideal for the DIA format, namely so-called stencil matrices, 
we analyze the performance of the SpMV kernel using the DIA format in the cases with and without the cache blocking technique. 
Finally, for a general matrices, we analyze the performance of the cache blocked SpMV kernel using the HDC format and that using the M-HDC format.

\subsection{Preliminary}
According to the paper by \citet{2014_kreutzer}, we construct performance models for SpMV kernels in the following approach. 
Let $P^\textrm{(*)}$ be the performance (i.e. Flop/s) of a SpMV kernel ($*$ represents the kernel's name), 
and $T^\textrm{(*)}$ be the execution time (i.e. sec) of the kernel. 
Then, we define 
\begin{equation}
	P^\textrm{(*)} \coloneqq \frac{2 \nnz}{T^\textrm{(*)}}. \label{eq:def_P}
\end{equation}
\par
In this analysis, we focus the case that an input matrix is sufficiently large, 
that is, the whole data of the matrix and vectors in SpMV cannot be accommodated in the cache memory. 
Accordingly, we assume that the execution time of SpMV is determined by the amount of data transferred from/to the main memory. 
Letting $V^\textrm{(*)}$ be the amount (i.e. byte) of the data, we write
\begin{equation}
	T^\textrm{(*)} = \frac{V^\textrm{(*)}}{w_\textrm{mem}}, \label{eq:def_T}
\end{equation}
where $w_\textrm{mem}$ is the effective bandwidth (i.e. byte/s) of the main memory. 
We assume that $w_\textrm{mem}$ does not depend on SpMV kernels. 
In this modeling, we ignore the latency in the main memory access and all of the cost in the data movement from/to the cache memory. 
\par
From Equations~\ref{eq:def_P} and \ref{eq:def_T}, we can obtain the relative performance, i.e. the speedup, of the kernel A over B as 
\begin{equation}
	\frac{P^\textrm{(A)}}{P^\textrm{(B)}} = \frac{T^\textrm{(B)}}{T^\textrm{(A)}} 
	= \frac{V^\textrm{(B)}}{V^\textrm{(A)}}
	= 1 + \frac{V^\textrm{(B)} - V^\textrm{(A)}}{V^\textrm{(A)}}. \label{eq:speedup}
\end{equation}
\par
Now, we explain how to model $V^\textrm{(*)}$ for each SpMV kernel. 
We break down $V^\textrm{(*)}$ as
\begin{equation}
	V^\textrm{(*)} = V^\textrm{(*)}_A + V^\textrm{(*)}_{\vec{x}} + V^\textrm{(*)}_{\vec{y}},
\end{equation}
where $V^\textrm{(*)}_A$, $V^\textrm{(*)}_{\vec{x}}$, and $V^\textrm{(*)}_{\vec{y}}$ are the amount of the data corresponding to the matrix $A$, the right-hand side vector $\vec{x}$, and the left-side vector $\vec{y}$, respectively. 
Since the data of $A$ and $\vec{y}$ are continuously accessed in all kernels, 
$V^\textrm{(*)}_A$ and $V^\textrm{(*)}_{\vec{y}}$ are obtained by simply counting the number of the data access. 
On the other hand, the data access for $\vec{x}$ is generally discontinuous in the CSR kernel but continuous in the DIA kernel. 
Hence, $V^\textrm{(*)}_{\vec{x}}$ is need to be carefully modeled. 

\subsection{Analysis for stencil matrices}
\label{sec:analysis_stencil}
First, we analyze the performance in an ideal case for the DIA format, precisely, the case where an input matrix is perfectly diagonally structured. 
We consider a matrix appearing in the finite difference analysis, which is so-called stencil matrix. 
For the simplicity, we assume that all elements in $\ndiag (\in \mathbb{N})$ diagonal lines are nonzero but other elements are zero, which implies 
\begin{equation}
	\nnz \simeq \ndiag n.\label{eq:asm_diag}
\end{equation}

\subsubsection{Model for the CSR kernel}
From Figure~\ref{fig:spmv_csr}, we can write
\begin{align}
	V^\textrm{(CSR)}_A 	&\simeq \underbrace{b_\textrm{fp} \nnz}_{\text{val}} + \underbrace{b_\textrm{int} \nnz}_{\text{col\_ind}} + \underbrace{b_\textrm{int} n}_{\text{row\_ptr}}, \nonumber \\
	&\simeq b_\textrm{fp}  (\ndiag + b \ndiag + b) n, \nonumber \\
	V^\textrm{(CSR)}_{\vec{x}} &\simeq b_\textrm{fp} \gamma^\textrm{(CSR)} \nnz \nonumber \\
	&\simeq b_\textrm{fp} \gamma^\textrm{(CSR)} \ndiag n, \nonumber \\
	V^\textrm{(CSR)}_{\vec{y}} &\simeq b_\textrm{fp} n, \nonumber
\end{align}
where $b_\textrm{fp}$ and $b_\textrm{int}$ are the data size for a floating point number and an integer number, respectively, 
and $b$ is their ratio:
\begin{equation}
	b \coloneqq \frac{b_\textrm{int}}{b_\textrm{fp}}.
\end{equation}
For example, $b_\textrm{fp} = 8$ in FP64 (double precision), and $b_\textrm{int} = 4$ in INT32. 
\par
In the above modeling of $V^\textrm{(CSR)}_{\vec{x}}$, 
we introduce $\gamma^\textrm{(CSR)}$ as a parameter that represents the effect of the cache memory in the access for $\vec{x}$. 
$\gamma^\textrm{(CSR)}$ belongs to the range of
\begin{equation}
	1 \ge \gamma^\textrm{(CSR)} \ge \frac{n}{\nnz} \simeq \frac{1}{\ndiag} \label{eq:gamma_csr}. 
\end{equation}
Here, $\gamma^\textrm{(CSR)} = 1$ means the worst situation, in which the main memory is accessed every time, 
and $\gamma^\textrm{(CSR)} \simeq 1/\ndiag$ means the best situation, in which each element of $\vec{x}$ is loaded only once from the main memory 
(and from the cache memory in other times). 
On a current processor, all of the data in the same cache line is ordinarily together loaded from the main memory,
and discontinuous access thus tends to cause an excess transfer due to unnecessary data in the same cache line. 
We need to take this into account when modeling $V^\textrm{(CSR)}_{\vec{x}}$ for a general sparse matrix, 
however it is ignorable under the assumption with stencil matrices; the next element of an element of $\vec{x}$ is used in the calculation for the next row. 

\subsubsection{Model for the DIA kernel}
Under the assumption that $n$ is sufficient large, 
we presume every access to $\vec{x}$ and $\vec{y}$ require the data transfer from/to the main memory in the DIA kernel.
Thus, from Figure~\ref{fig:spmv_dia}, we can write
\begin{align}
	V^\textrm{(DIA)}_A &\simeq b_\textrm{fp} \ndiag n, \nonumber \\
	V^\textrm{(DIA)}_{\vec{x}} &\simeq b_\textrm{fp} \ndiag n, \nonumber \\
	V^\textrm{(DIA)}_{\vec{y}} &\simeq b_\textrm{fp} (\underbrace{1}_{\mathclap{\text{y[i]=0}}} + \underbrace{2 \ndiag}_{\mathclap{\text{y[i]+=$\cdots$}}}) n. \nonumber
\end{align}

\subsubsection{Model for the B-DIA kernel}
With an appropriate block length, 
we suppose that only the final result of $\vec{y}$ is stored to the main memory;
all of the intermediate results are loaded from (or stored to) the cache memory. 
The cache hit rate in loading $\vec{x}$ is also expected to be improved, 
which is modeled in the same way as in the CSR kernel, with the parameter $\gamma^\textrm{(B-DIA)}$. 
Taking these considerations into account, from Figure~\ref{fig:spmv_dia_block}, we can write
\begin{align}
	V^\textrm{(B-DIA)}_A &\simeq b_\textrm{fp} \ndiag n, \nonumber \\
	V^\textrm{(B-DIA)}_{\vec{x}} &\simeq b_\textrm{fp} \gamma^\textrm{(B-DIA)} \ndiag n, \nonumber \\
	V^\textrm{(B-DIA)}_{\vec{y}} &\simeq b_\textrm{fp} n, \nonumber
\end{align}
where 
\begin{equation}
	1 \ge \gamma^\textrm{(B-DIA)} \gtrsim \frac{1}{\ndiag}. \label{eq:asm_gamma_b_dia}
\end{equation}

\subsubsection{Comparison between the DIA and CSR kernels}
We have 
\begin{equation}
	\frac{V^\textrm{(CSR)} - V^\textrm{(DIA)}}{V^\textrm{(DIA)}}
	\simeq \frac{b(\ndiag+1) - (3-\gamma^\textrm{(CSR)})\ndiag}{4\ndiag + 1}. \label{eq:dia_over_csr}
\end{equation}
From Equation~\ref{eq:gamma_csr} and $\ndiag \ge 1$, we can bound
\begin{equation}
	\frac{2b}{5} \ge \frac{b(\ndiag+1)}{4\ndiag+1}, 
\end{equation}
and
\begin{equation}
	\frac{(3-\gamma^\textrm{(CSR)})\ndiag}{4\ndiag+1} \ge \frac{2}{5}.
\end{equation}
Thus, from Equation~\ref{eq:speedup}, we can obtain
\begin{equation}
	\frac{3 + 2b}{5} \gtrsim \frac{P^\textrm{(DIA)}}{P^\textrm{(CSR)}}.
\end{equation}
Generally, we can assume
\begin{equation}
	b_\textrm{fp} \ge b_\textrm{int} \quad \Leftrightarrow \quad 1 \ge b,   
\end{equation}
and under this assumption, we have 
\begin{equation}
	1 \gtrsim \frac{P^\textrm{(DIA)}}{P^\textrm{(CSR)}}. \label{eq:dia_over_csr_max}
\end{equation}
This result indicates that the DIA kernel (i.e. straightforward use of the DIA format in SpMV) cannot provide the performance improvement over the CSR kernel even if the input matrix is ideal for the DIA formant. 
\par
The first term in the numerator in Equation~\ref{eq:dia_over_csr} represents the positive effect in the DIA kernel; 
no array (excepting the small array {\tt offset[]}) is necessary for storing the position of the nonzero elements of $A$ in the DIA format. 
On the other hand, the second term represents the negative effect; 
the data of $\vec{x}$ and $\vec{y}$ need to be moved from/to the main memory much more than in the CSR kernel. 
The result in Equation~\ref{eq:dia_over_csr_max} tells us that the negative effect is usually larger than the positive effect. 

\subsubsection{Comparison between the B-DIA and CSR kernels}
We first consider the difference between $\gamma^\textrm{(CSR)}$ and $\gamma^\textrm{(B-DIA)}$. 
Regardless the B-DIA or CSR kernel, the cache hit rate in the access for $\vec{x}$ is high if diagonal lines are close to each other, and low if far from. 
From this fact, we assume that the rate depends only on the sparse pattern (i.e. the position of each diagonal lines), which implies 
\begin{equation}
	\gamma^\textrm{(CSR)} \simeq \gamma^\textrm{(B-DIA)} \simeq \gamma, \quad 1 \ge \gamma \gtrsim \frac{1}{\ndiag}. \label{eq:asm_gamma}
\end{equation}
\par
Then, we compare the B-DIA kernel with the CSR kernel; we have 
\begin{align}
	\frac{V^\textrm{(CSR)} - V^\textrm{(B-DIA)}}{V^\textrm{(B-DIA)}} 
	&\simeq \frac{b(\ndiag+1) + (\gamma^\textrm{(CSR)} - \gamma^\textrm{(B-DIA)})\ndiag}{(1+\gamma^\textrm{(B-DIA)})\ndiag+1} \nonumber \\
	&\simeq \frac{b(\ndiag+1)}{(1+\gamma)\ndiag+1}. \label{eq:b_dia_over_csr}
\end{align}
From Equation~\ref{eq:asm_gamma} and $\ndiag \ge 1$, we can bound 
\begin{equation}
	b > \frac{b(\ndiag+1)}{(1+\gamma)\ndiag+1} > \frac{b}{2}. 
\end{equation}
Thus, we can obtain
\begin{equation}
	1 + b \gtrsim \frac{P^\textrm{(B-DIA)}}{P^\textrm{(CSR)}} \gtrsim 1 + \frac{b}{2}. \label{eq:b_dia_over_csr_bound}
\end{equation}
This result assures that the B-DIA kernel provides a speedup over the CSR kernel; the speedup increases as $b$ becomes large. 
\par
Comparing Equation~\ref{eq:b_dia_over_csr} with \ref{eq:dia_over_csr}, 
only the positive effect provided by the DIA format remains; 
the negative effect in the DIA kernel is removed by the cache blocking technique introduced in the B-DIA kernel. 

\subsubsection{Comparison between the B-DIA and DIA kernels}
We also compare the B-DIA kernel with the DIA kernel; we have 
\begin{equation}
	\frac{V^\textrm{(DIA)} - V^\textrm{(B-DIA)}}{V^\textrm{(B-DIA)}}
	\simeq \frac{(3-\gamma^\textrm{(B-DIA)})\ndiag}{(1+\gamma^\textrm{(B-DIA)})\ndiag+1}. \label{eq:b_dia_over_dia}
\end{equation}
From Equation~\ref{eq:asm_gamma_b_dia} and $\ndiag \ge 1$, we can bound 
\begin{equation}
	3 > \frac{(3-\gamma^\textrm{(B-DIA)})\ndiag}{(1+\gamma^\textrm{(B-DIA)})\ndiag+1} \ge \frac{2}{3}, 
\end{equation}
and we can obtain
\begin{equation}
	4 \gtrsim \frac{P^\textrm{(B-DIA)}}{P^\textrm{(DIA)}} \gtrsim \frac{5}{3}. 
\end{equation}
This result clearly shows the impact of the cache blocking technique in the B-DIA kernel. 

\subsection{Analysis for general matrices}
Next, we analyze the performance of the B-HDC and M-HDC kernels for general matrices. 
Hereafter, we denote the average of the number of nonzero elements per row as
\begin{equation}
	c \coloneqq \frac{\nnz}{n}.
\end{equation}

\subsubsection{Model for the B-HDC kernel}
Let $\beta$ be the rate of the nonzero elements still stored in the CSR part in the HDC format, which we call the CSR rate. 
We define the filling rate of the DIA part as 
\begin{equation}
	\alpha \coloneqq \frac{(1 - \beta) \nnz}{\ndiag n} = \frac{(1-\beta)c}{\ndiag}, 
\end{equation}
where $\ndiag$ is the number of diagonal lines in the DIA part. 
\par
Then, we can write
\begin{align}
	V^\textrm{(B-HDC)}_A 
	&\simeq \underbrace{b_\textrm{fp} \beta \nnz}_{\mathclap{\text{csr\_val}}} + \underbrace{b_\textrm{int} \beta \nnz}_{\mathclap{\text{csr\_col\_ind}}} + \underbrace{b_\textrm{int} n}_{\mathclap{\text{csr\_row\_ptr}}} + \underbrace{b_\textrm{fp} \ndiag n}_{\mathclap{\text{dia\_val}}} \nonumber \\
	&= b_\textrm{fp} \left ( \beta (c + bc) + b + \frac{(1 - \beta)c}{\alpha} \right ) n.
\end{align}
Since modeling $V^\textrm{(*)}_{\vec{x}}$ is difficult for a general matrix, we simply write
\begin{align}
	V^\textrm{(CSR)}_{\vec{x}} &\simeq b_\textrm{fp} v_{\vec{x}} n, \\
\intertext{and}
	V^\textrm{(B-HDC)}_{\vec{x}} &\simeq b_\textrm{fp} (v_{\vec{x}} + \Delta v_{\vec{x}}) n, \; \Delta v_{\vec{x}} \ge 0, 
\end{align}
where $\Delta v_{\vec{x}}$ represents the difference from the CSR kernel.
The reason why we assume $\Delta v_{\vec{x}} \ge 0$ is that additional access will be required in the B-HDC kernel due to the zero elements explicitly stored in the DIA part. 
On the other hand, we can write
\begin{equation}
	V^\textrm{(B-HDC)}_{\vec{y}} \simeq b_\textrm{fp} n.
\end{equation}

\subsubsection{Comparison between the B-HDC and CSR kernels}
Our target here is deriving the upper bound of the performance improvement of the B-HDC kernel over the CSR kernel. 
In this situation, letting $\Delta v_{\vec{x}} = 0$ (for the simplicity) is acceptable. 
Then, we have
\begin{align}
	\frac{V^\textrm{(CSR)} - V^\textrm{(B-HDC)}}{V^\textrm{(B-HDC)}}
	&\simeq \frac{b(1-\beta)c - (1 - \beta) \left(\dfrac{1}{\alpha} - 1 \right)c}{\beta (c + bc) + b + \dfrac{(1 - \beta)c}{\alpha} + v_{\vec{x}} + 1}.  \label{eq:b_hdc_over_csr}
\end{align}
Since $v_{\vec{x}} \ge 1$, $\alpha \le 1$, and $\beta \ge 0$, we can bound
\begin{equation}
	b > \frac{b(1-\beta)c - (1 - \beta) \left(\dfrac{1}{\alpha} - 1 \right)c}{\beta (c + bc) + b + \dfrac{(1 - \beta)c}{\alpha} + v_{\vec{x}} + 1}, 
\end{equation}
where the upper bound is obtained when $v_{\vec{x}}=1$, $\alpha = 1$, $\beta = 0$, and $c \to \infty$. 
Finally, we have 
\begin{equation}
	1 + b > \frac{P^\textrm{(B-HDC)}}{P^\textrm{(CSR)}}.
\end{equation}
This result is consistent with the upper bound in Equation~\ref{eq:b_dia_over_csr_bound}.
\par
We give an interpretation for Equation~\ref{eq:b_hdc_over_csr}; 
the first term in the numerator represents the amount of the reduced index data in the CSR part in the HDC format, 
and the second term represents the amount of additionally required value data (for zero elements in diagonal lines) in the DIA part of the HDC format.
From Equation~\ref{eq:b_hdc_over_csr}, we can find that 
\begin{equation}
	b(1-\beta)c \ge (1 - \beta)\bigl(\dfrac{1}{\alpha} - 1 \bigr)c
	\; \Leftrightarrow \; \alpha \ge \frac{1}{b+1} \label{eq:cond_alpha}
\end{equation}
is required for the efficient use of the B-HDC kernel; 
for example, if we use FP64 and INT32, 
\begin{equation}
	\alpha \ge \frac{1}{\dfrac{1}{2} + 1} = \frac{2}{3}
\end{equation}
is required. 

\subsubsection{Model for the M-HDC kernel}
Now, we consider the M-HDC kernel. 
Let $\tilde{\beta}$ be the CSR rate in the M-HDC format.  
Instead of $\alpha$ in the B-HDC kernel, we introduce the average filling rate as 
\begin{equation}
	\tilde{\alpha} \coloneqq \frac{(1 - \tilde{\beta}) \nnz}{\sum_{ib=0}^{\nbl-1}\ndiag^{(ib)}bl} = \frac{(1 - \tilde{\beta}) cn}{\sum_{ib=0}^{\nbl-1}\ndiag^{(ib)}bl},  
\end{equation}
where $\nbl (= n / bl)$ and $\ndiag^{(ib)}$ are the number of row blocks and the number of diagonal lines in the $ib$-th row block, respectively.
Then, we can write 
\begin{align}
	V^\textrm{(M-HDC)}_A 
	&\simeq \underbrace{b_\textrm{fp} \beta \nnz}_{\mathclap{\text{csr\_val}}} + \underbrace{b_\textrm{int} \beta \nnz}_{\mathclap{\text{csr\_col\_ind}}} + \underbrace{b_\textrm{int} n}_{\mathclap{\text{csr\_row\_ptr}}} + \underbrace{b_\textrm{fp} \sum_{ib=0}^{\nbl-1}\ndiag^{(ib)}bl }_{\mathclap{\text{dia\_val}}} \nonumber \\
	&= b_\textrm{fp} \left ( \beta (c + bc) + b + \frac{(1 - \tilde{\beta})c}{\tilde{\alpha}} \right ) n. 
\end{align}
As well as in the case of the B-HDC kernel, we write 
\begin{equation}
	V^\textrm{(M-HDC)}_{\vec{x}} \simeq b_\textrm{fp} (v_{\vec{x}} + \Delta \tilde{v}_{\vec{x}})n, \; \Delta \tilde{v}_{\vec{x}} \ge 0, 
\end{equation}
where $\Delta \tilde{v}_{\vec{x}}$ is the difference from the CSR kernel.
We also write
\begin{equation}
	V^\textrm{(M-HDC)}_{\vec{y}} \simeq b_\textrm{fp} n.
\end{equation}

\subsubsection{Comparison between the M-HDC and CSR kernels}
The only difference between the B-HDC and M-HDC kernels is the definition of the filling rate and the CSR rate. 
Therefore, by replacing $\alpha$ and $\beta$ in the B-HDC kernel with $\tilde{\alpha}$ and $\tilde{\beta}$, respectively, we can obtain the same results for the M-HDC kernel as for the B-HDC kernel. 
\par
We expect that the M-HDC format more efficiently picks up partial diagonal structures, 
which means $\tilde{\alpha} > \alpha$ (less zero elements in the DIA part) and $\tilde{\beta} < \beta$ (more nonzero elements in the DIA part). 
From Equation~\ref{eq:b_hdc_over_csr}, we can observe that the performance improvement over the CSR kernel becomes larger in this situation, 
which indicates the superiority of the M-HDC kernel to the B-HDC kernel.

\subsubsection{Examples}
\begin{figure*}[t]
\begin{center}
\includegraphics[scale=0.4]{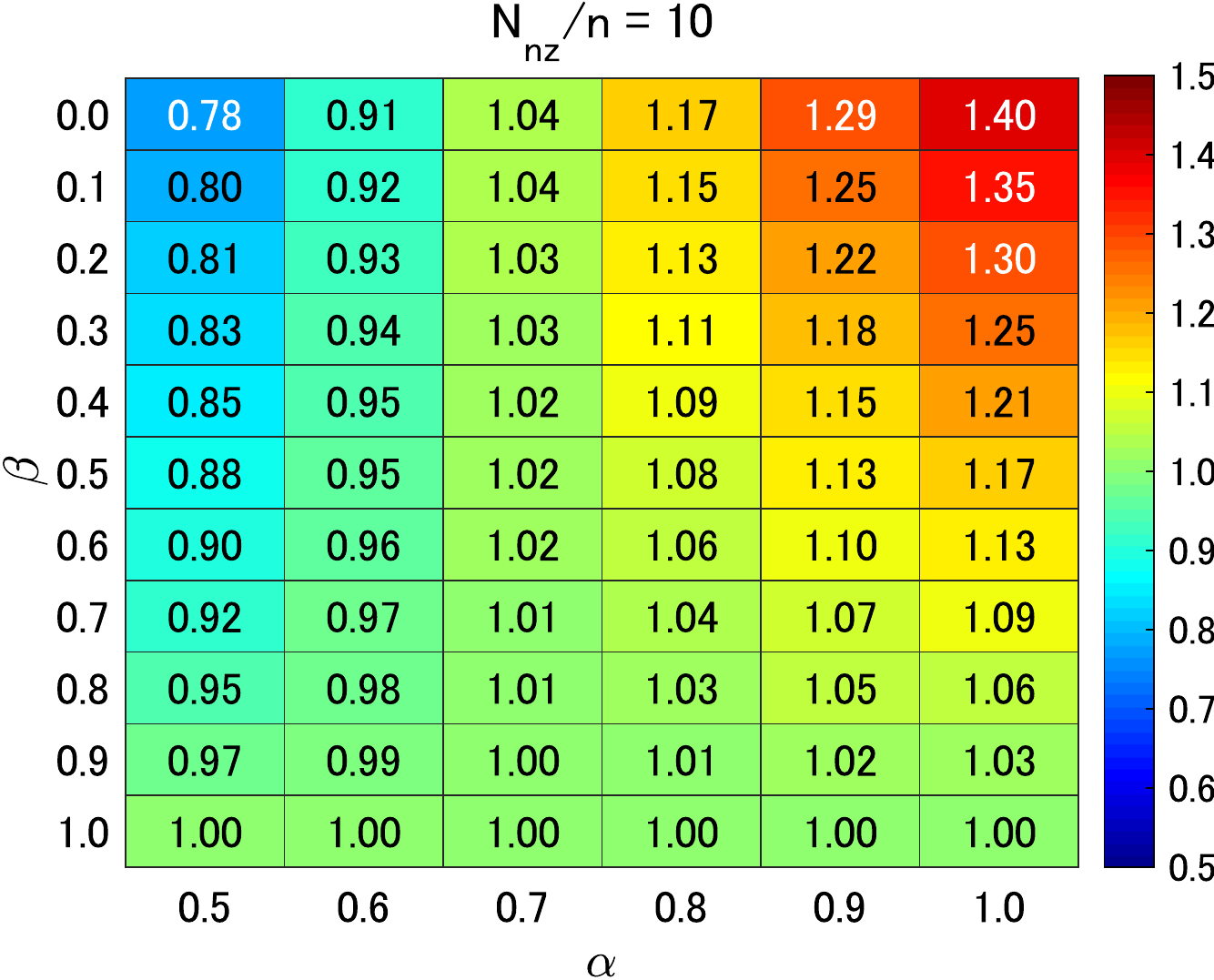}
\includegraphics[scale=0.4]{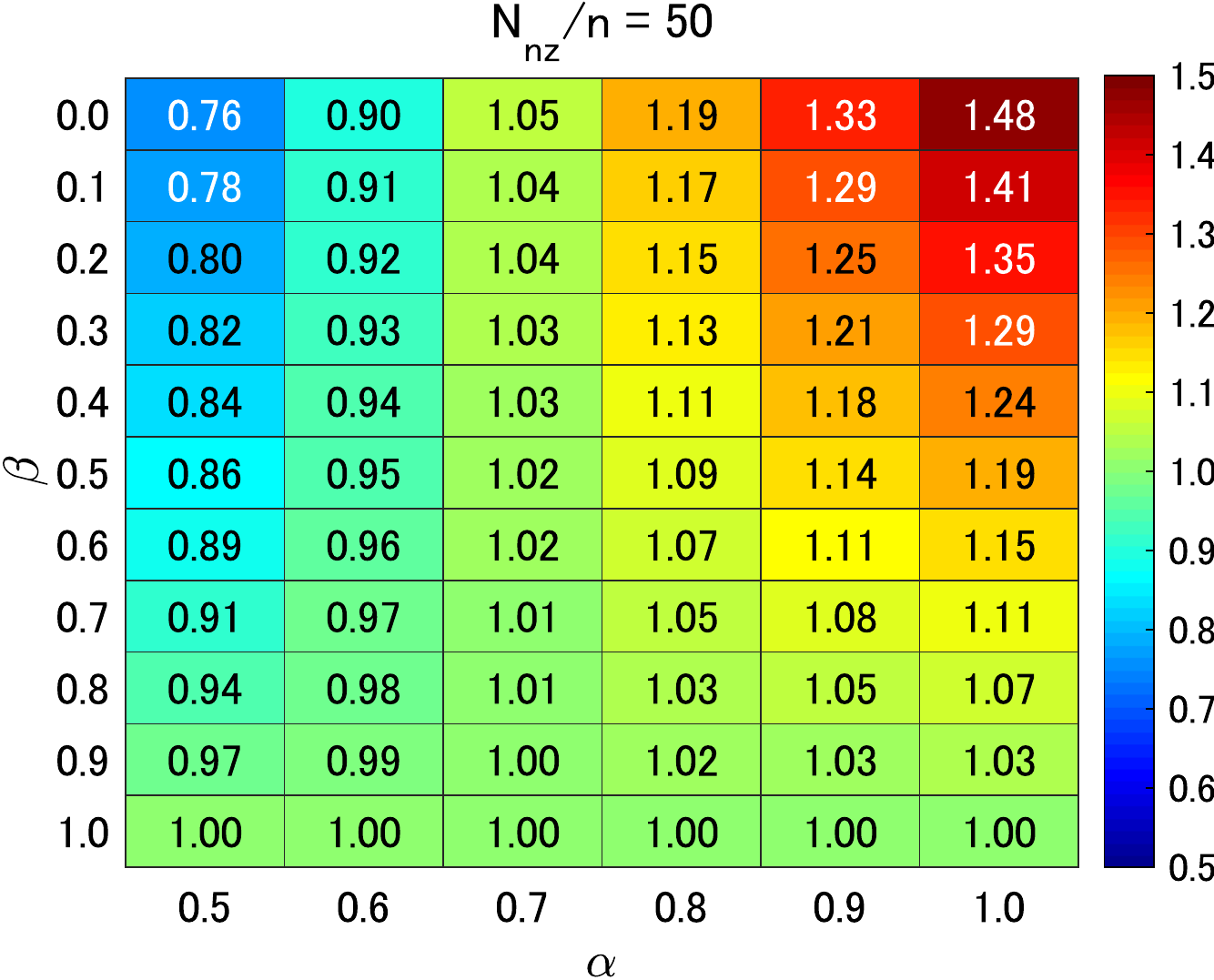}
\includegraphics[scale=0.4]{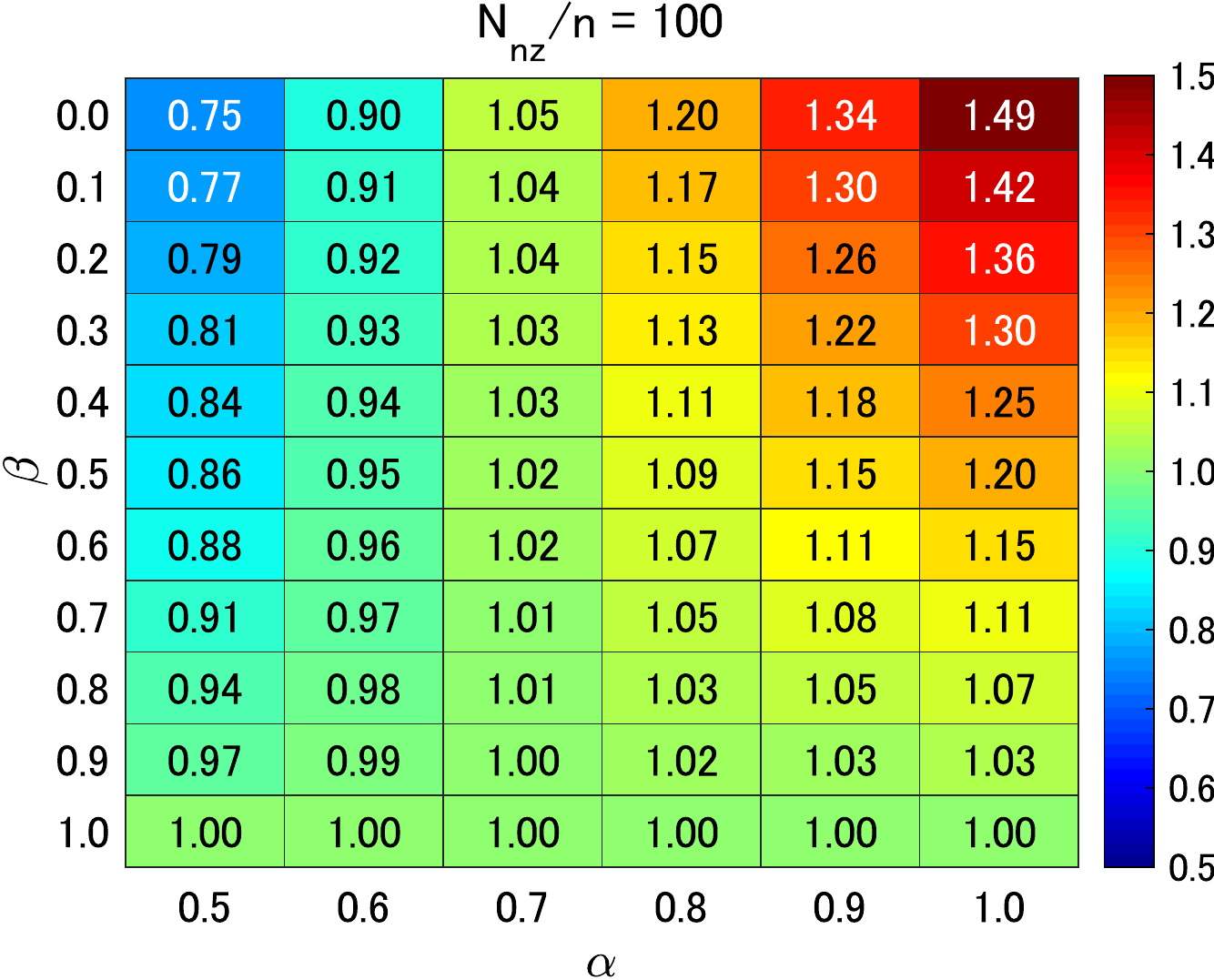}
\end{center}
\caption{Estimated upper bound of the speedup of the B-HDC (or M-HDC) kernel over the CSR kernel based on the performance models, 
which is calculated by Equation~\ref{eq:b_hdc_over_csr}.}
\label{fig:upper_bound_estimate}
\end{figure*}

Using Equation~\ref{eq:b_hdc_over_csr}, we demonstrate $P^\textrm{(B-HDC)}/P^\textrm{(CSR)}$ (or $P^\textrm{(M-HDC)}/P^\textrm{(CSR)}$). 
We set $v_{\vec{x}}=1$ and $c (= \nnz/n) = 10, 50, 100$, and the calculated results are shown in Figure~\ref{fig:upper_bound_estimate}. 
Since there is no essential difference between $\alpha$ and $\tilde{\alpha}$ and between $\beta$ and $\tilde{\beta}$, 
we hereafter do not distinguish between them and describe $\alpha$ and $\beta$ for the simplicity.
From Figure~\ref{fig:upper_bound_estimate}, for significant speedup (e.g. $1.1 \times$ speedup), both sufficient small $\beta$ (e.g. $\beta \le 0.5$) and large $\alpha$ (e.g. $\alpha \ge 0.8$) are required. 
On the other hand, $c$ makes almost no differences; 
the theoretical upper bound is $1.5$, and it is almost achieved when $c=50$. 

\section{Experimental results}
\label{sec:evaluation}
In this section, we report the experimental results. 
Our goal here is evaluating the effectiveness of the cache blocking techniques, namely the B-DIA, B-HDC, and M-HDC kernels, on recent multi-core CPUs.
First, we explain the experimental settings including the computational environments and report the effective memory performance. 
Next, we presents the results for stencil matrices, which correspond to the analysis in Section~\ref{sec:analysis_stencil}. 
Finally, we show the results for general matrices. 

\subsection{Settings}
\subsubsection{Computational environments}
We used a single computational node of three different supercomputer systems, 
each of which equips multi-core CPUs with different microarchitectures: Intel Broadwell, Skylake, and Cascade Lake. 
The specifications of these environments are listed in Table~\ref{tbl:spec}. 
In every environment, we occupied a node during our program execution, which means that programs by other users did not run on the node at the same time. 
We assigned a thread per core; we used totally 36 cores on Broadwell, 40 cores on Skylake, and 56 cores on Cascade Lake. 
In all environments, we set the thread affinity {\tt compact}. 

\begin{table*}[t]
\begin{center}
\caption{Specifications of the computational environments used in the performance evaluation.}
\label{tbl:spec}
\begin{tabular}{llccc}
\toprule
\multicolumn{2}{l}{Notation in this paper} & Broadwell            & Skylake                 & Cascade Lake    \\
\multicolumn{2}{l}{System name}          & Laurel~2           & Grand Chariot         & Oakbridge-CX  \\
\multicolumn{2}{l}{Site}                 & Kyoto Univ., Japan & Hokkaido Univ., Japan & The Univ. of Tokyo, Japan \\
\midrule 
CPU  & No.                  & Xeon E5-2695 v4         & Xeon Gold 6148          & Xeon Platinum 8280          \\
     & Microarchitecture    & Broadwell               & Skylake                 & Cascade Lake                \\
     & Frequency            & 2.10GHz                 & 2.4GHz                  & 2.7GHz                      \\
     & \#cores              & 18                      & 20                      & 28                          \\ 
     & L1 cache size (data) & 32KB/core               & 32KB/core               & 32KB/core                   \\ 
     & L2 cache size        & 256KB/core              & 1MB/core                & 1MB/core                    \\
     & L3 cache size        & 2.5MB/core              & 1.375MB/core            & 1.375MB/core                \\
\midrule
Node & \#CPUs               & 2                       & 2                       & 2                           \\
     & Memory size          & 128GB                   & 384GB                   & 192GB                       \\
     & Peak FLOPS           & 1.21TFLOPS              & 3.07TFLOPS              & 4.84 TFLOPS                 \\
     & Peak Memory BW       & 153.6GB/s               & 255.9GB/s               & 275.0GB/s                   \\
\midrule
Software & Compiler         & icc 18.0.5              & icc 19.0.5.281          & icc 19.0.5.281              \\
         & Math Library     & MKL 2018.0.4            & MKL 2019.0.5            & MKL 2019.0.5                \\
\bottomrule
\end{tabular}
\end{center}
\end{table*}

\subsubsection{Implementation}
We implemented the following SpMV kernels: 
\begin{itemize}
	\item the CSR kernel (Figure~\ref{fig:spmv_csr}),
	\item the DIA kernel (Figure~\ref{fig:spmv_dia}),
	\item the B-DIA kernel (Figure~\ref{fig:spmv_dia_block}),
	\item the HDC kernel (Figure~\ref{fig:spmv_hdc}),
	\item the B-HDC kernel (Figure~\ref{fig:spmv_hdc_block}),
	\item the M-HDC kernel (Figure~\ref{fig:spmv_mhdc}).
\end{itemize}
Programs of all kernels were written in C language and thread parallelized using OpenMP. 
FP64 (double precision) and INT32 were used in the programs. 
\par
Programs were complied by the Intel compiler ({\tt icc}), whose version in each environment is listed in Table~\ref{tbl:spec}, 
with the following options:
\begin{itemize}
	\item Broadwell: {\tt -O3} {\tt -qopenmp} {\tt -xHost} {\tt -no-vec} {\tt -no-simd}
	\item Skylake: {\tt -O3} {\tt -qopenmp} {\tt -xCORE-AVX512} {\tt -no-vect} {\tt -no-simd}
	\item Cascade Lake: {\tt -O3} {\tt -qopenmp} {\tt -xCORE-AVX512} {\tt -no-vec} {\tt -no-simd}
\end{itemize}
\par
We prepared two routines for each kernel: with the SIMD vectorization and without the SIMD vectorization. 
A routine with the SIMD vectorization was generated by using the OpenMP directive (i.e. {\tt omp\_simd}), 
where some loops (for details, see the pseudo-code of each kernel) were explicitly vectorized by the directive. 
Otherwise, loops were not vectorized because complier's vectorization was prevented by the options ({\tt -no-vec} {\tt -no-simd}).

\subsubsection{Timing}
\label{sec:timing}
We used {\tt omp\_get\_wtime} for timing. 
The execution time of a kernel was measured in the manner shown in Figure~\ref{fig:timing}; 
measuring the execution time for repeating a kernel $n\_ites$ times, and calculating the average time for a single kernel execution. 
The above measurement was conducted $n\_loops$ times, and the best result was chosen to be evaluate. 

\begin{figure}[t]
\begin{center}
\begin{lstlisting}
for(loop = 0; loop < n_loops; loop++){
	t0 = omp_get_wtime();
	for(ite = 0; ite < n_ites; ite++){
		kernel(...);
	}
	t1 = omp_get_wtime();
	time = (t1 - t0)/n_ites;
	save_timing_result(time, ...);
}
\end{lstlisting}
\end{center}
\setlength\abovecaptionskip{-5pt}
\caption{The outline of how to measure the execution time of a kernel.}
\label{fig:timing}
\end{figure}

\subsection{Benchmark on the memory performance}
Since SpMV kernels are memory bound, it is worth understanding the effective memory performance in recent multi-core systems. 
We prepared two kernels that measured the effective memory performance: 
a kernel with only direct indexing (Figure~\ref{fig:stream_direct}, we call \textit{direct}) 
and a kernel with indirect indexing (Figure~\ref{fig:stream_indirect}, we call \textit{indirect}). 
Here, arrays {\tt A[]}, {\tt B[]}, and {\tt C[]} were FP64, and {\tt I[]} was INT32. 
Since {\tt I[i]~=~i}, the access pattern to the array {\tt C[]} was not changed in the kernel with indirect indexing. 
\par
We measured the execution time of these two kernels for various array size (i.e. $N$); 
the execution time was measured in the same manner as for SpMV kernels (see Figure~\ref{fig:timing}). 
We set $n\_ites = 1,000$ and $n\_loops = 20$. 
The performance of each kernel was calculated as 
\begin{equation}
	\textrm{(byte/s)} \coloneqq \frac{M N}{\textrm{(execution time)}},
\end{equation}
where $M = 32$ (3 loads and 1 store for 8 byte data) in the direct kernel, and $M = 36$ (additional 1 load for 4 byte integer data).
\par
\begin{figure*}[t]
\begin{center}
\includegraphics[scale=0.35]{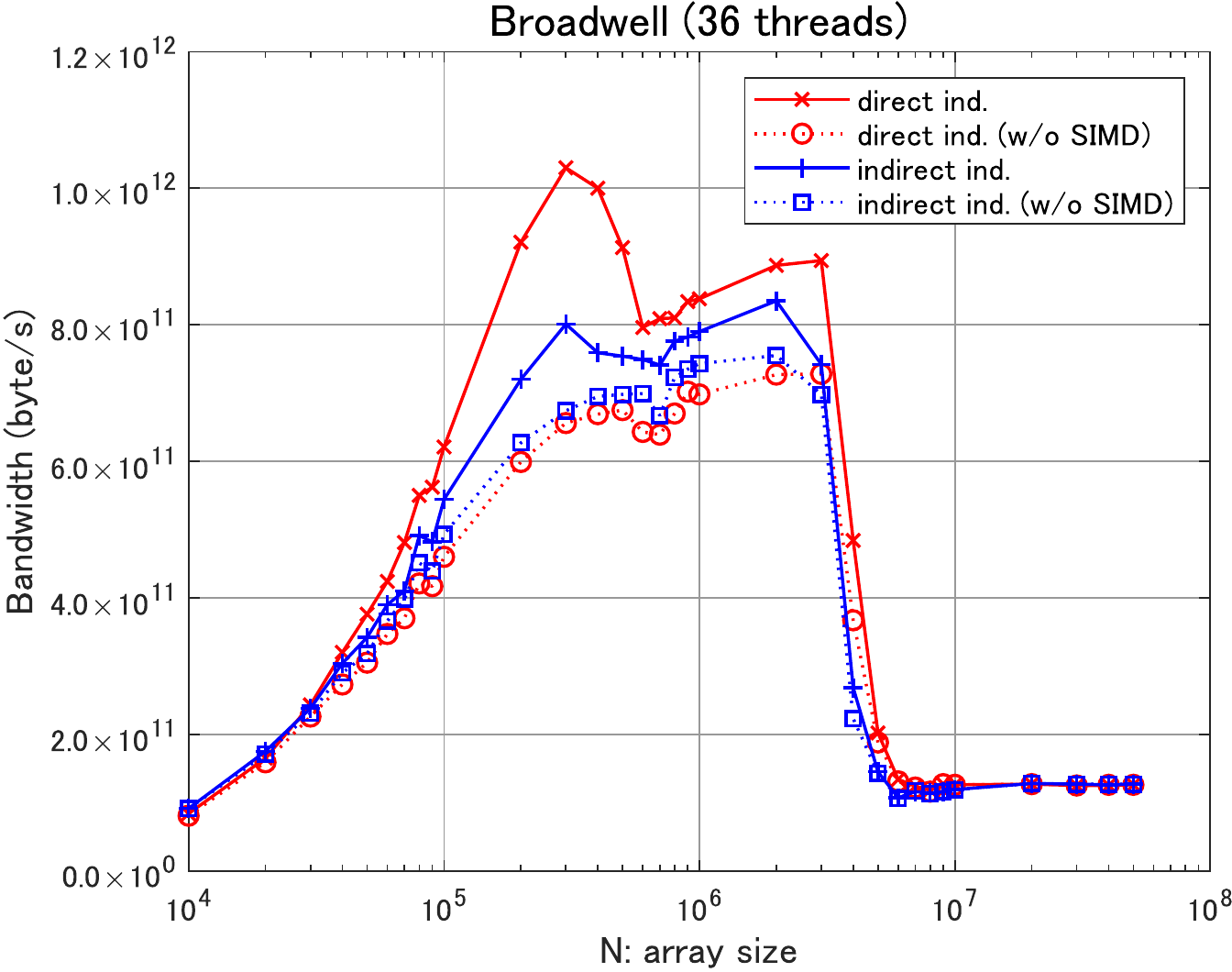}
\includegraphics[scale=0.35]{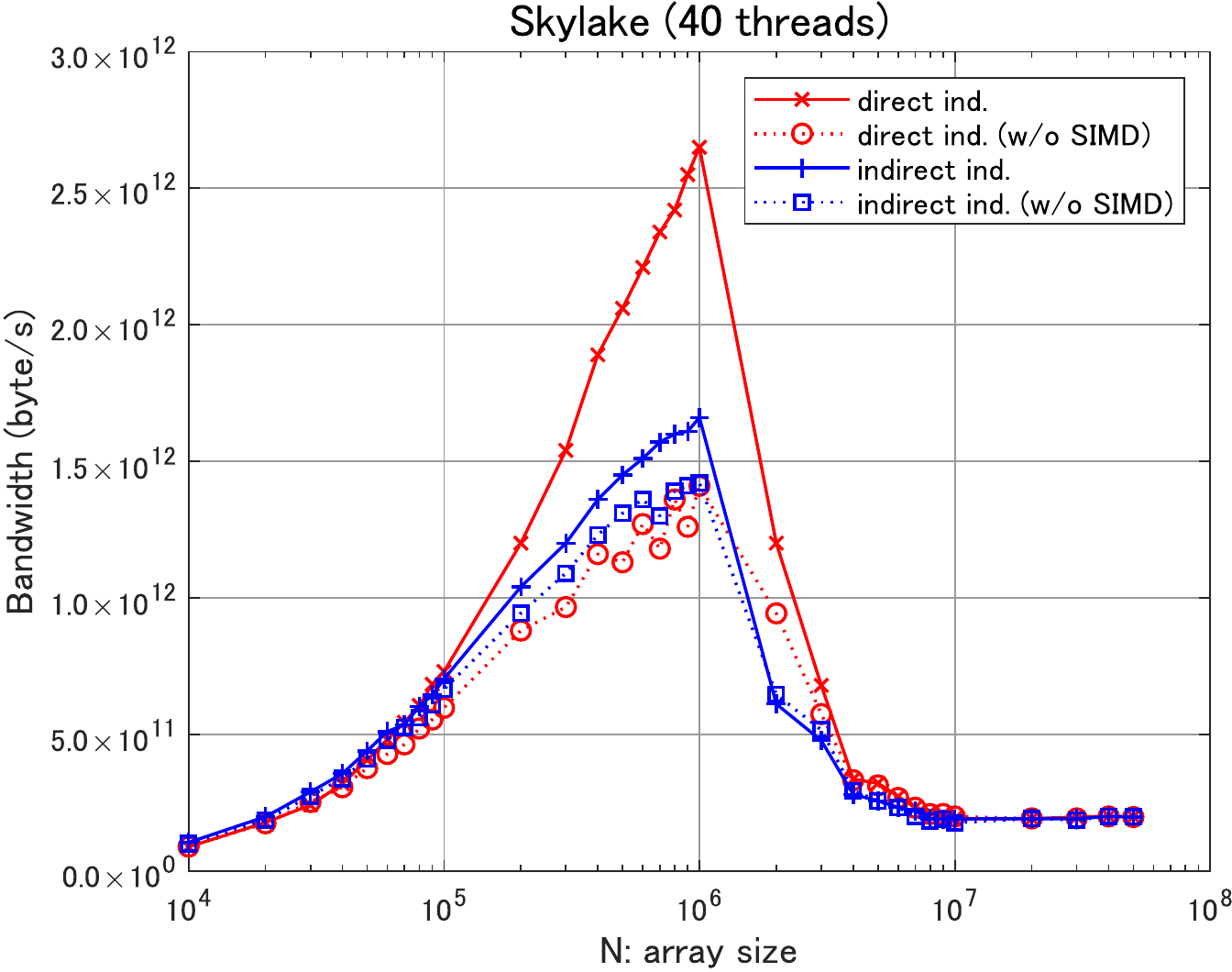}
\includegraphics[scale=0.35]{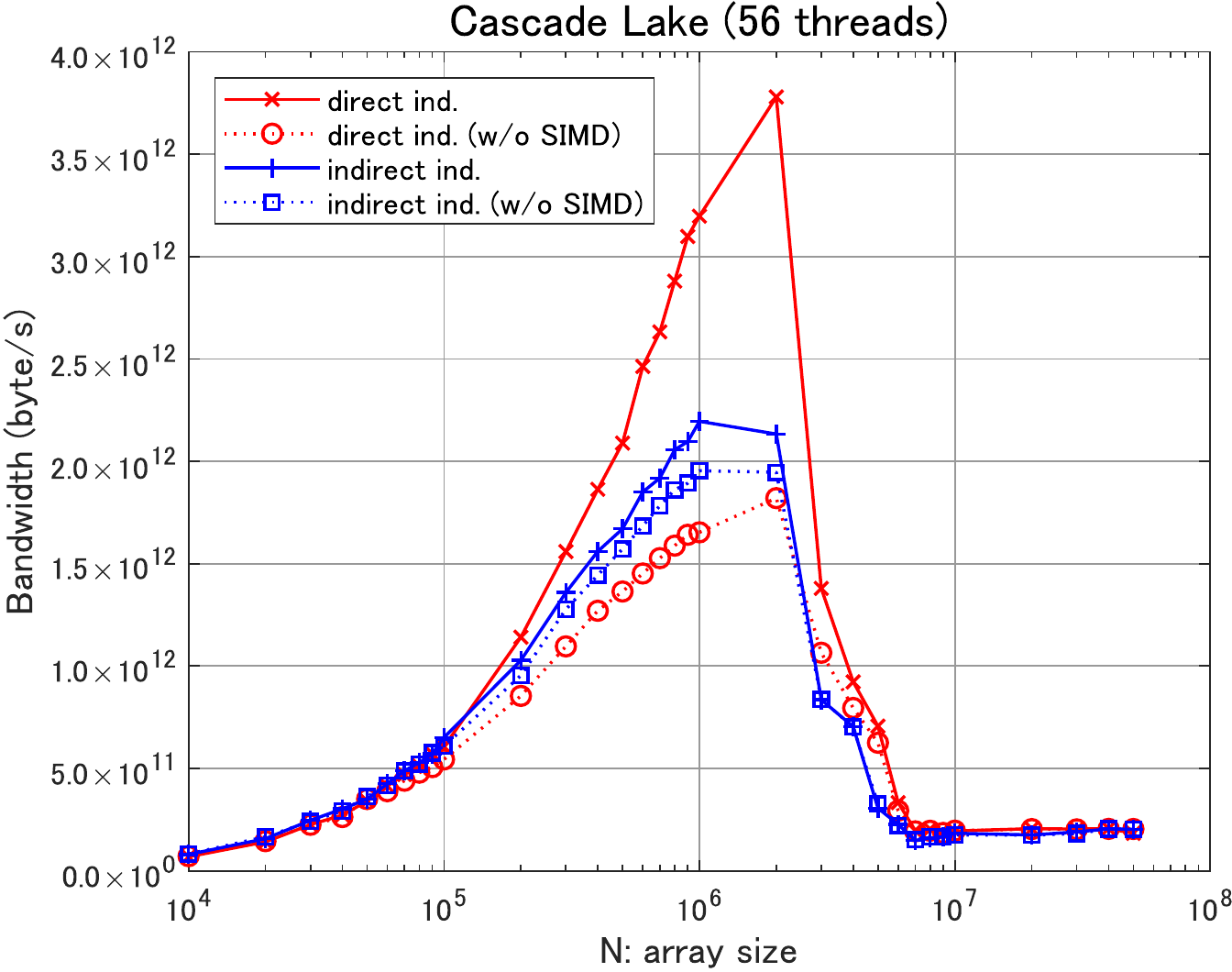}
\end{center}
\caption{Effective memory performance: comparison between the direct and indirect indexing, and between with and without the SIMD vectorization.}
\label{fig:results_stream}
\end{figure*}

The obtained results are presented in Figure~\ref{fig:results_stream}. 
Here, in each environment, we compare four kernels: direct with SIMD, direct without SIMD, indirect with SIMD, and indirect without SIMD. 
From the graphs, we obtain the following important observations: 
\begin{itemize}
\item When $N$ is sufficiently large, i.e. the total data amount is larger than the size of the cache memory, there is no difference in the performance between among the four kernels. 
\item If $N$ is not large, the SIMD vectorization makes a remarkable performance increase for the direct kernels but little for the indirect kernels. 
\item As already known, the performance and its behavior are drastically different between the cases that $N$ is sufficiently large (i.e. the out-of-cache case) and not large (i.e. the in-cache case). 
\end{itemize}

\begin{figure}[t]
\begin{center}
\begin{lstlisting}
#pragma omp parallel for simd private(i)
for(i = 0; i < N; i++) {
	C[i] += A[i] * B[i];
}
\end{lstlisting}
\end{center}
\setlength\abovecaptionskip{-5pt}
\caption{Memory benchmark kernel only with direct indexing.}
\label{fig:stream_direct}
\begin{center}
\begin{lstlisting}
#pragma omp parallel for simd private(i)
for(i = 0; i < N; i++) {
	C[i] += A[i] * B[I[i]]; // I[i] = i
}
\end{lstlisting}
\end{center}
\setlength\abovecaptionskip{-5pt}
\caption{Memory benchmark kernel with indirect indexing.}
\label{fig:stream_indirect}
\end{figure}

\subsection{Experiments for stencil matrices}
With stencil matrices, we evaluated the performance of the CSR, DIA, and B-DIA kernels; this part corresponds to the analysis in Section~\ref{sec:analysis_stencil}.
We prepared the following three types of stencil matrices: given the matrix size $n$, 
\begin{itemize}
\item 1D-3Point stencil matrix: 
$a_{ij} \ne 0$ if $j = i, i \pm 1$.
\item 2D-5Point stencil matrix: 
$a_{ij} \ne 0$ if $j = i, i \pm 1, i \pm n_x$, where $n_x \coloneqq \lfloor \sqrt{n} \rfloor$.
\item 3D-7Point stencil matrix: 
$a_{ij} \ne 0$ if $j = i, i \pm 1, i \pm n_x, i \pm n_x^2$, where $n_x \coloneqq \lfloor \sqrt[3]{n} \rfloor$.
\end{itemize}
We generated test matrices with various $n \in [1 \times 10^4, 5 \times 10^7]$, 
and measured the performance of the kernels under the settings $n\_ite = 1,000$ and $n\_loops = 20$. 
\par
\begin{figure*}[t]
\begin{center}
\subfloat[1D-3Point stencil matrix]
{
	\includegraphics[scale=0.42]{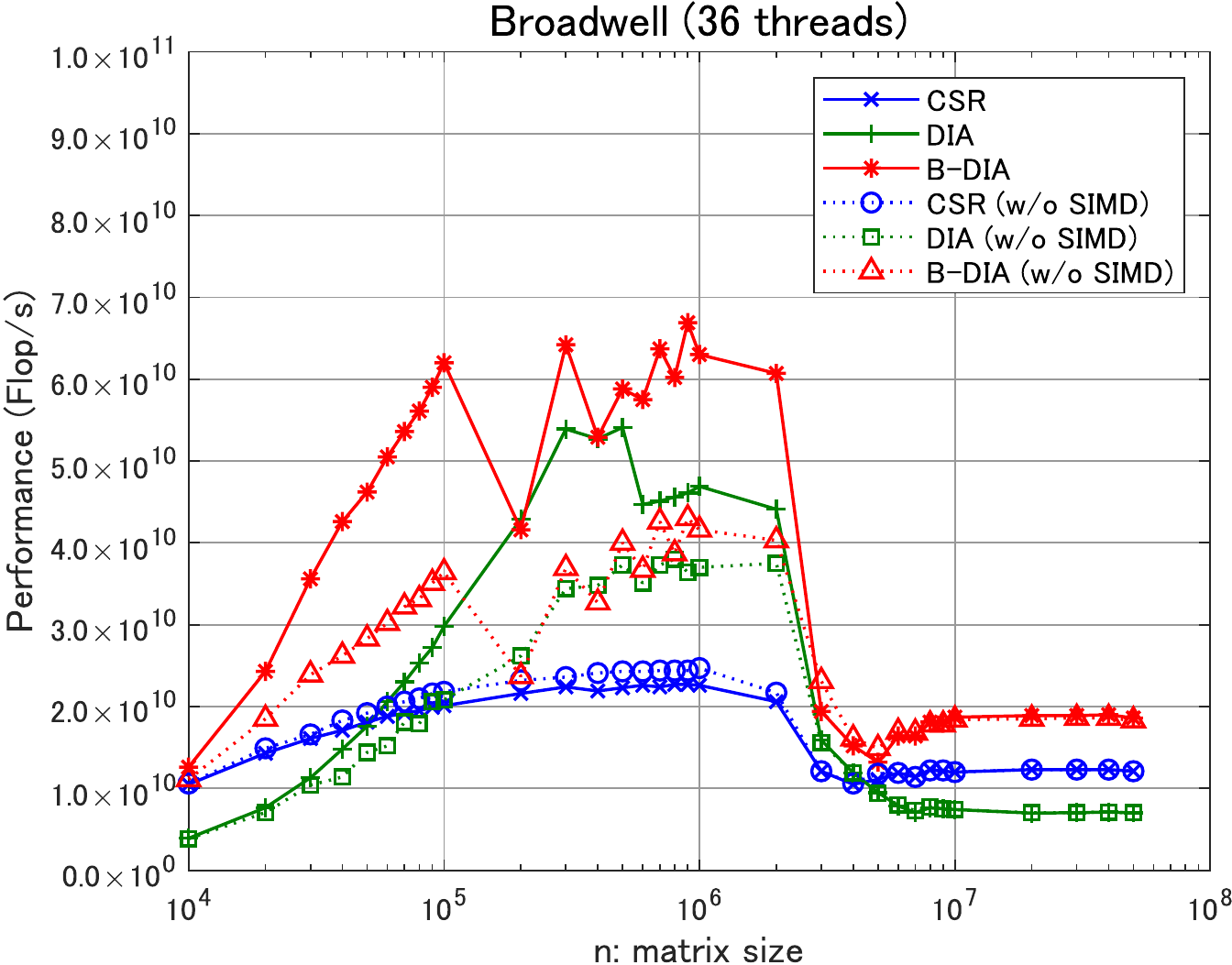}
	\includegraphics[scale=0.42]{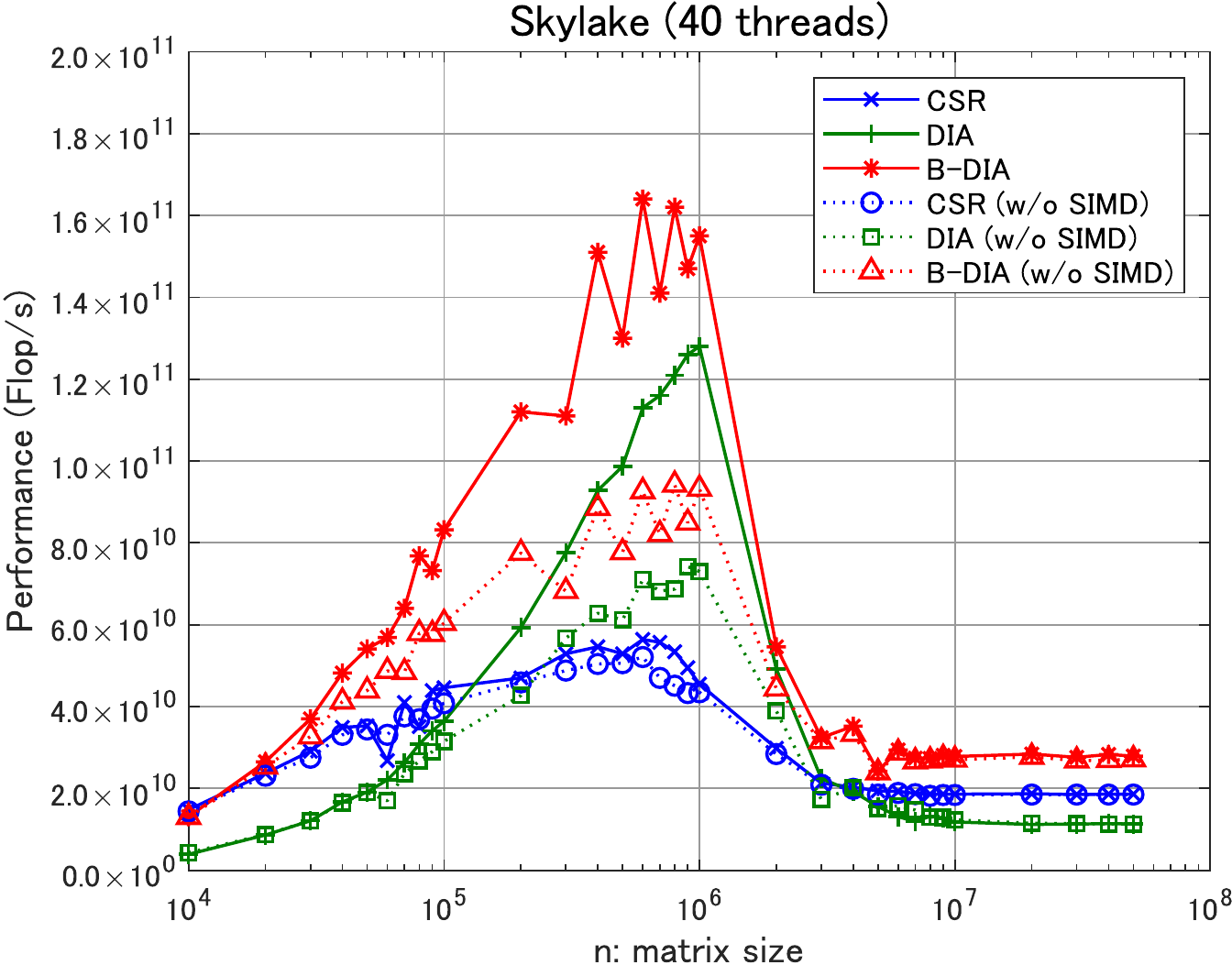}
	\includegraphics[scale=0.42]{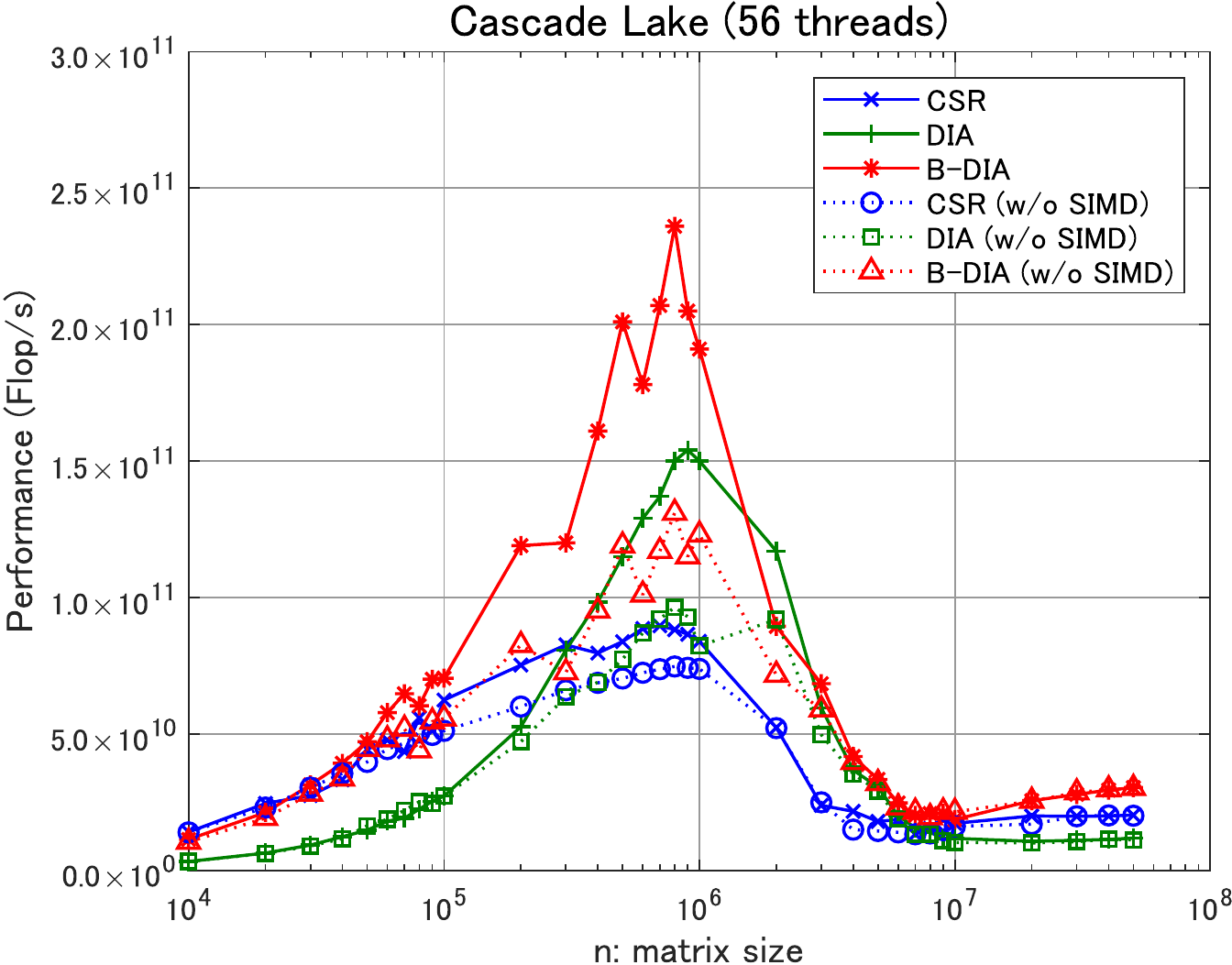}
	\label{fig:flops_diag3}
}
\hfill
\subfloat[2D-5Point stencil matrix]
{
	\includegraphics[scale=0.42]{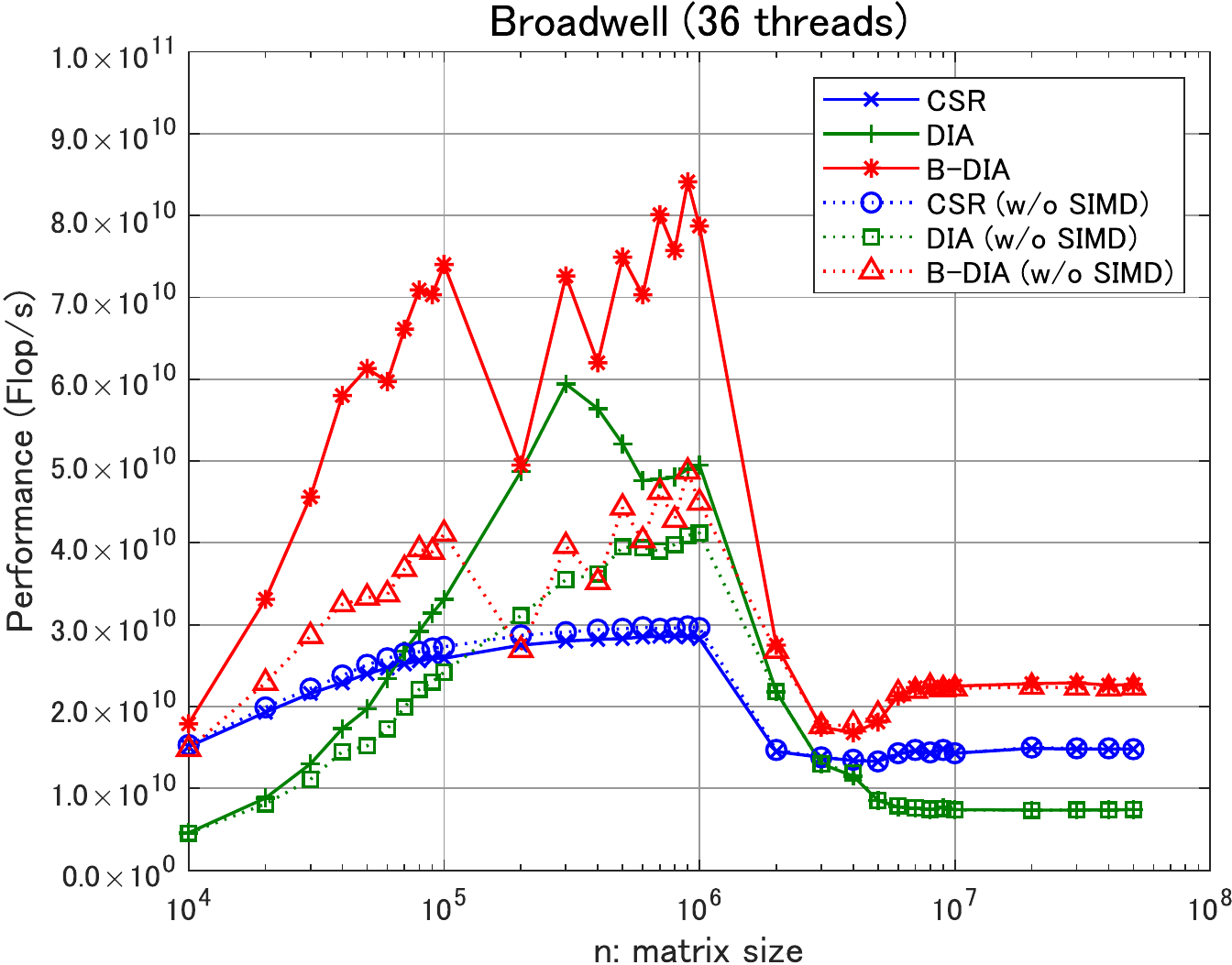}
	\includegraphics[scale=0.42]{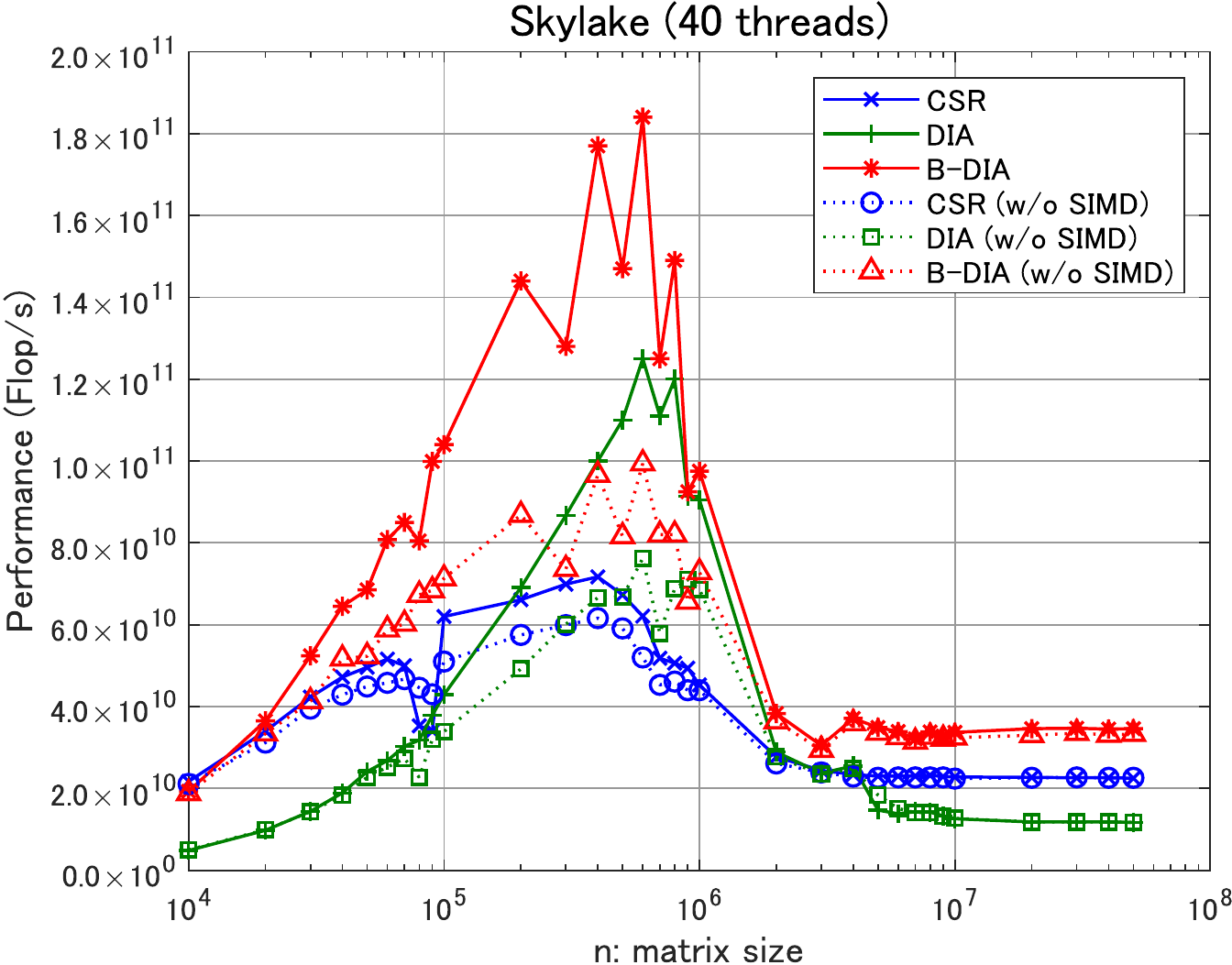}
	\includegraphics[scale=0.42]{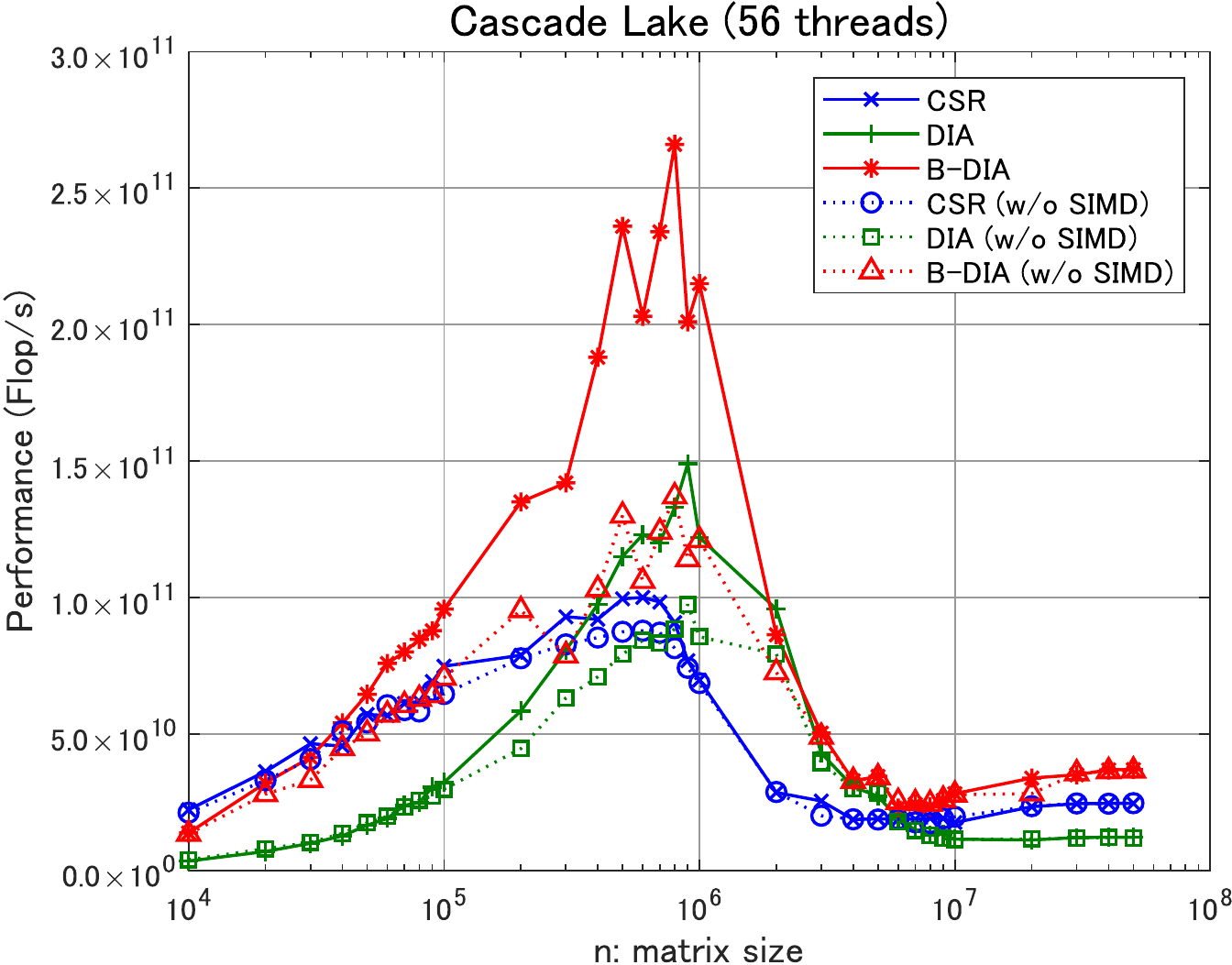}
	\label{fig:flops_diag5}
}
\hfill
\subfloat[3D-7Point stencil matrix]
{
	\includegraphics[scale=0.42]{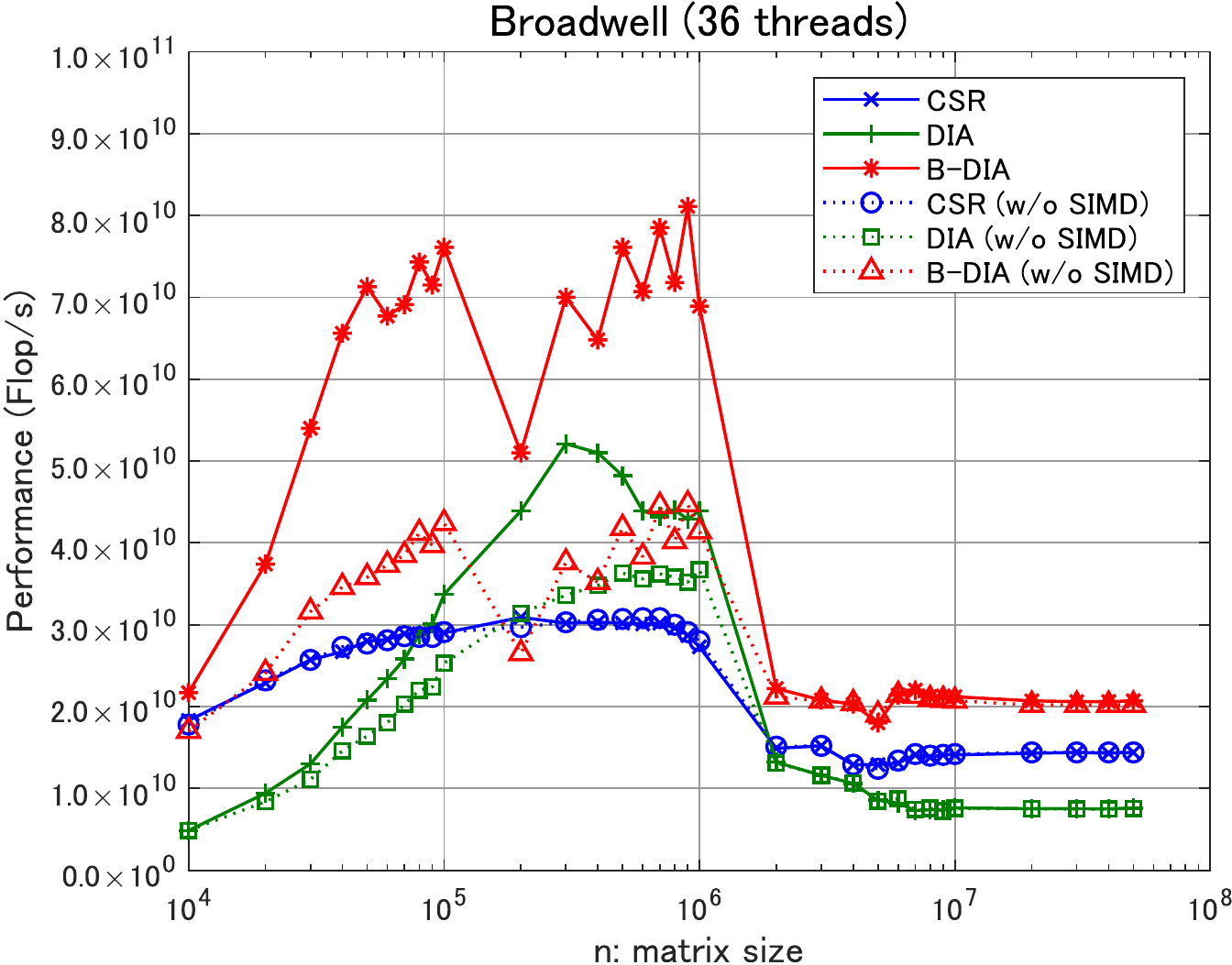}
	\includegraphics[scale=0.42]{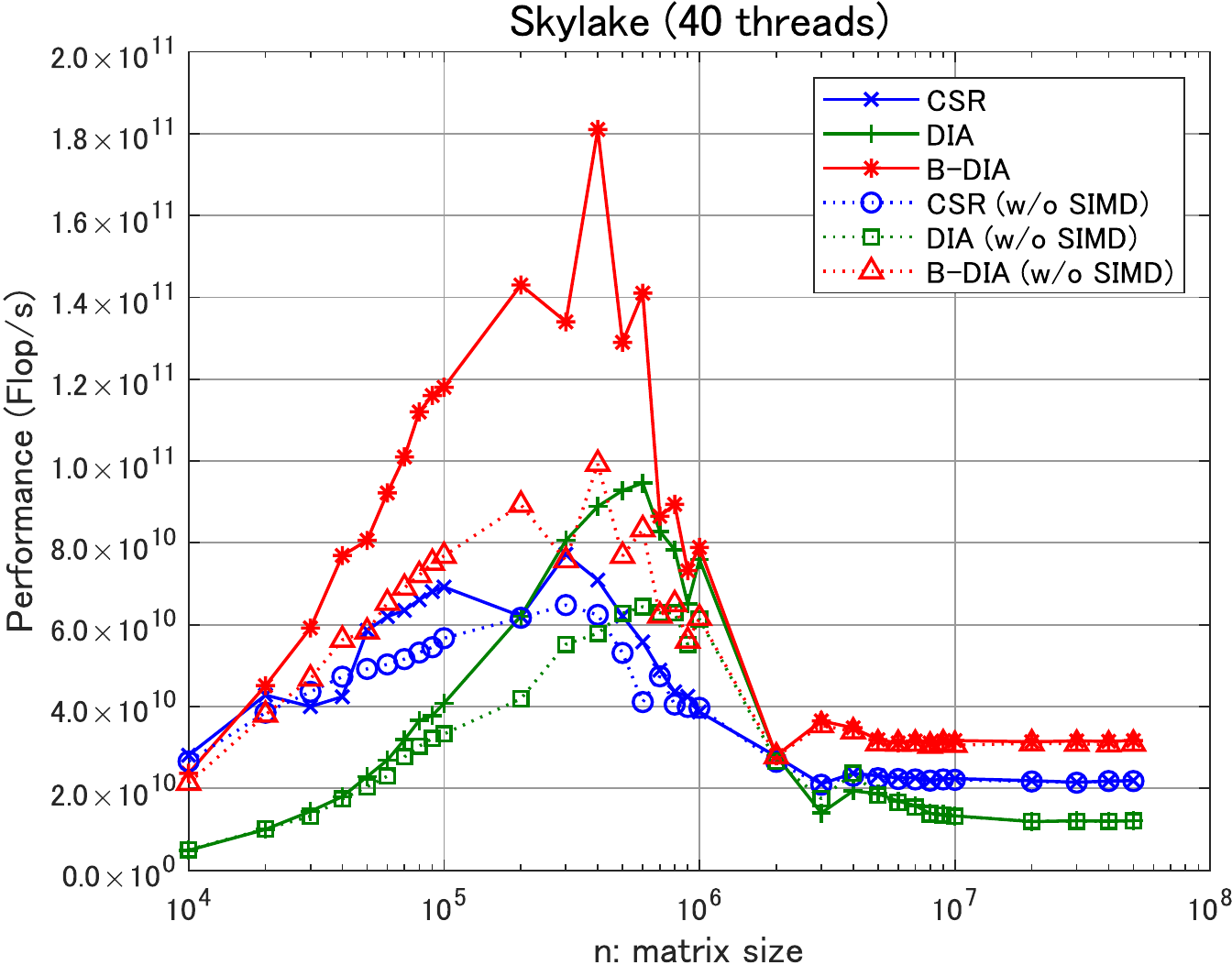}
	\includegraphics[scale=0.42]{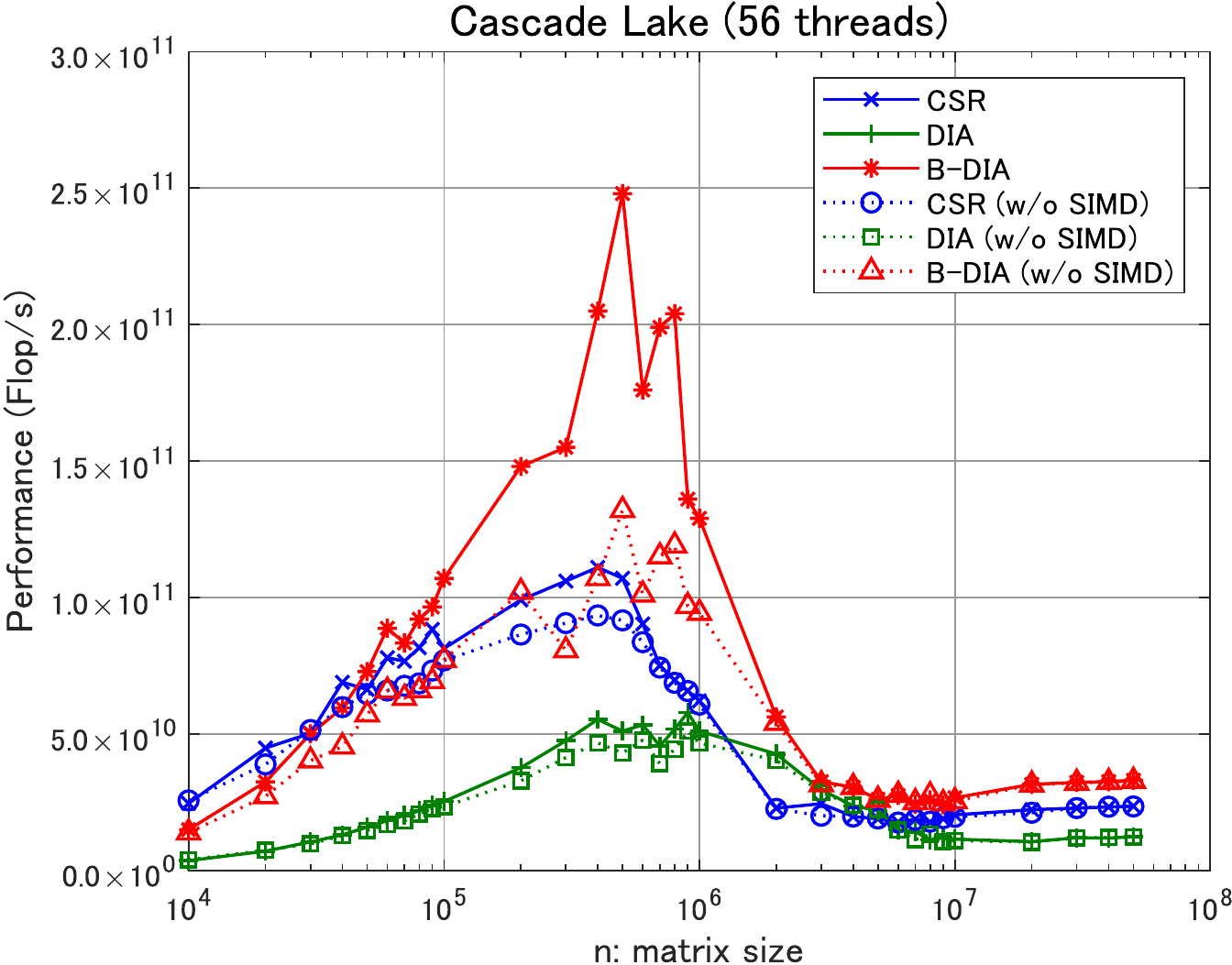}
	\label{fig:flops_diag7}
}
\end{center}
\caption{Measured performance of the CSR, DIA, and B-DIA kernels for stencil matrices.}
\label{fig:flops_diag}
\end{figure*}

The obtained results are presented in Figure~\ref{fig:flops_diag}.
From the graphs, we have the following observations: 
\begin{itemize}
\item When $n$ is sufficiently large (i.e. the out-of-cache case), there is no difference between with and without the SIMD vectorization. 
In this case, the B-DIA kernel outperforms the CSR kernel, but the DIA kernel underperforms. 
These results are consistent with the theoretical analysis in Section~\ref{sec:analysis_stencil}. 
The obtained performance differences are due to the differences in the amount of the data access cost to the main memory. 
\item When $n$ is not large (i.e. the in-cache case), whether with or without the SIMD vectorization makes a remarkable performance difference for the DIA and B-DIA kernels but almost no difference for the CSR kernel. 
This is easily expected from the memory performance presented in the previous section; the DIA and B-DIA kernels have no indirect indexing, but the CSR kernel has it. 
Additionally, the CSR kernel is less suitable for the SIMD vectorization because its most inner loop is usually really short. 
In this case, the B-DIA kernel (with SIMD) outperforms other kernels. 
However, the performance of each kernel more fluctuates than that in the out-of-cache case, and more careful investigation is needed. 
\end{itemize}
\par
\begin{figure*}[t]
\begin{center}
\includegraphics[scale=0.4]{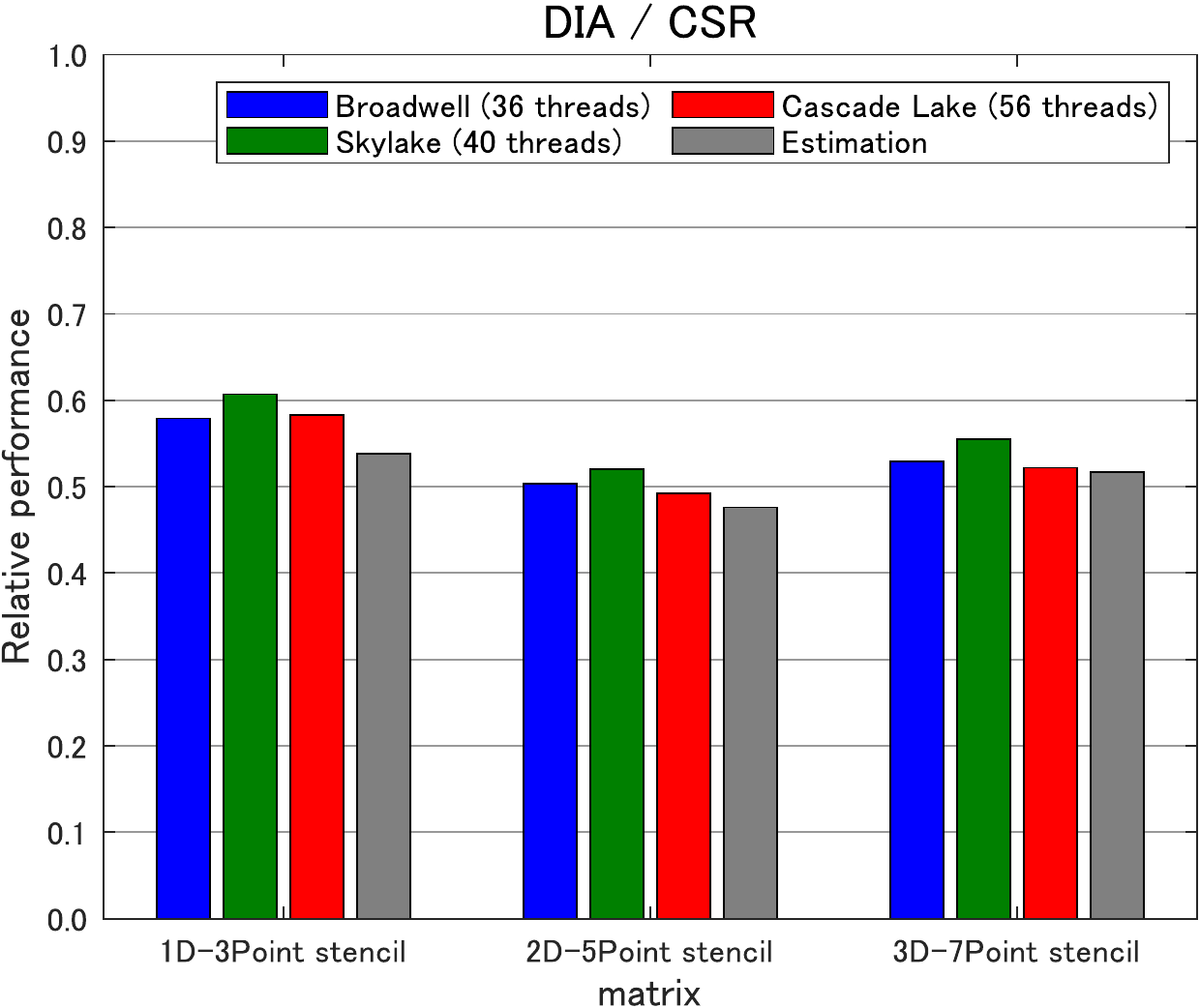}
\includegraphics[scale=0.4]{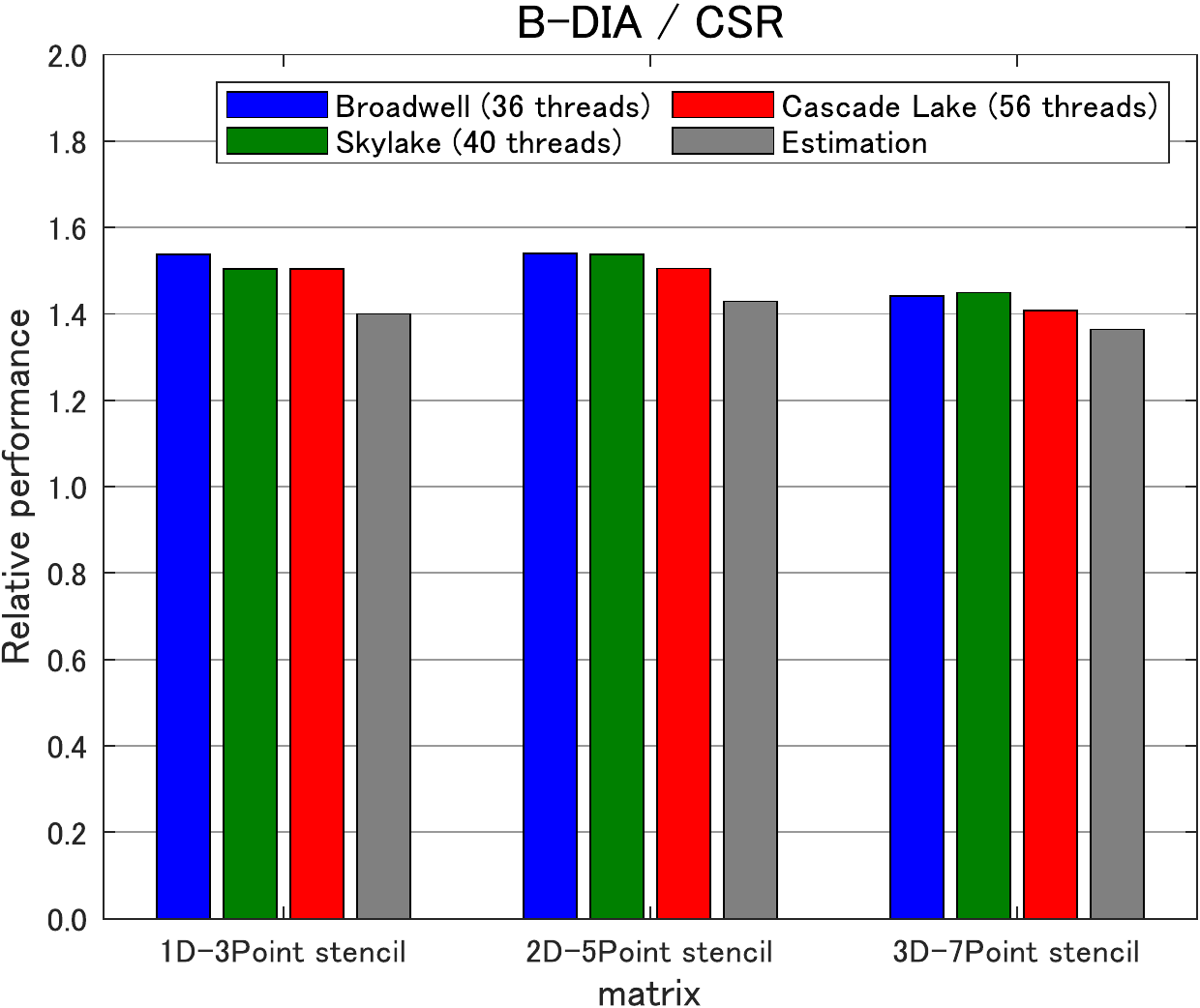}
\includegraphics[scale=0.4]{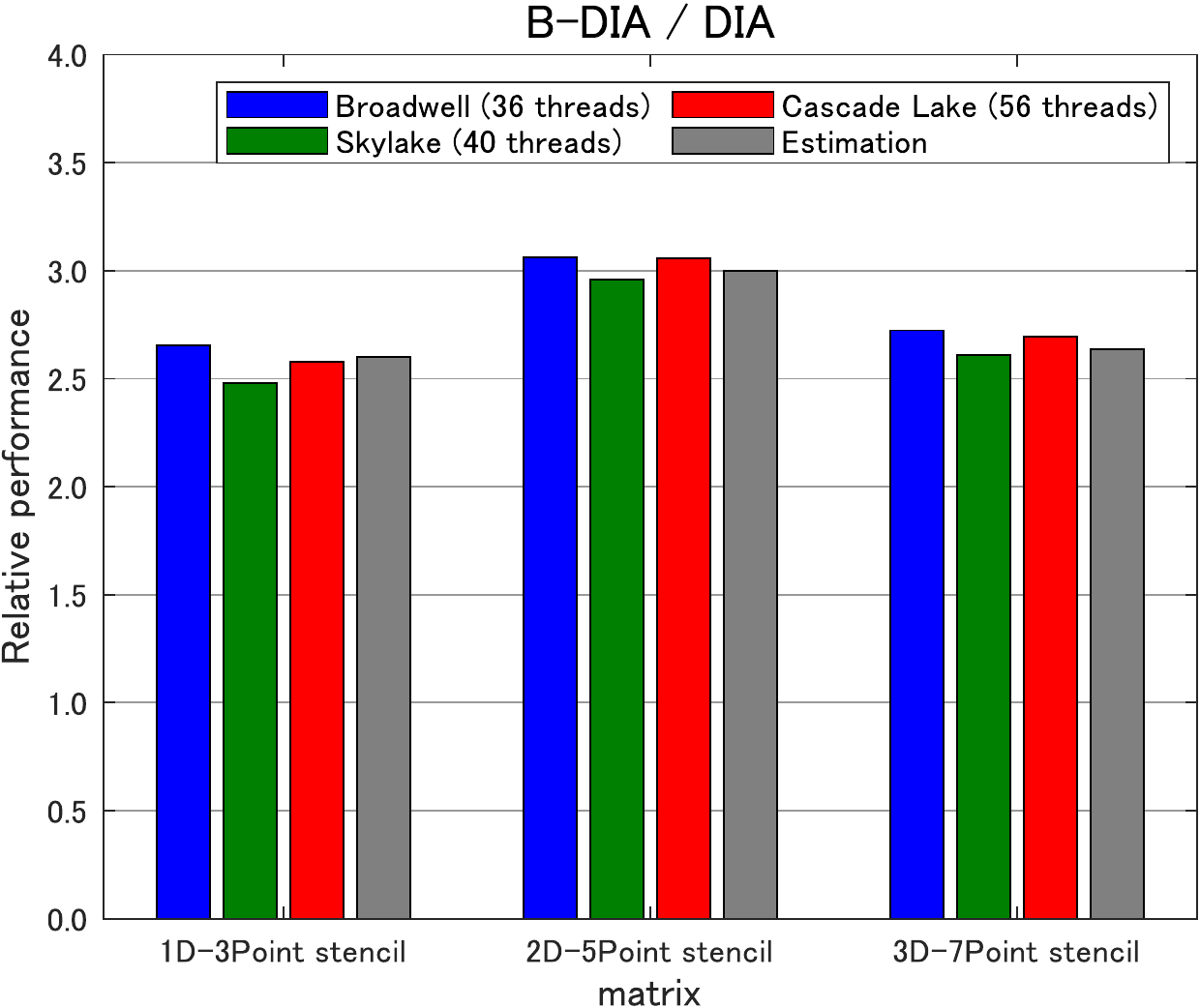}
\end{center}
\caption{Performance comparison for stencil matrices with $n=5 \times 10^7$.}
\label{fig:diag_comparison}
\end{figure*}

Now, we compare the obtained performance differences among the CSR, DIA, and B-DIA kernels (with the SIMD vectorization).
In Figure~\ref{fig:diag_comparison}, we present the obtained relative performance for $n = 5 \times 10^7$ together with their estimation from Equations~\ref{eq:dia_over_csr}, \ref{eq:b_dia_over_csr}, and \ref{eq:b_dia_over_dia}. 
In the estimation, we set $\gamma \;(= \gamma^\textrm{(CSR)} = \gamma^\textrm{(B-DIA)})$ $1/3$ ($\ndiag=3$, 1D-3Point), $1/5$ ($\ndiag=5$, 2D-5Point), and $3/7$ ($\ndiag=7$, 3D-7Point). 
When $\ndiag = 7$, the two diagonal lines with offset $\pm n_x^2$ are far from the other diagonal lines, and their corresponding elements of $\vec{x}$ are expected to be loaded from the main memory. 
This is the reason why $\gamma = 3/7$ when $\ndiag=7$. 
From the graphs, we can find that the obtained relative performance are almost equivalent among the three environments. 
In addition, the obtained results are basically close to the estimation, which supports the reasonability of the analysis in Section~\ref{sec:analysis_stencil}. 
\par
\begin{figure*}[t]
\begin{center}
\includegraphics[scale=0.35]{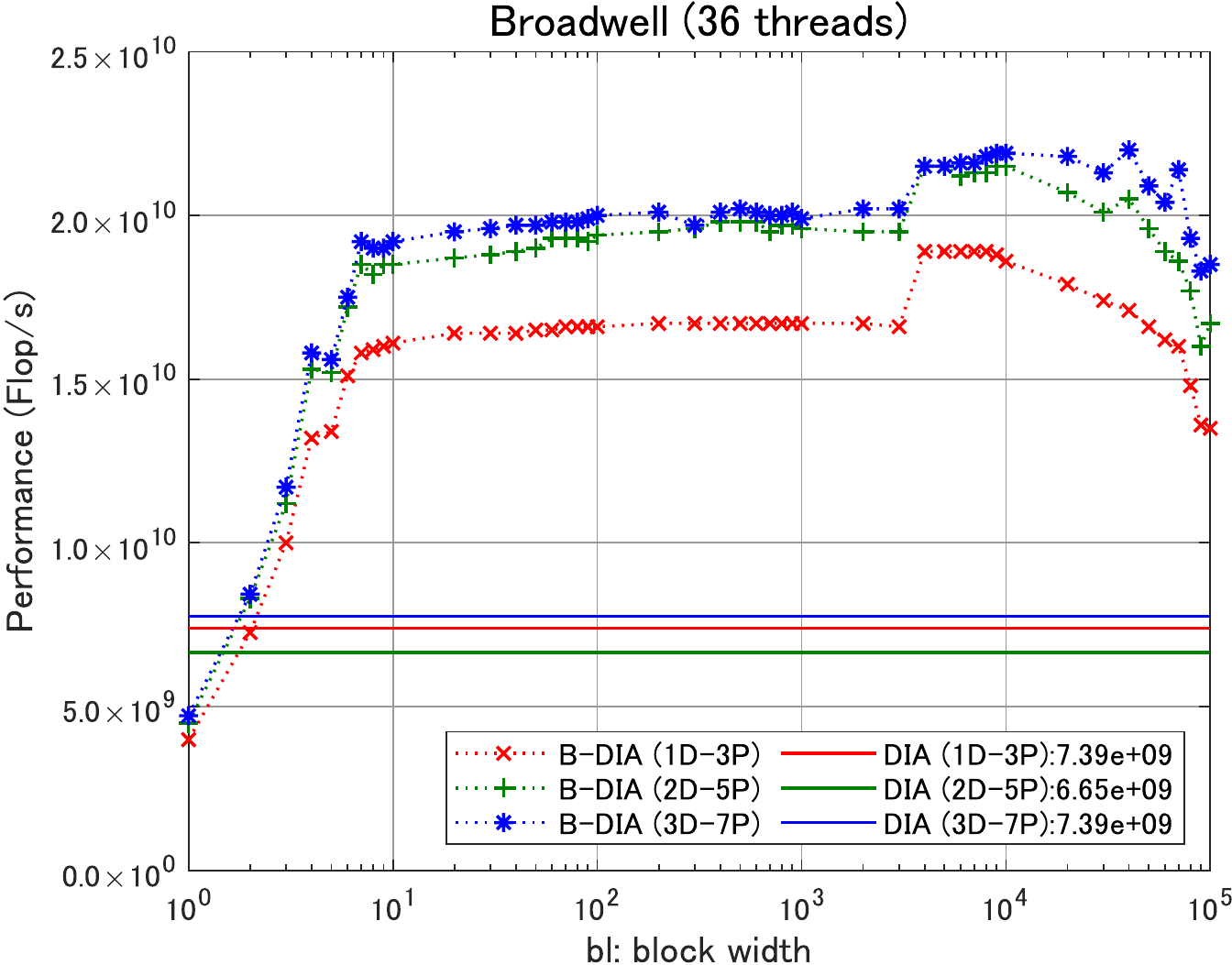}
\includegraphics[scale=0.35]{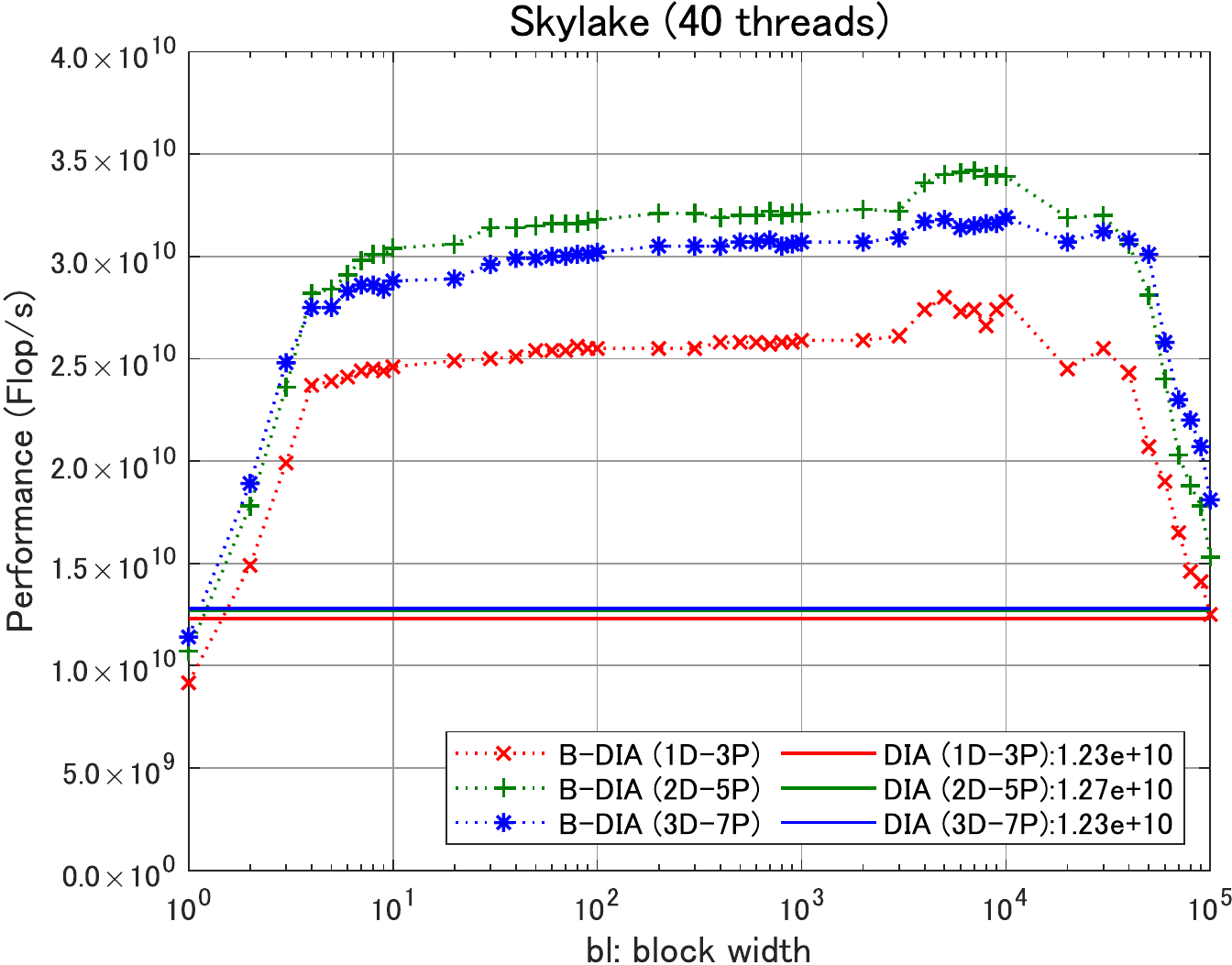}
\includegraphics[scale=0.35]{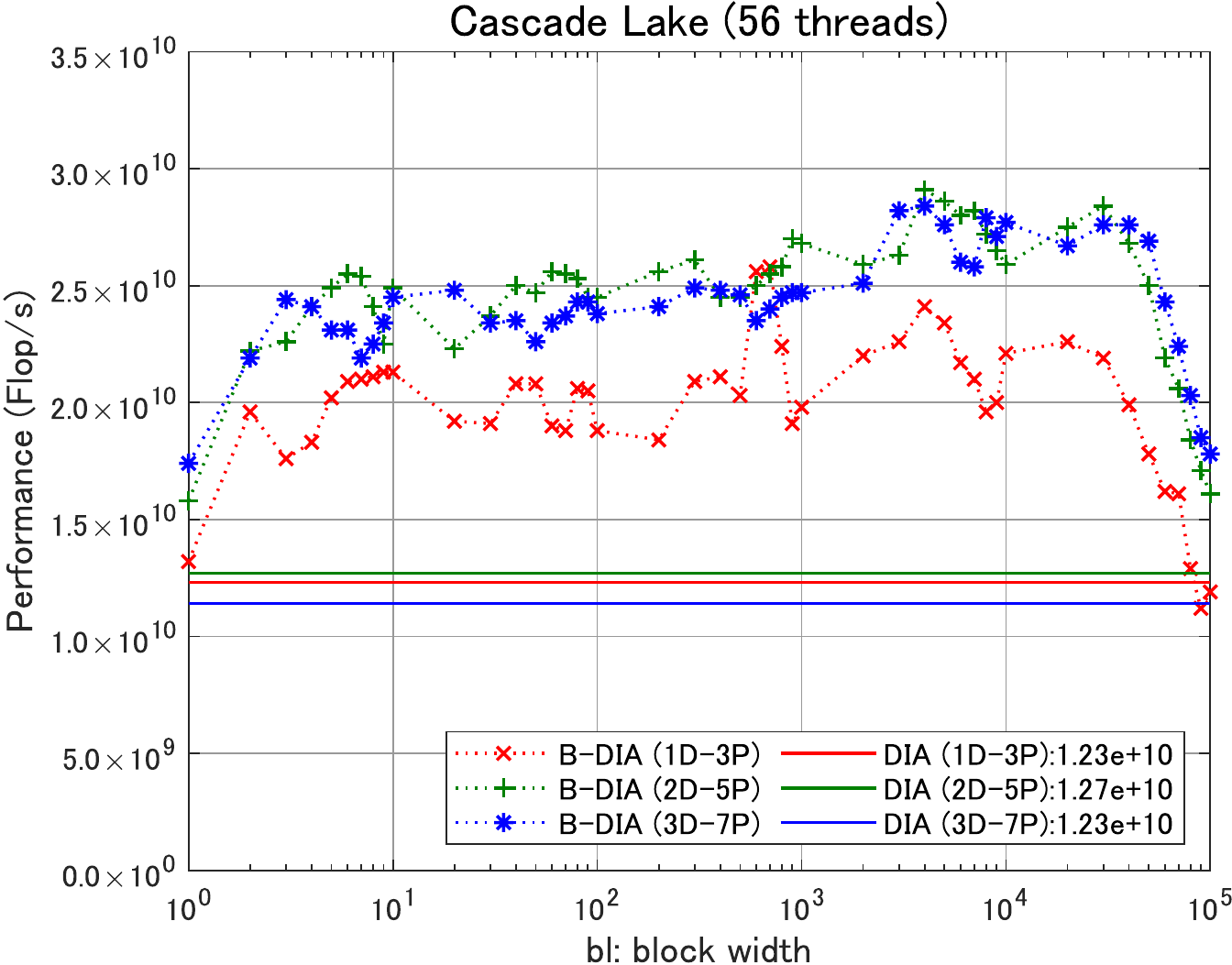}
\end{center}
\caption{Relationship between the performance of the B-DIA kernel and the block width for stencil matrices with $n = 1 \times 10^7$.}
\label{fig:diag_bl_impact}
\end{figure*}

Finally, we investigate the relationship between the performance of the B-DIA kernel and the block width $bl$. 
For the case of $n = 1 \times 10^7$, we show the performance of the B-DIA kernel with each $bl$ in Figure~\ref{fig:diag_bl_impact}, 
together with the performance of the DIA kernel. 
From the graphs, we can find the effect of the cache blocking technique excepting the cases that the block width is too small (roughly $n < 10$) or too large ($n > 10^4$).
In all environments and for all matrices, block width around $bl = 5000$ provides the highest performance, 
however sensitive optimization is not necessarily required as the graphs show; 
in the range of $10 \le bl \le 10^4$, significant performance differences are not observed. 

\subsection{Experiments for matrices in practical applications}
We next conduct experiments using matrices provided in the SuiteSparse Matrix Collection~\citep{2011_davis}, which represent the characteristics of practical applications. 
Using such matrices, we confirm the importance of the cache blocking techniques employed in the B-HDC kernel and the potential of the M-HDC kernel for SpMV with matrices appearing in practical applications. 

\subsubsection{Matrix selection and experimental settings}
As test matrices, from the SuiteSparse Matrix Collection, 
we selected 20 matrices that are nonsingular, real, unsymmetric, large (in terms of $\nnz$), and not derived from graph problems. 
Since our main interest is to examine the out-of-cache performance of each kernels, we selected large matrices.
In this research, we have more interests in computational science or engineering problems, 
e.g. matrices derived from the FDM or FEM discretization, than graph problems.
This is why we imposed the last condition. 
Table~\ref{tbl:matrix} lists the information of the selected matrices. 
\par
We performed four kernels: the CSR, HDC, B-HDC, and M-HDC kernels. 
For all kernels, the SIMD vectorization was applied. 
In the HDC, B-HDC, and M-HDC kernels, we set their parameters as follows: 
\begin{itemize}
	\item $\theta = 0.5, 0.6, 0.7, 0.8, 0.9$
	\item $bl = 10, 50, 100, 500, 1000, 5000, 10000$
\end{itemize}
We measured the execution time in the way described in Section~\ref{sec:timing}, and set $n\_loops = 20$ and $n\_ites = 1000$. 

\begin{table*}[t]
\begin{center}
\caption{Matrices selected from the SuiteSparse Matrix Collection as test matrices for the experiments.}
\label{tbl:matrix}
\begin{tabular}{rlrrrl}
\toprule
No. & Matrix name & $\nnz$ & $n$ & $\nnz/n$ & Kind \\
\midrule
 1 & HV15R            & 283,073,458 & 2,017,169 & 140 & Computational fluid dynamics problem \\
 2 & vas\_stokes\_4M  & 131,577,616 & 4,382,246 &  30 & Semiconductor process problem \\
 3 & ML\_Geer         & 110,879,972 & 1,504,002 &  74 & Structural problem \\
 4 & vas\_stokes\_2M  &  65,129,037 & 2,146,677 &  30 & Semiconductor process problem \\
 5 & nv2              &  52,728,362 & 1,453,908 &  36 & Semiconductor device problem \\
 6 & dgreen           &  38,259,877 & 1,200,611 &  32 & Semiconductor device problem \\
 7 & RM07R            &  37,464,962 &   381,689 &  98 & Computational fluid dynamics problem \\
 8 & vas\_stokes\_1M  &  34,767,207 & 1,090,664 &  32 & Semiconductor process problem \\
 9 & ss               &  34,753,577 & 1,652,680 &  21 & Semiconductor process problem \\
10 & ML\_Laplace      &  27,689,972 &   377,002 &  73 & Structural problem \\
11 & FullChip         &  26,621,990 & 2,987,012 &   9 & Circuit simulation problem \\ 
12 & Transport        &  23,500,731 & 1,602,111 &  15 & Structural problem \\
13 & CoupCons3D       &  22,322,336 &   416,800 &  54 & Structural problem \\
14 & rajat31          &  20,316,253 & 4,690,002 &   4 & Circuit simulation problem \\
15 & circuit5M\_dc    &  19,194,193 & 3,523,317 &   5 & Circuit simulation problem \\
16 & Freescale1       &  18,920,347 & 3,428,755 &   6 & Circuit simulation problem \\
17 & TSOPF\_RS\_b2383 &  16,171,169 &    38,120 & 424 & Power network problem \\
18 & memchip          &  14,810,202 & 2,707,524 &   5 & Semiconductor device problem \\
19 & test1            &  12,968,200 &   392,908 &  33 & Semiconductor device problem \\
20 & ohne2            &  11,063,545 &   181,343 &  61 & Semiconductor device problem \\
\bottomrule
\end{tabular}
\end{center}
\end{table*}

\begin{figure*}[t]
\begin{center}
\includegraphics[scale=0.5]{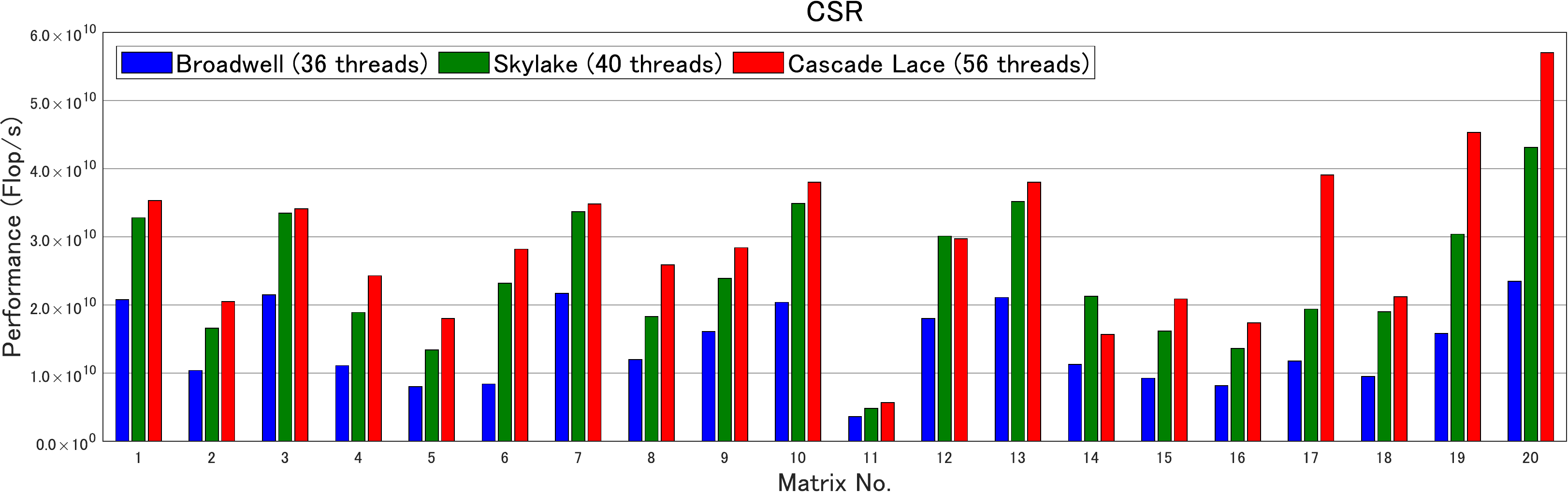}
\end{center}
\caption{Performance of the CSR kernel in each environment for the test matrices selected from the SuiteSparse Matrix Collection.}
\label{fig:ssmc_csr_gflops}
\end{figure*}

\begin{figure*}[t]
\begin{center}
\includegraphics[scale=0.5]{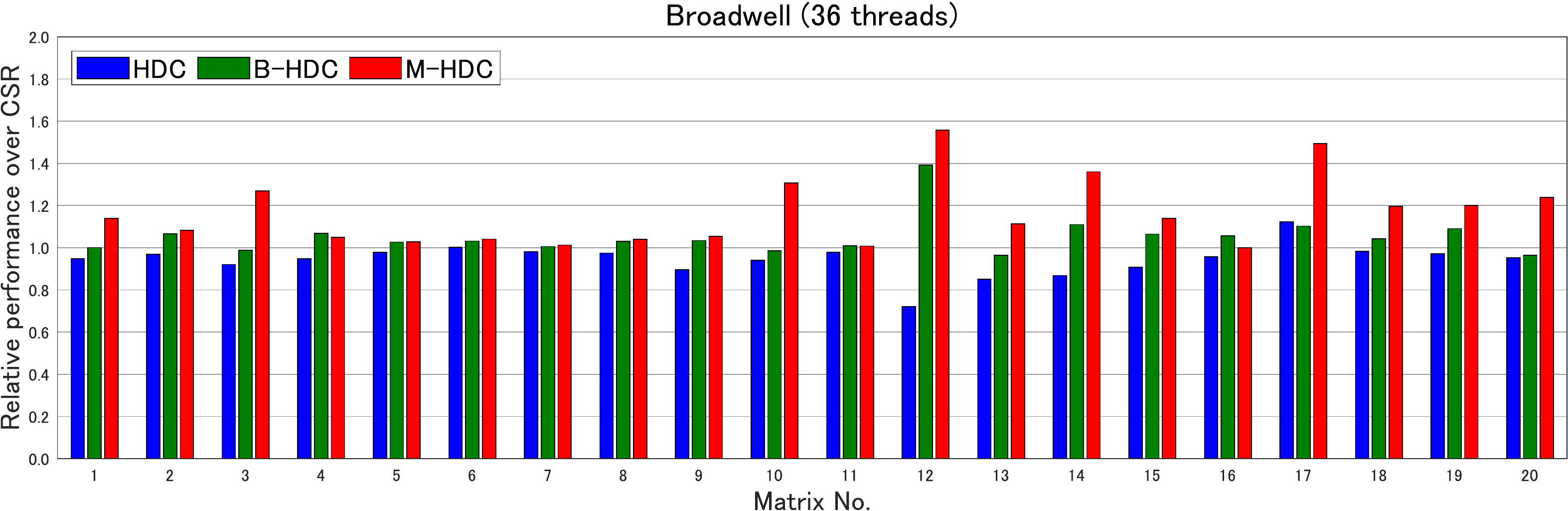}
\\
\vspace{5mm}
\includegraphics[scale=0.5]{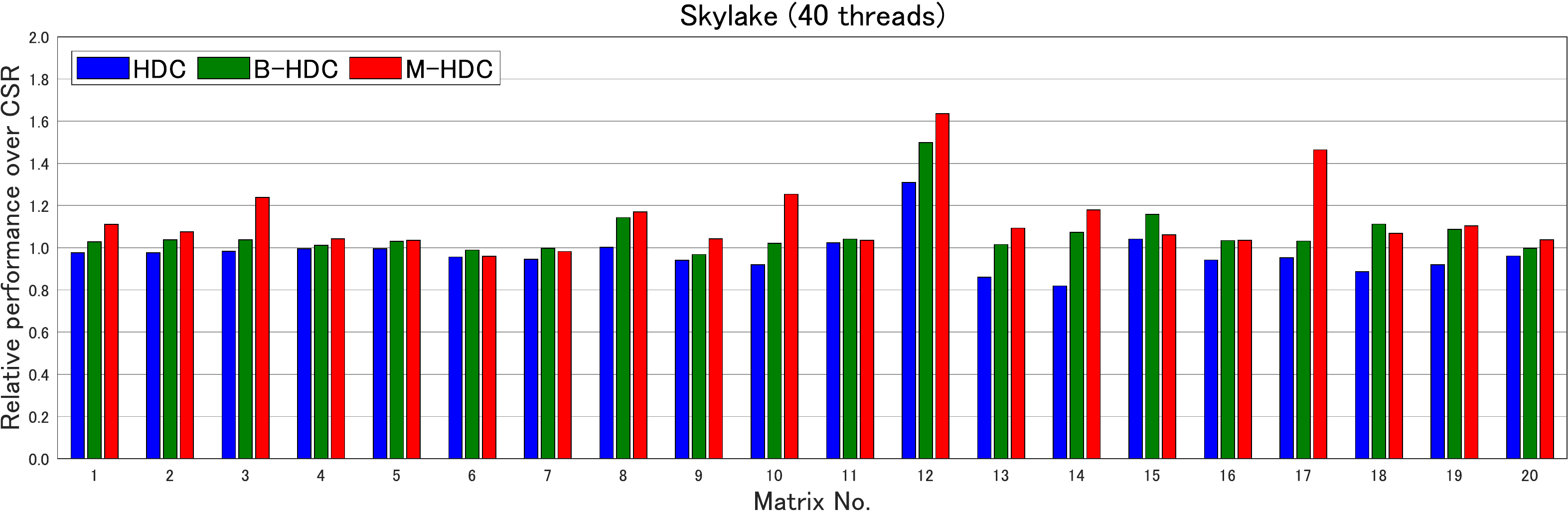}
\\
\vspace{5mm}
\includegraphics[scale=0.5]{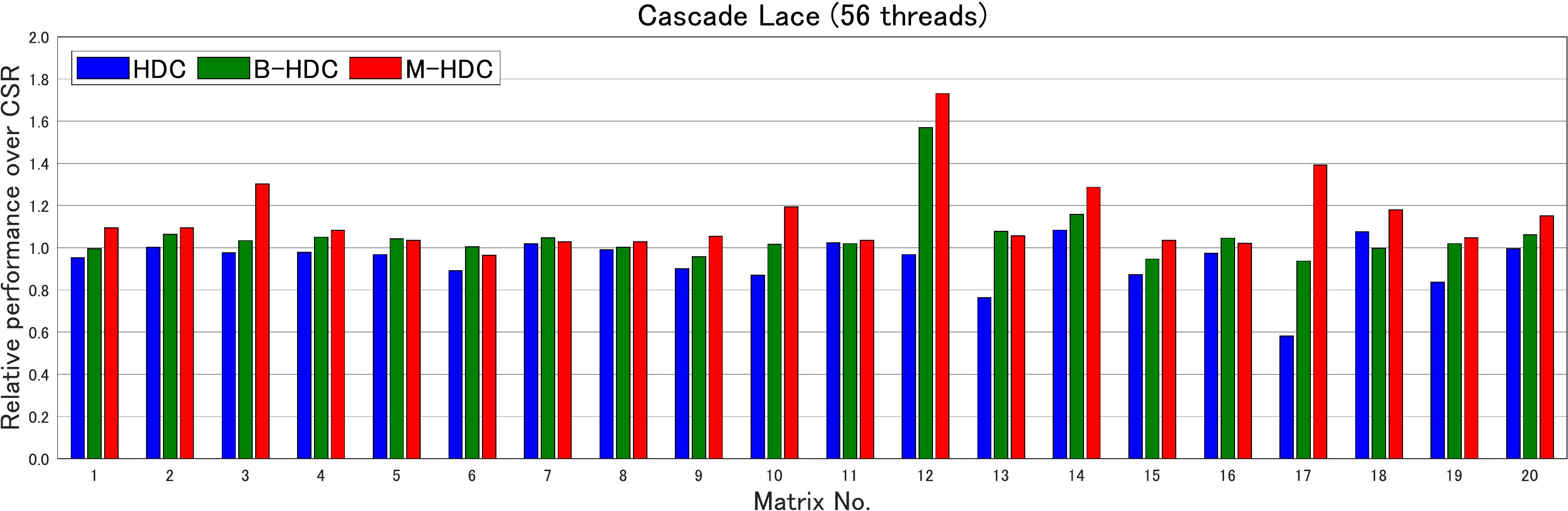}
\end{center}
\caption{Relative performance of the HDC, B-HDC, and M-HDC kernels over the CSR kernel for the test matrices selected from the SuiteSparse Matrix Collection.}
\label{fig:compare_ssmc}
\end{figure*}

\begin{figure*}[t]
\begin{center}
\includegraphics[scale=0.5]{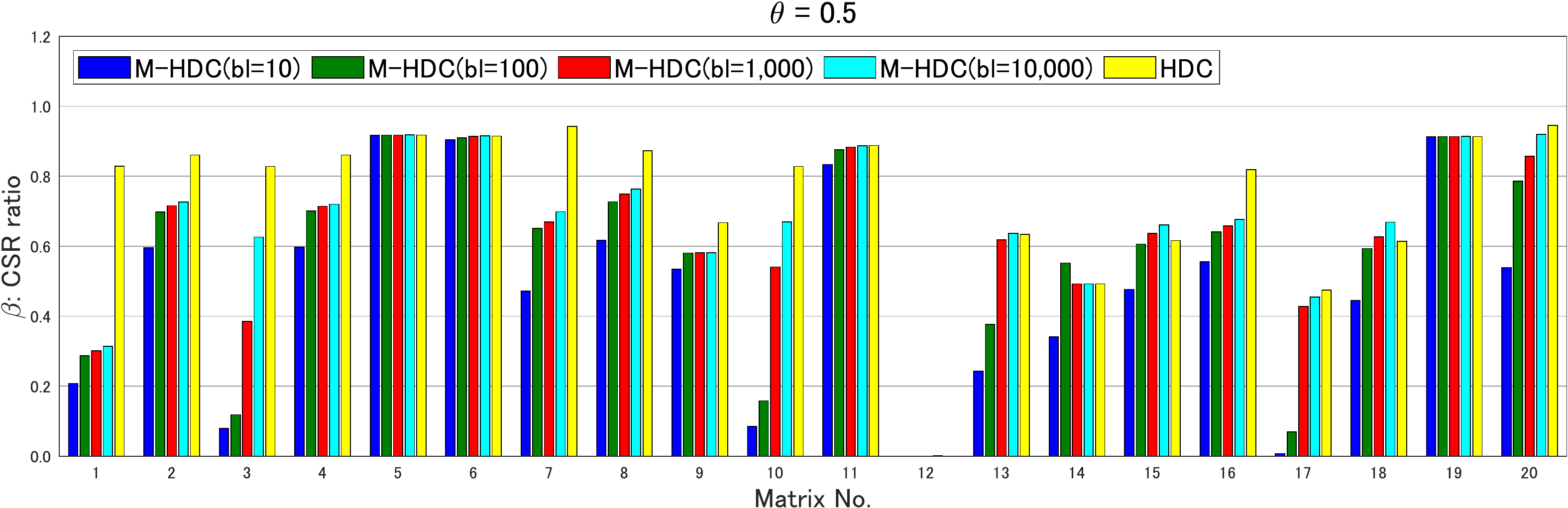}
\end{center}
\caption{The CSR rate in the HDC (or B-HDC) and M-HDC kernels for the test matrices selected from the SuiteSparse Matrix Collection.}
\label{fig:ssmc_csr_ratio}
\end{figure*}

\begin{figure*}[t]
\begin{center}
\subfloat[Matrix \#1]
{
\includegraphics[scale=0.24]{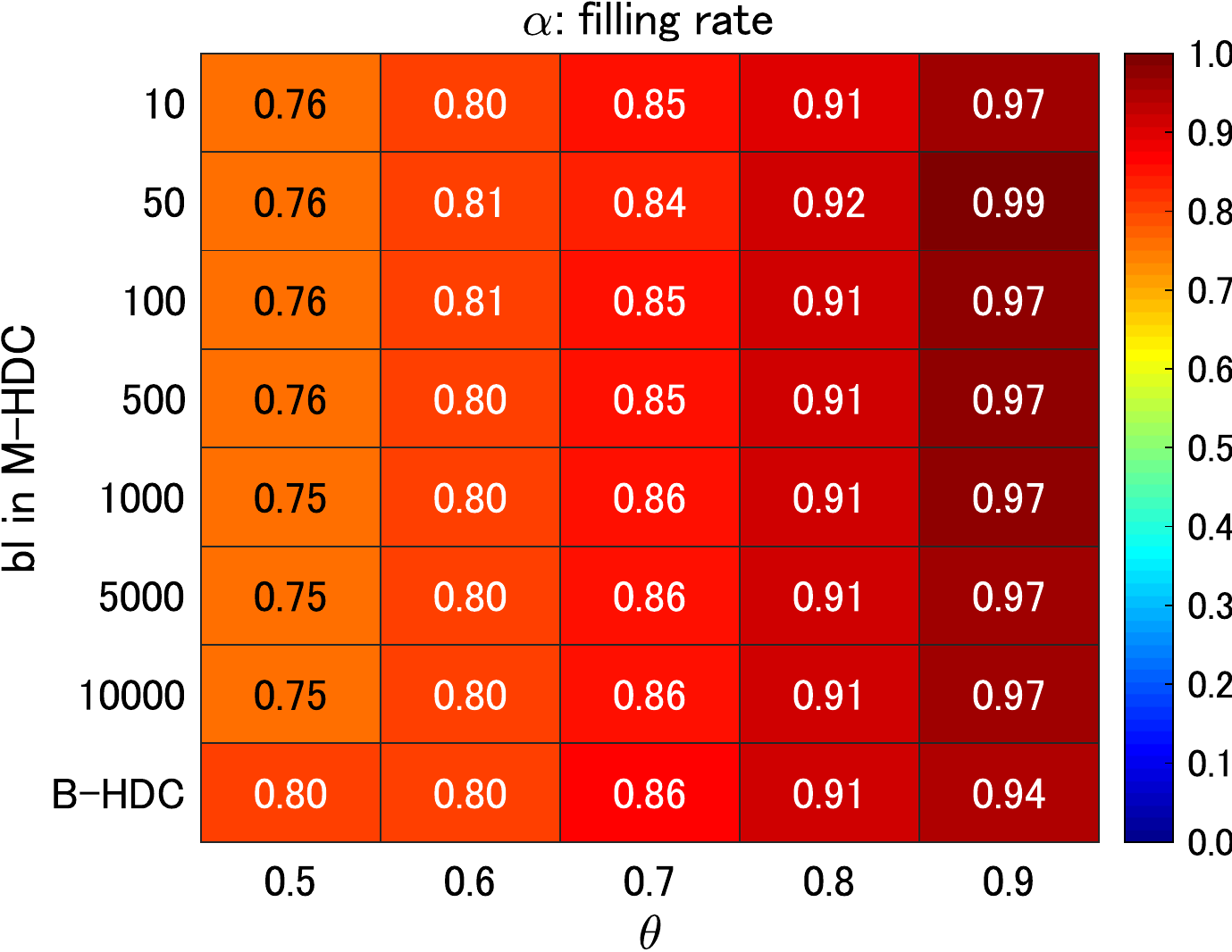}
\includegraphics[scale=0.24]{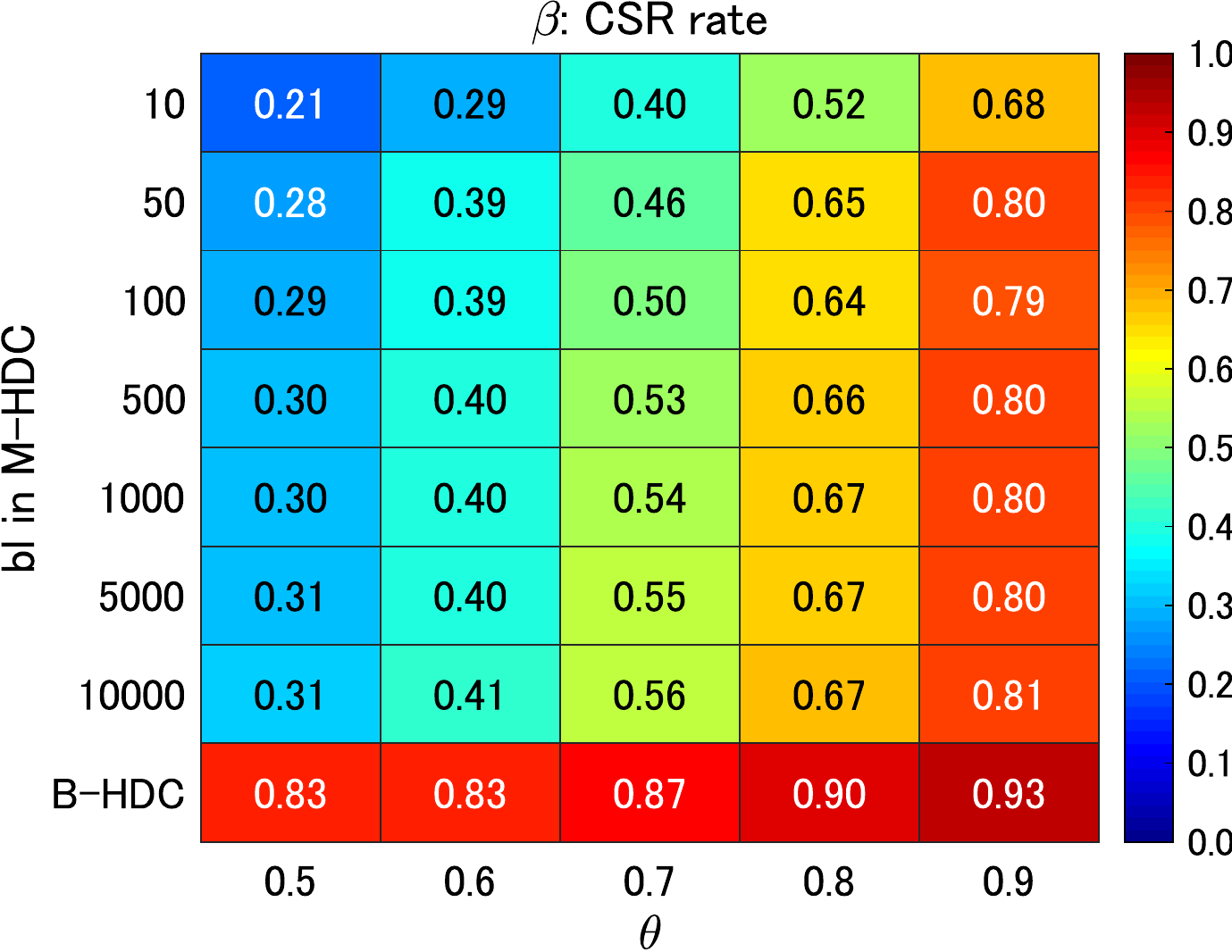}
\includegraphics[scale=0.24]{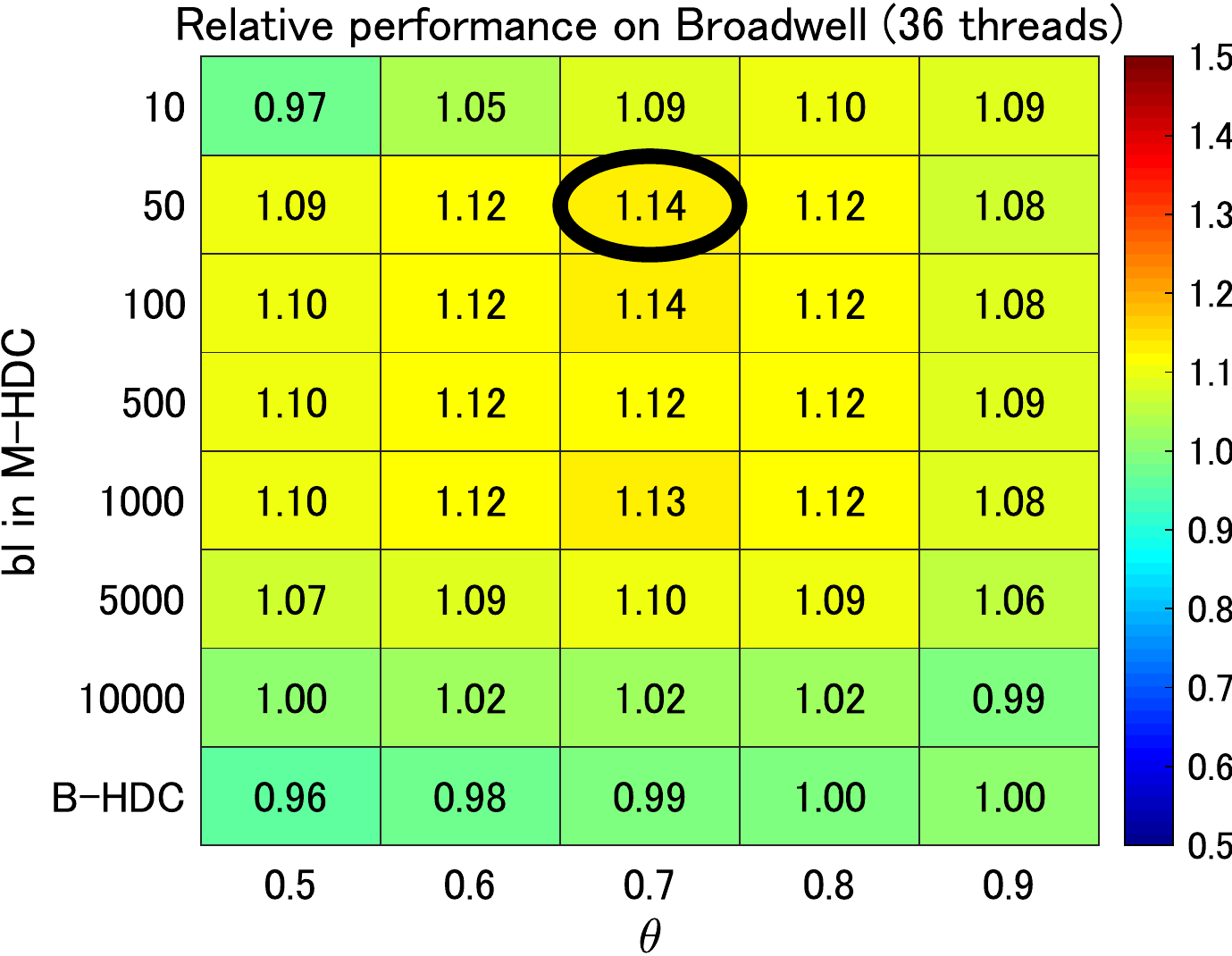}
\includegraphics[scale=0.24]{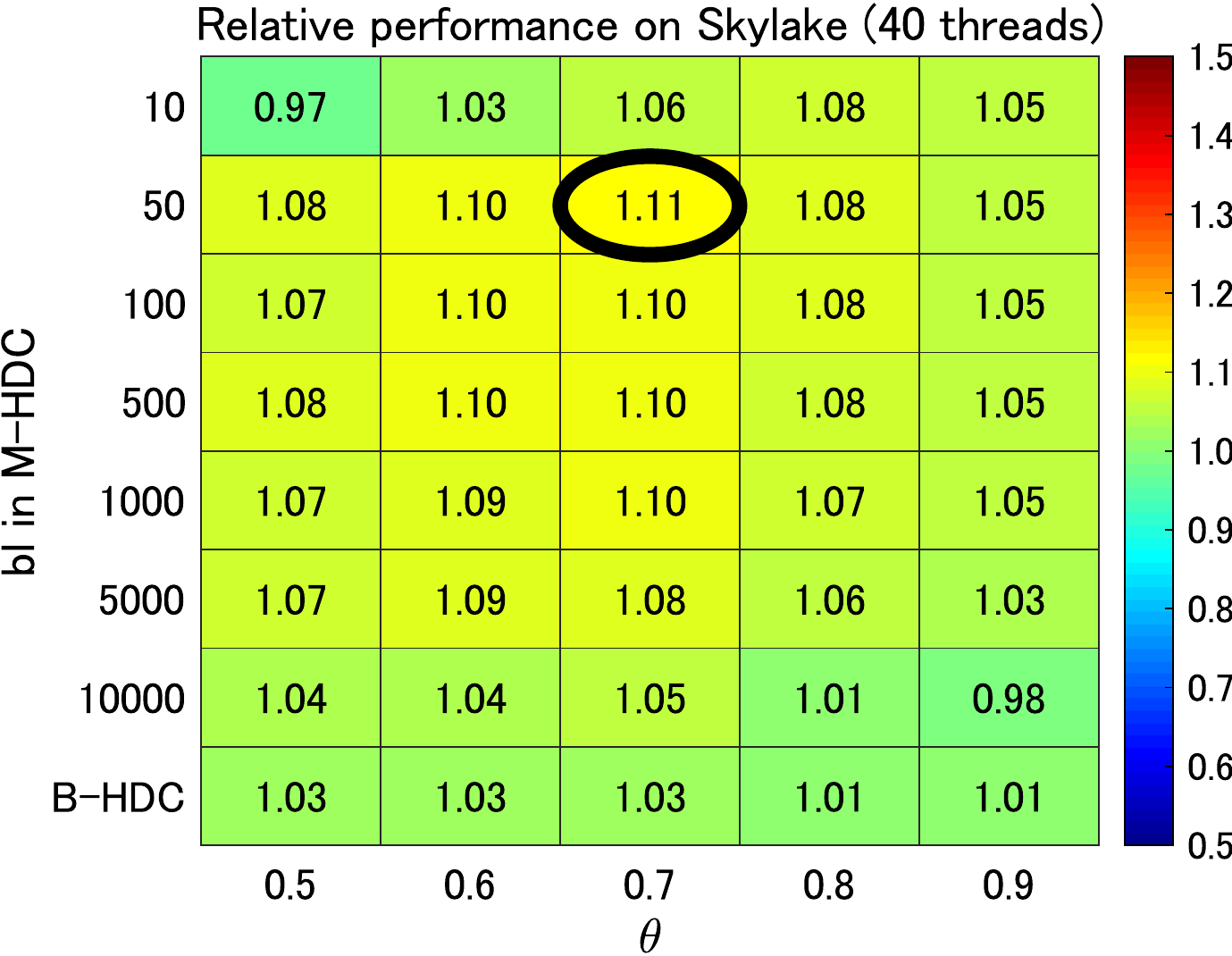}
\includegraphics[scale=0.24]{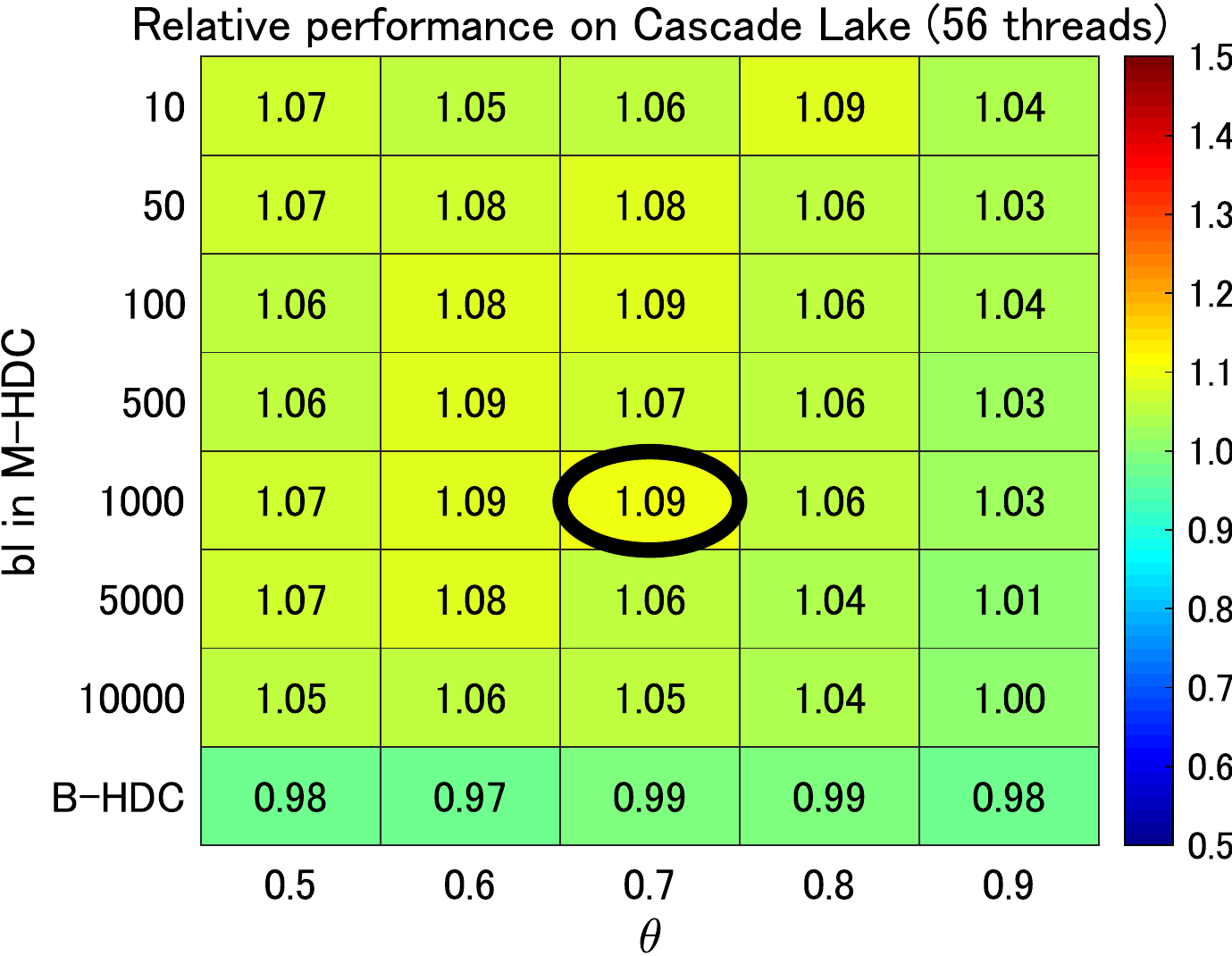}
}
\hfill
\subfloat[Matrix \#3]
{
\includegraphics[scale=0.24]{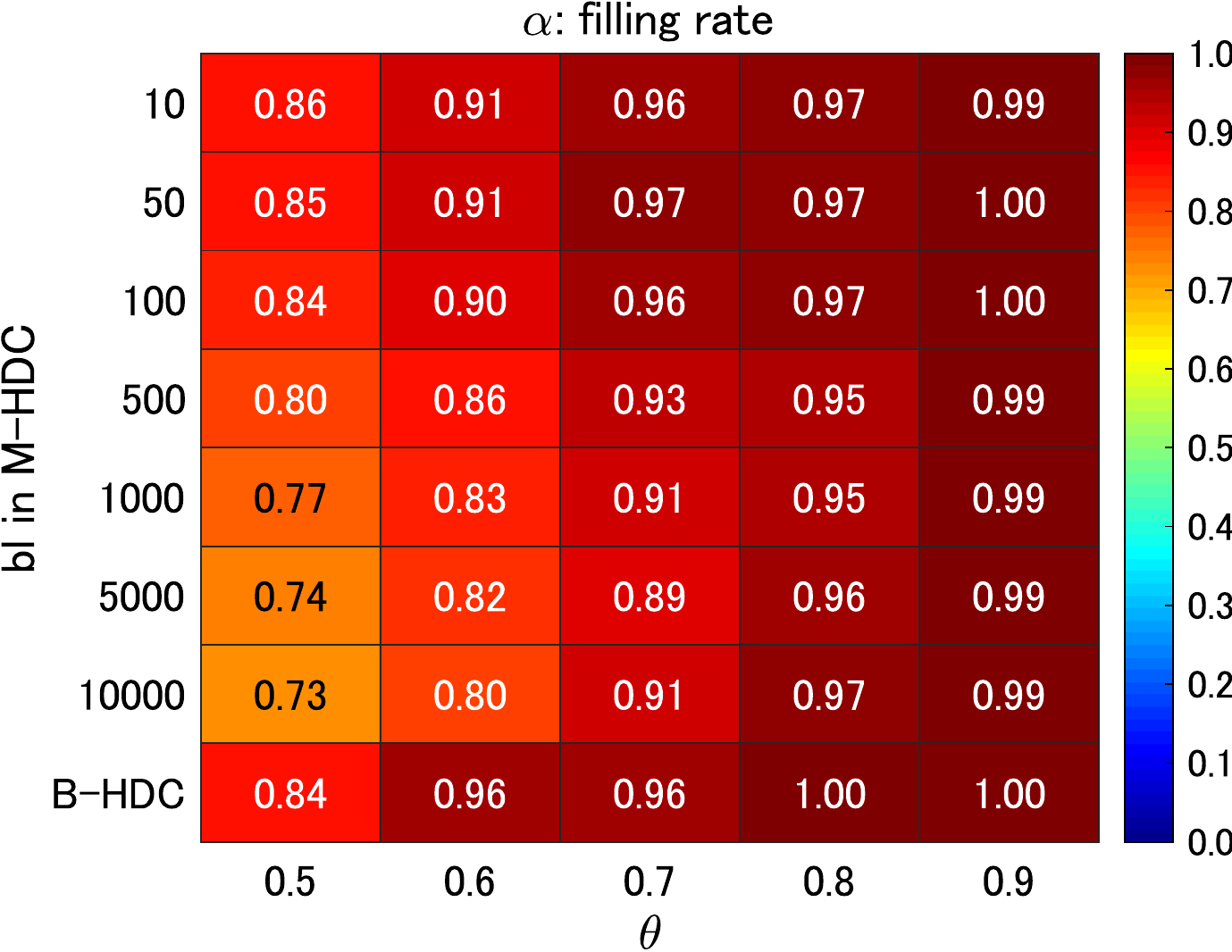}
\includegraphics[scale=0.24]{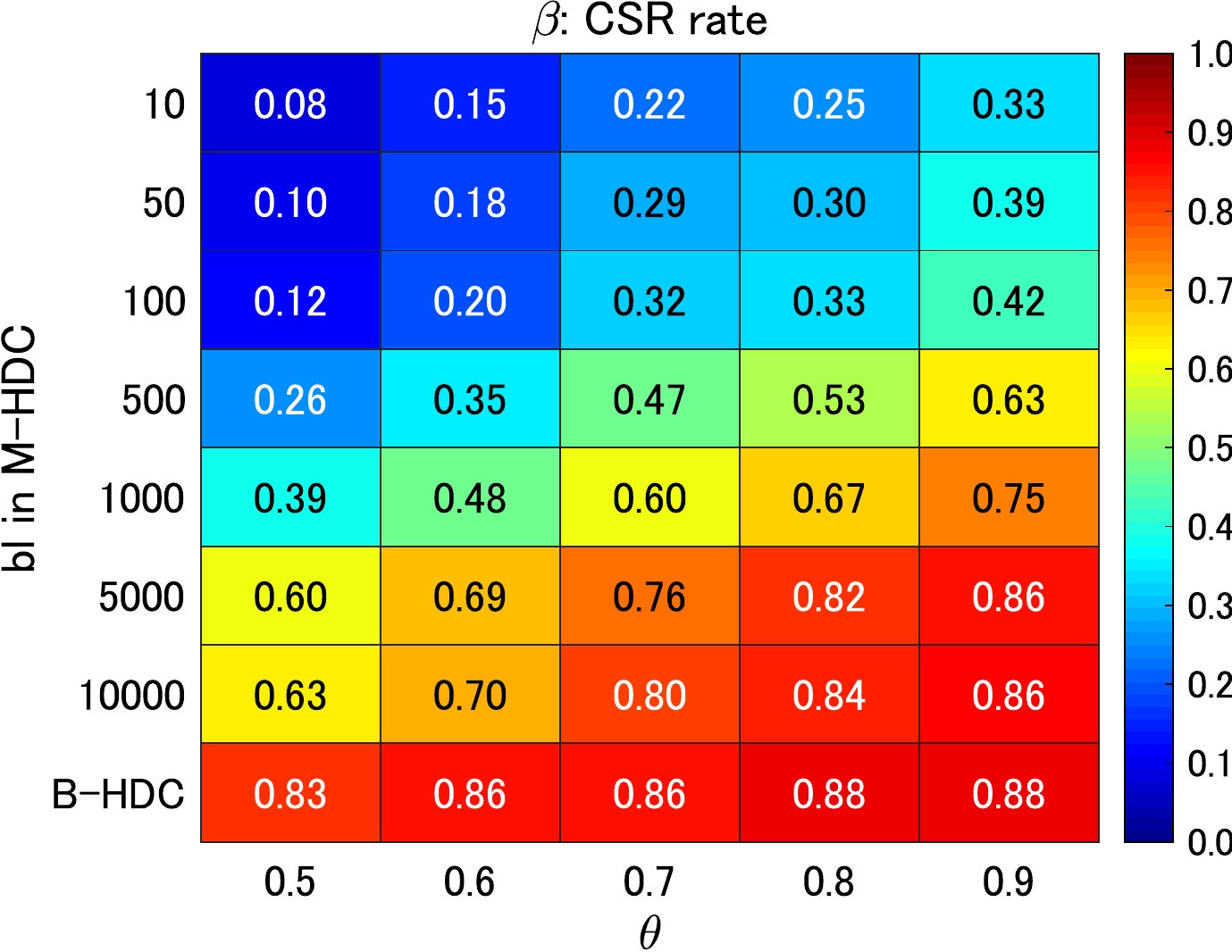}
\includegraphics[scale=0.24]{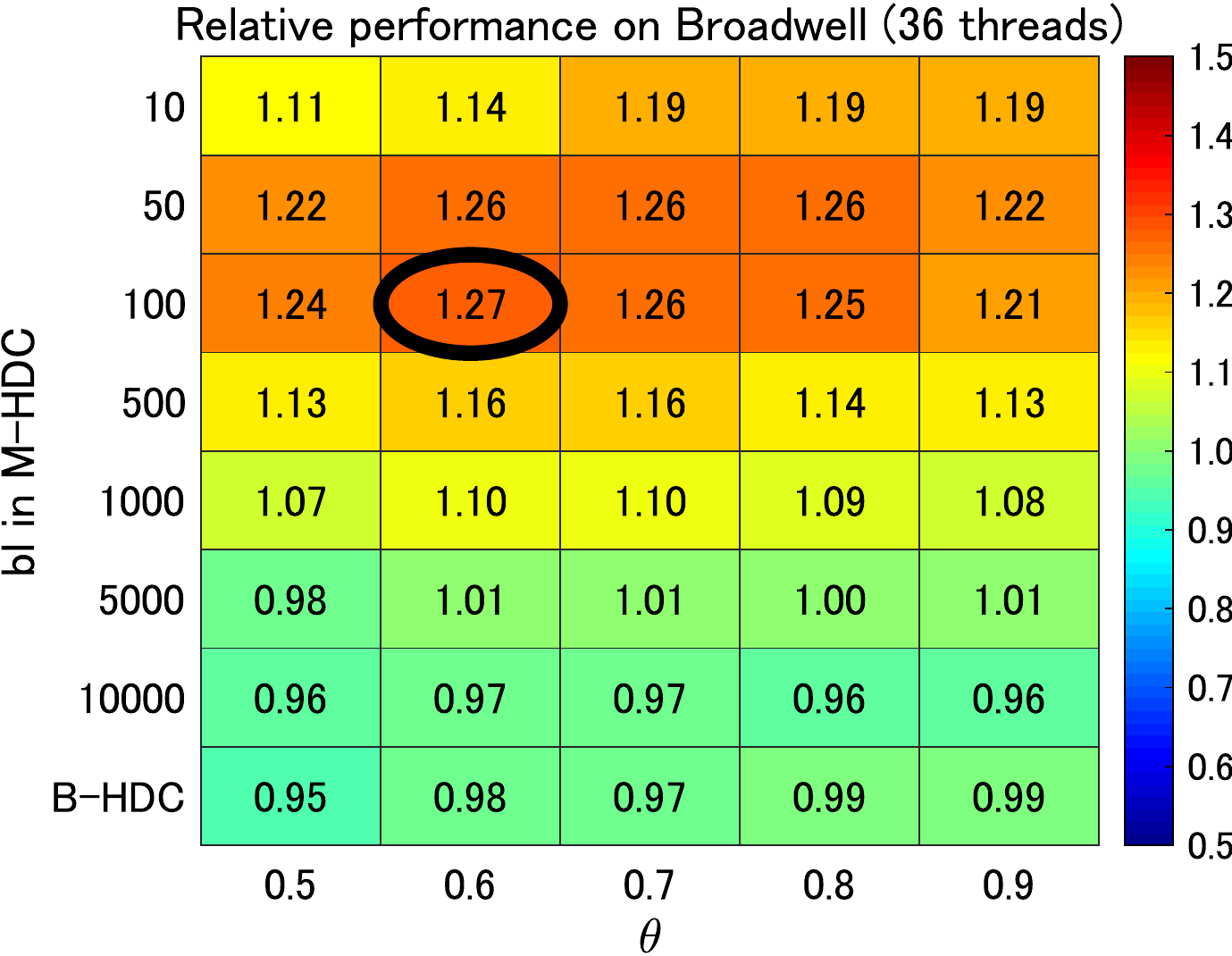}
\includegraphics[scale=0.24]{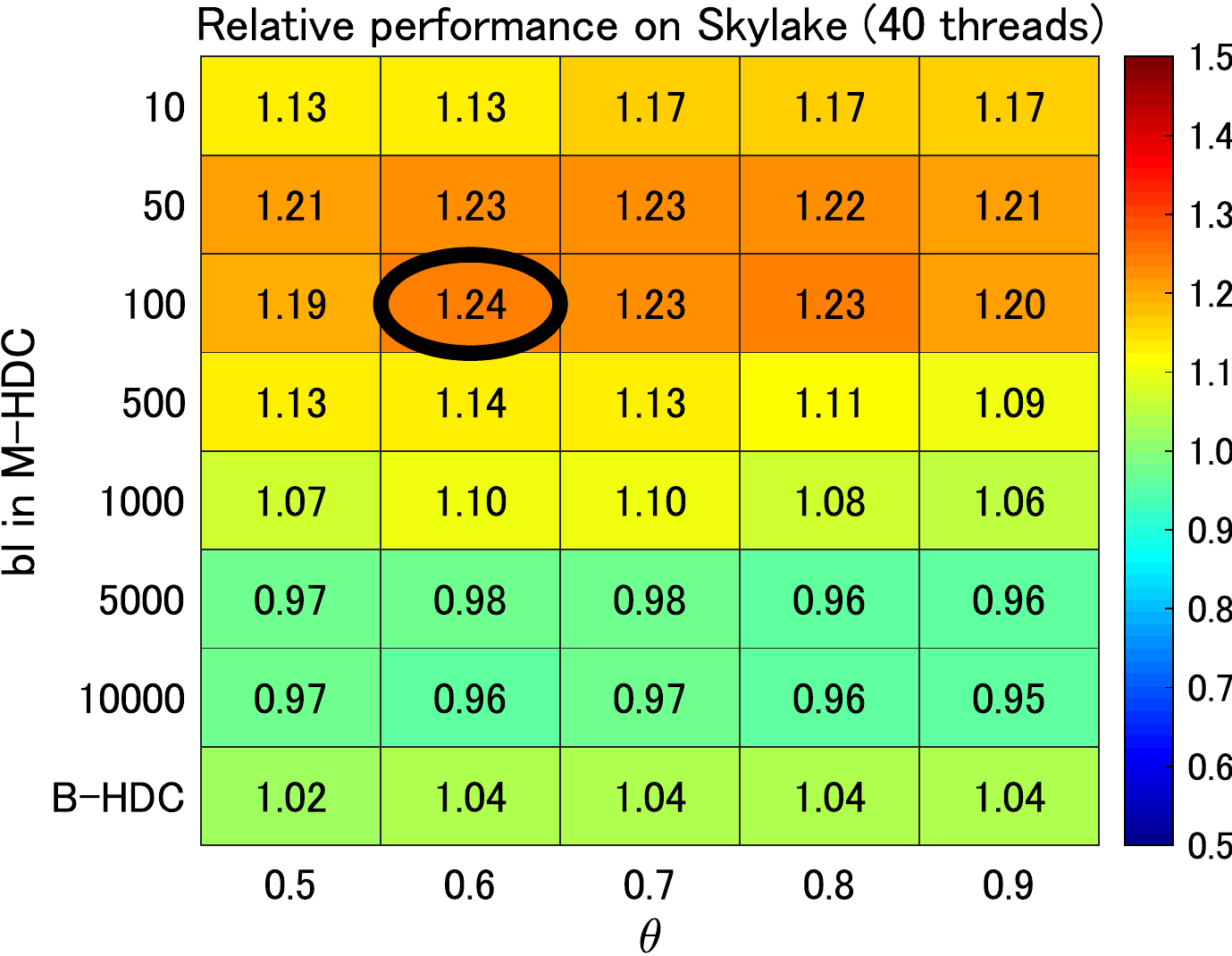}
\includegraphics[scale=0.24]{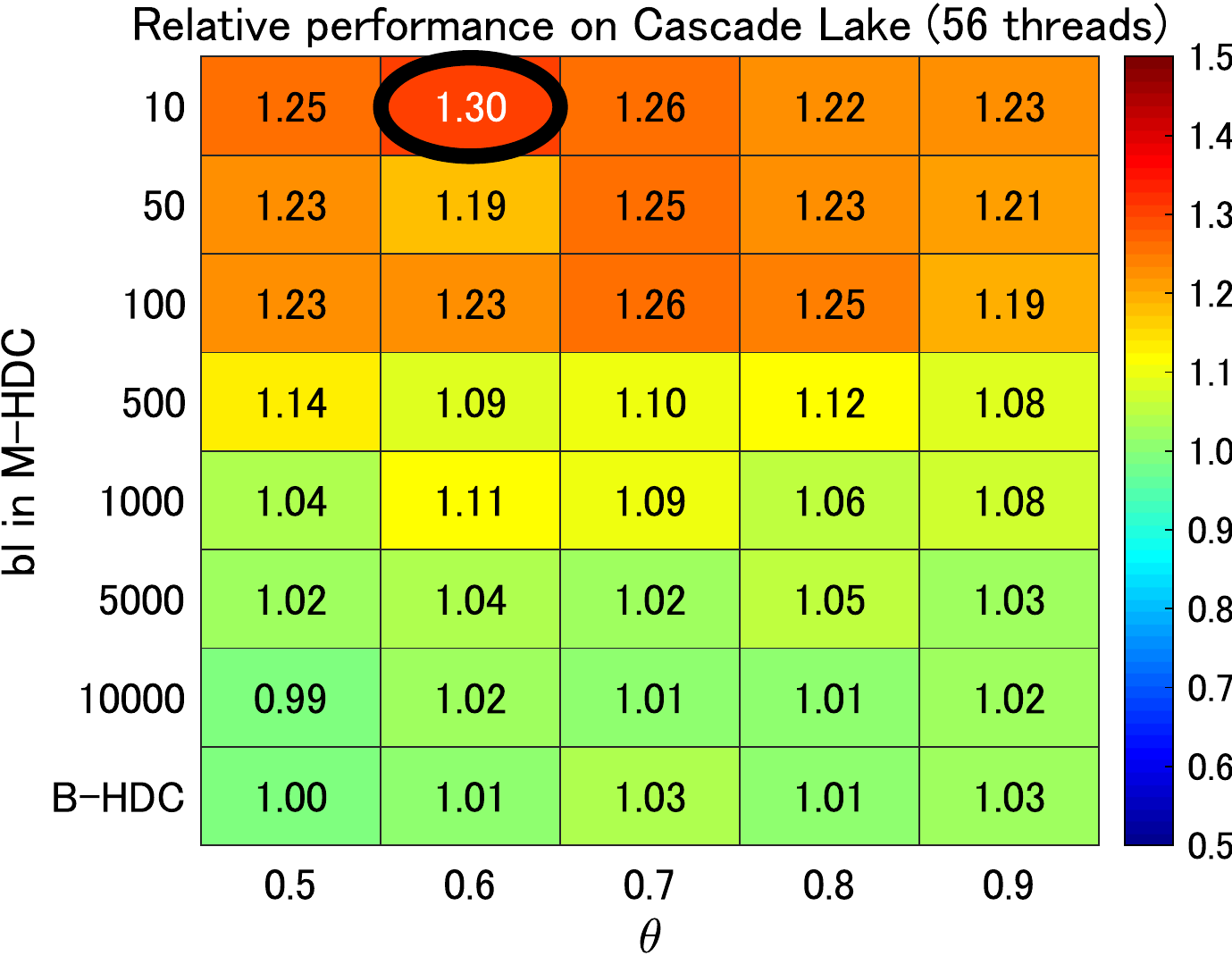}
}
\hfill
\subfloat[Matrix \#10]
{
\includegraphics[scale=0.24]{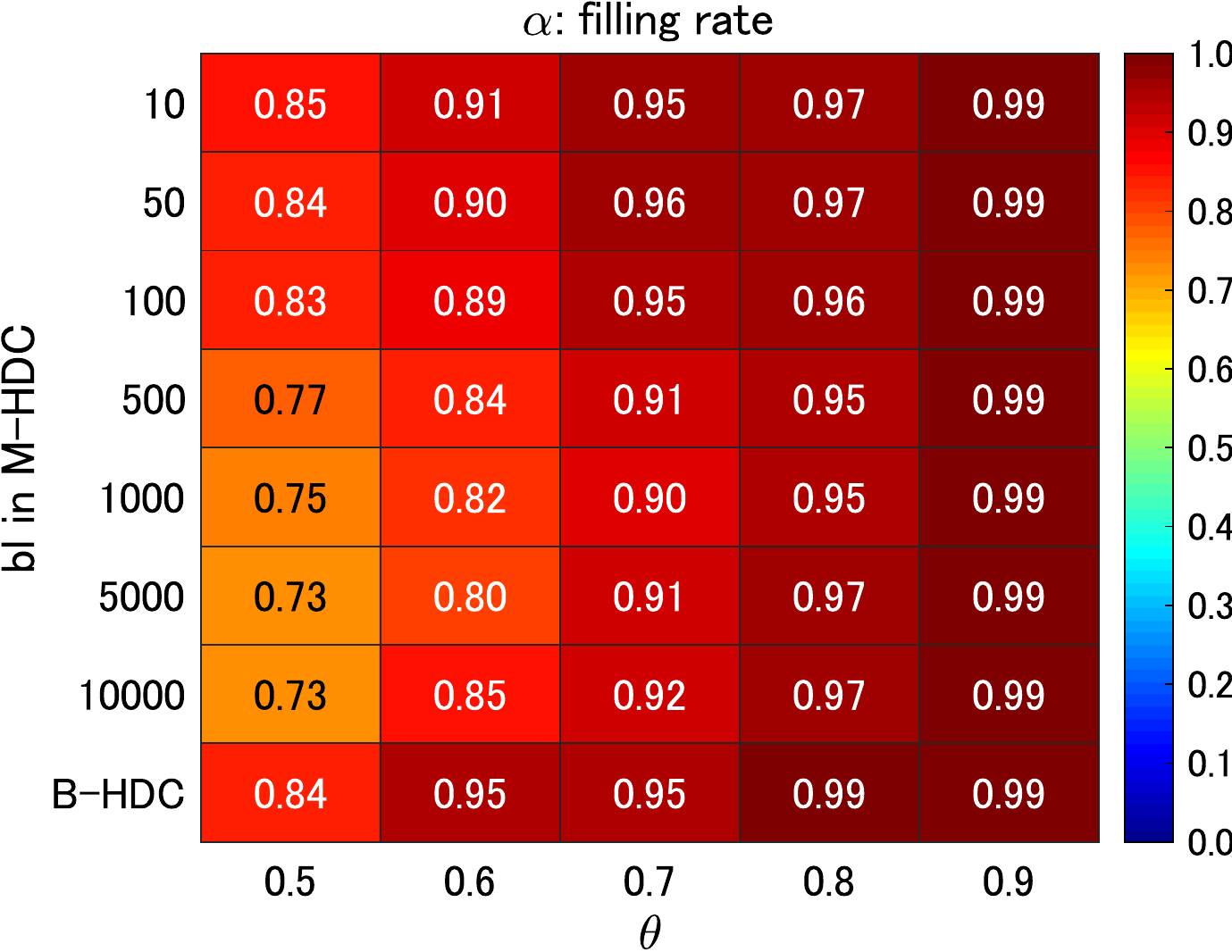}
\includegraphics[scale=0.24]{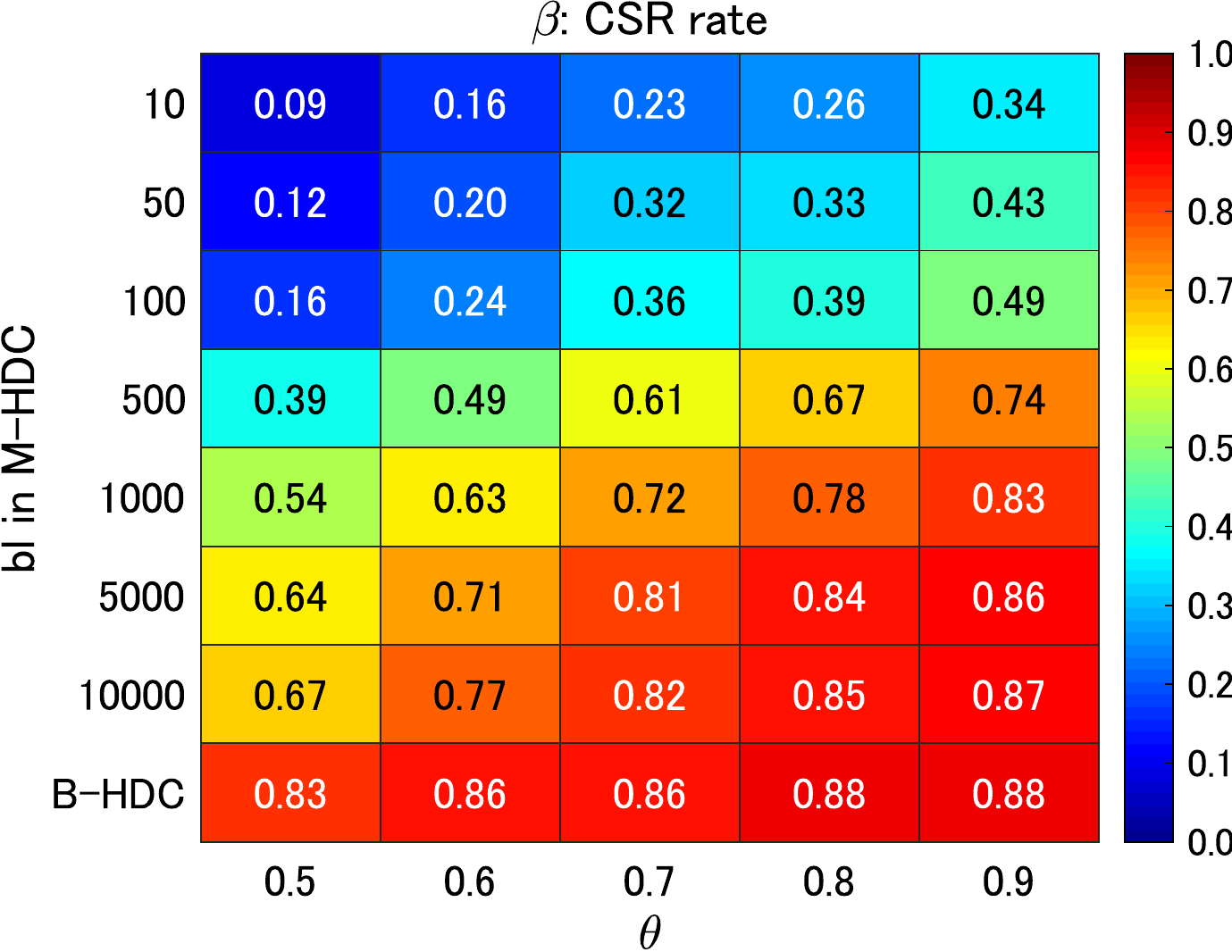}
\includegraphics[scale=0.24]{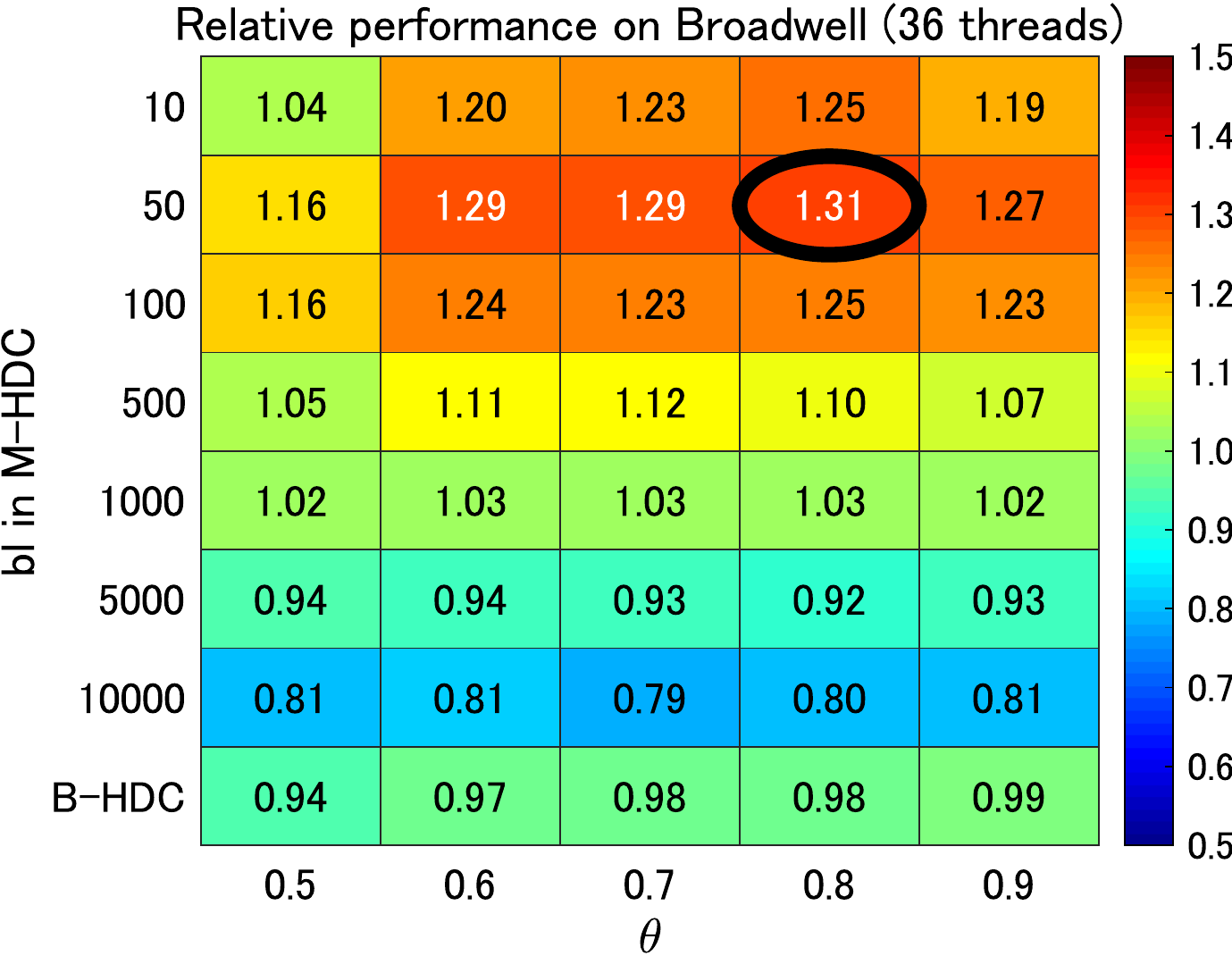}
\includegraphics[scale=0.24]{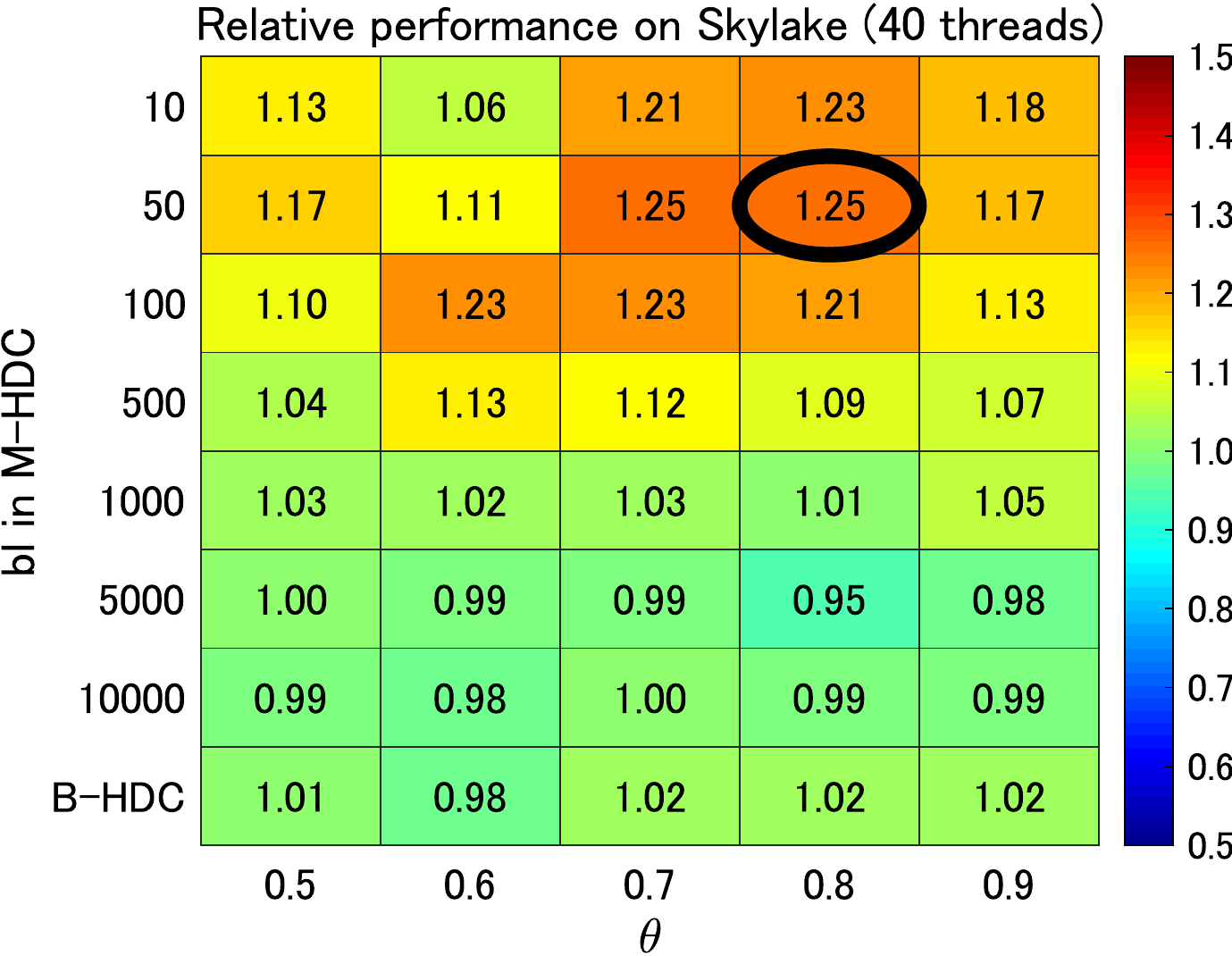}
\includegraphics[scale=0.24]{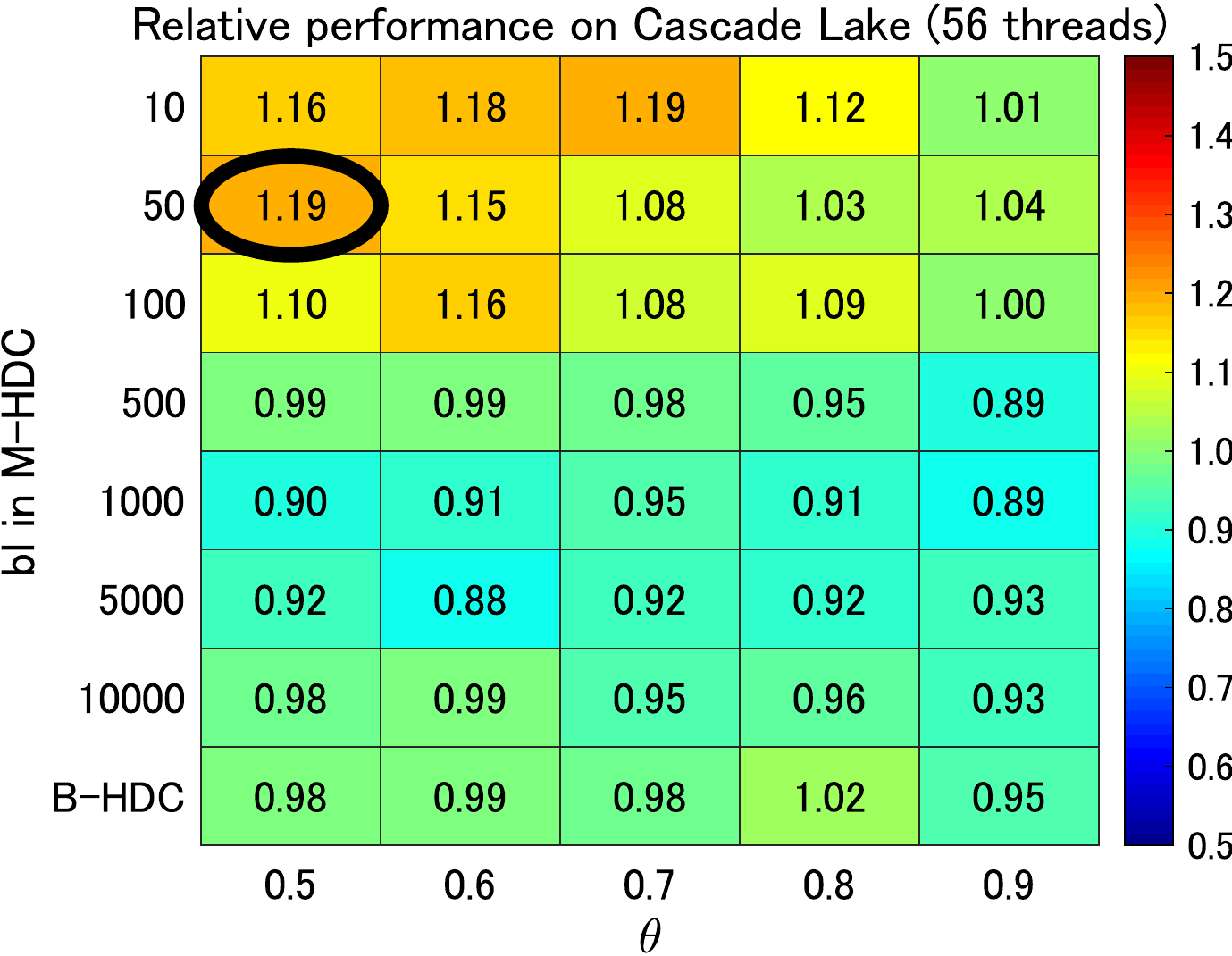}
}
\hfill
\subfloat[Matrix \#13]
{
\includegraphics[scale=0.24]{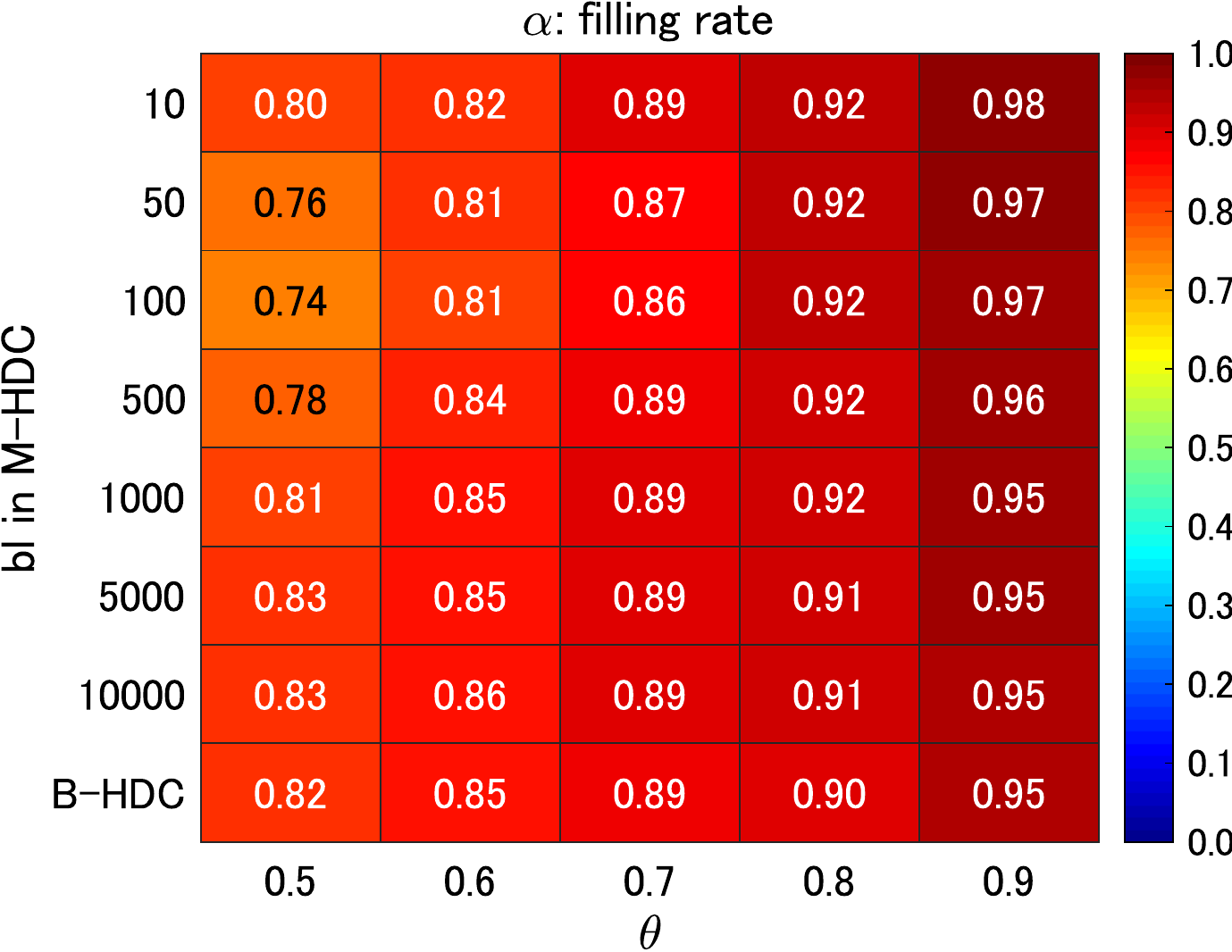}
\includegraphics[scale=0.24]{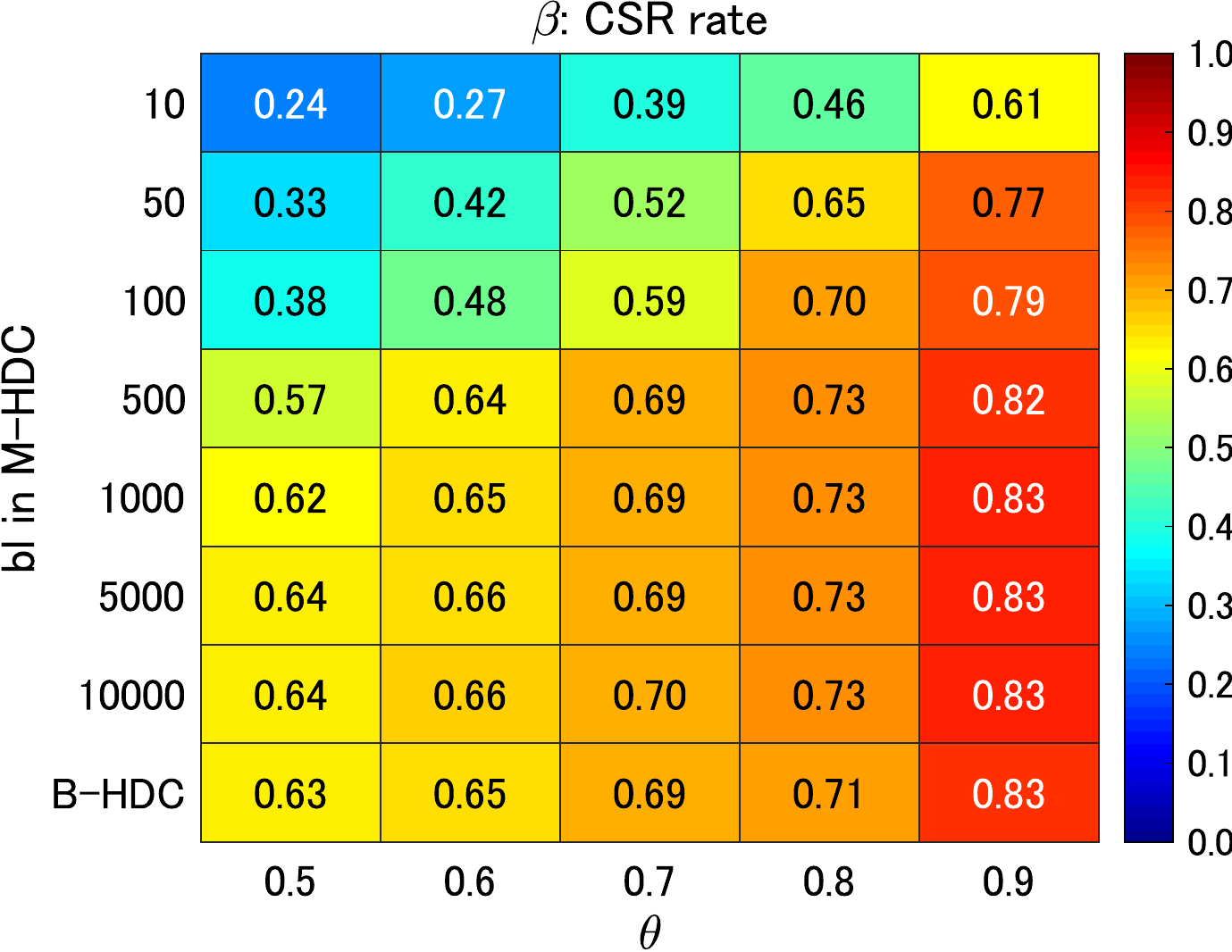}
\includegraphics[scale=0.24]{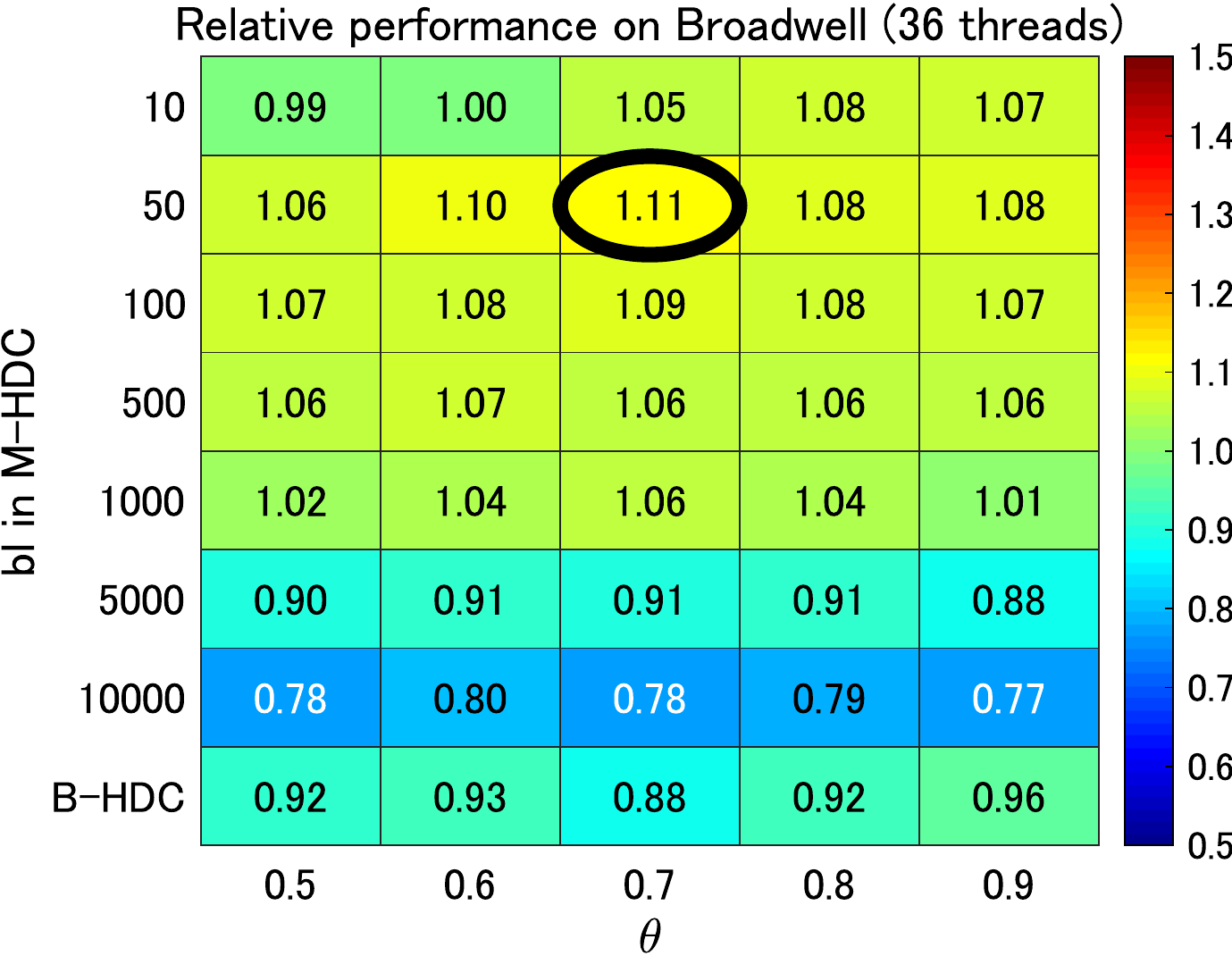}
\includegraphics[scale=0.24]{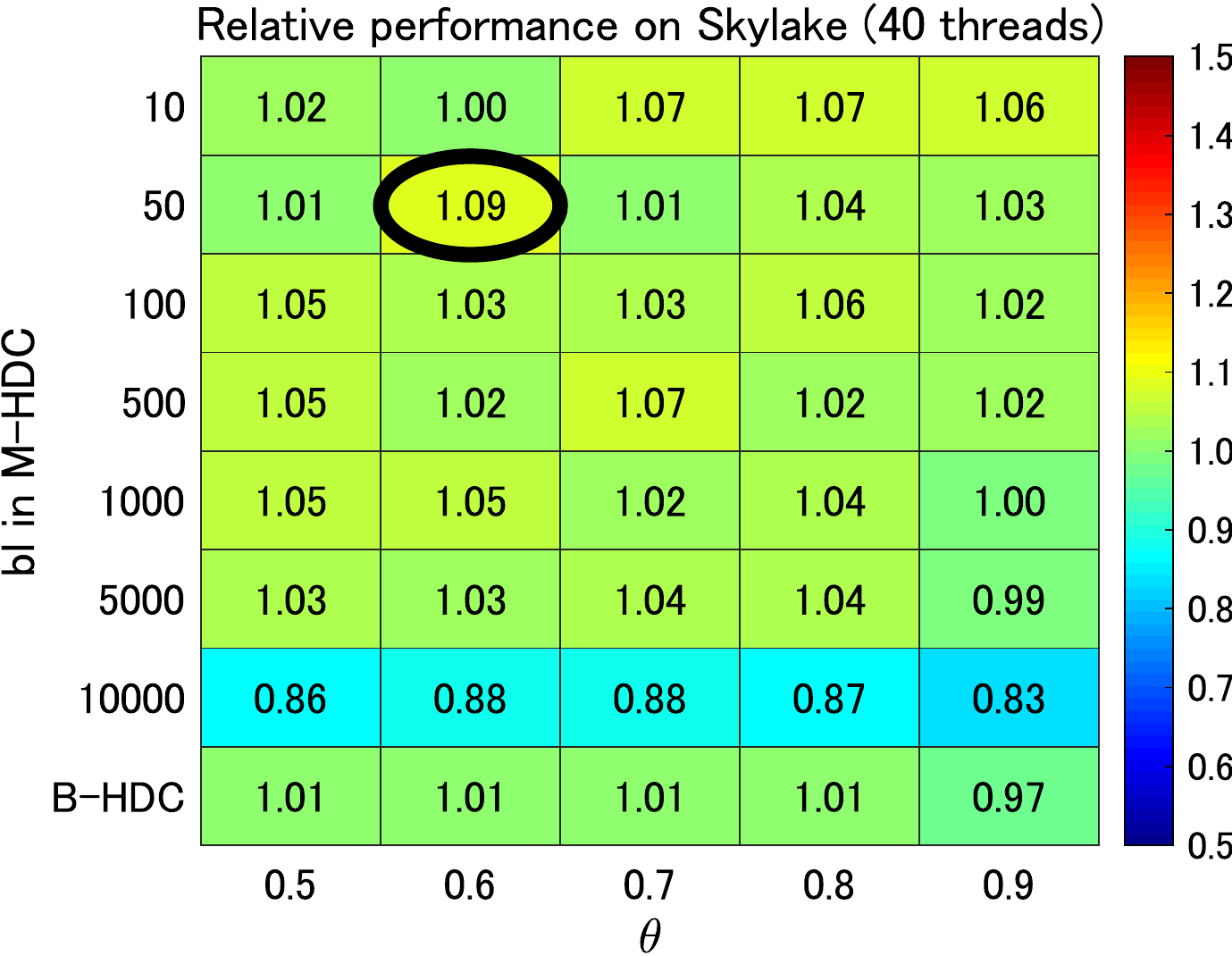}
\includegraphics[scale=0.24]{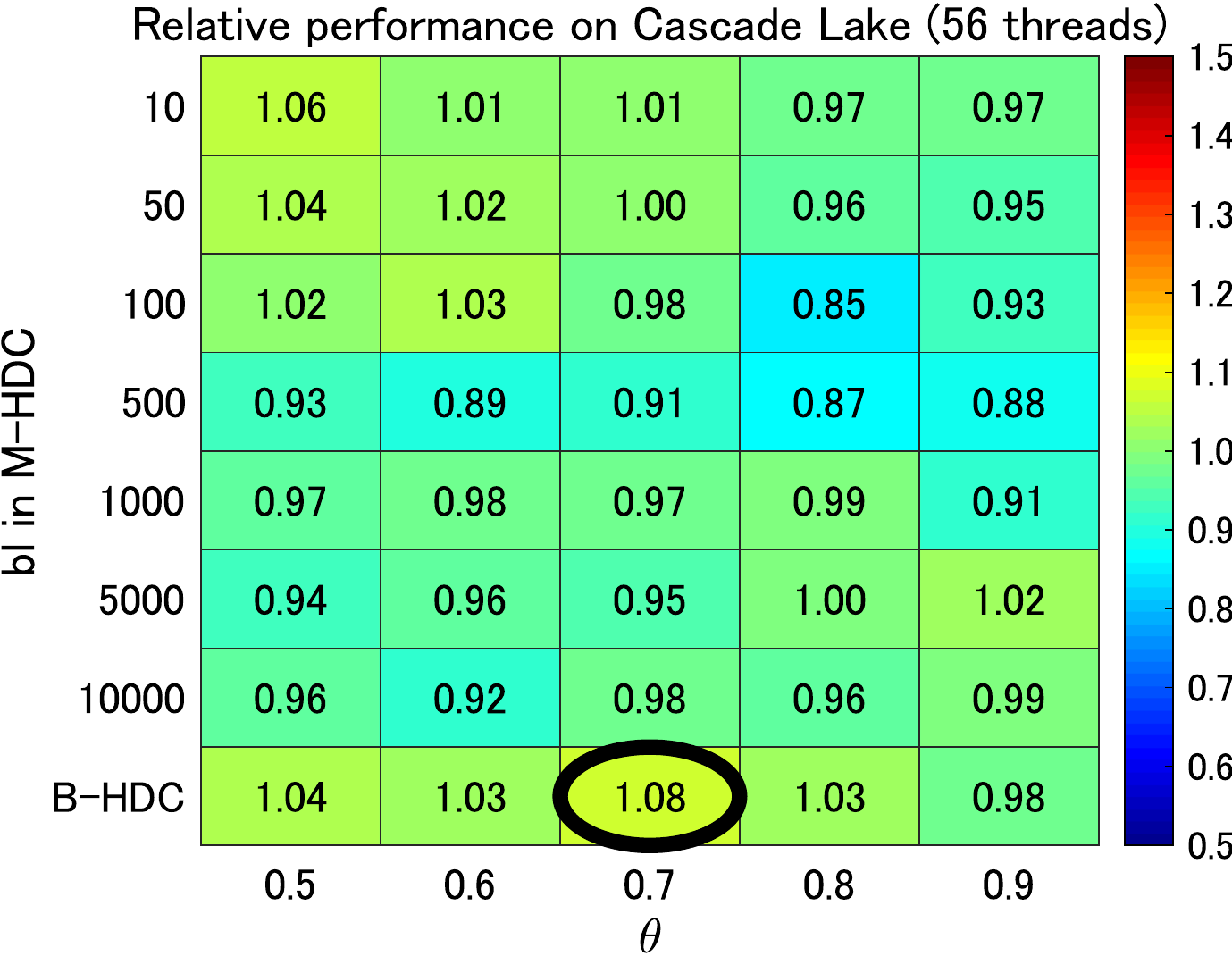}
}
\hfill
\subfloat[Matrix \#14]
{
\includegraphics[scale=0.24]{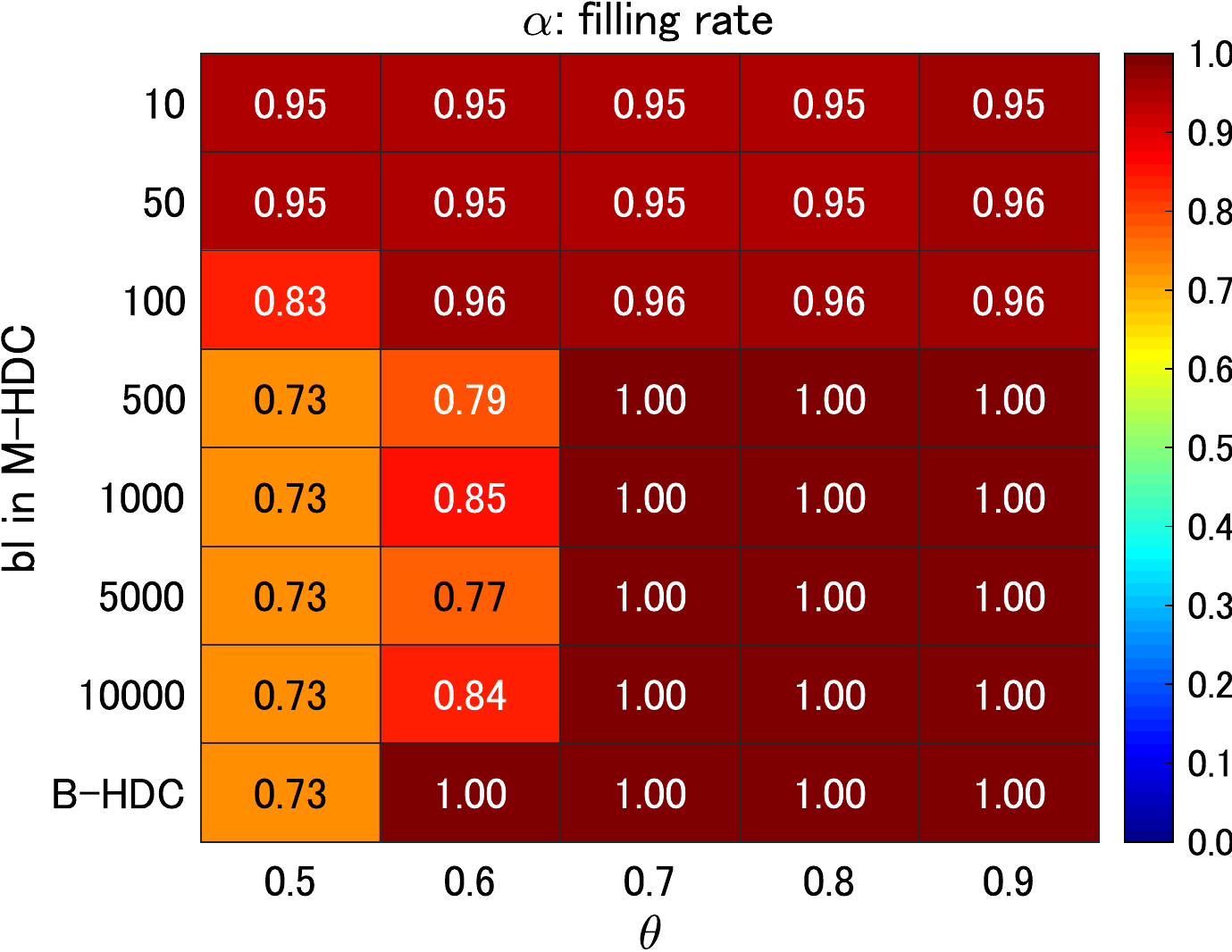}
\includegraphics[scale=0.24]{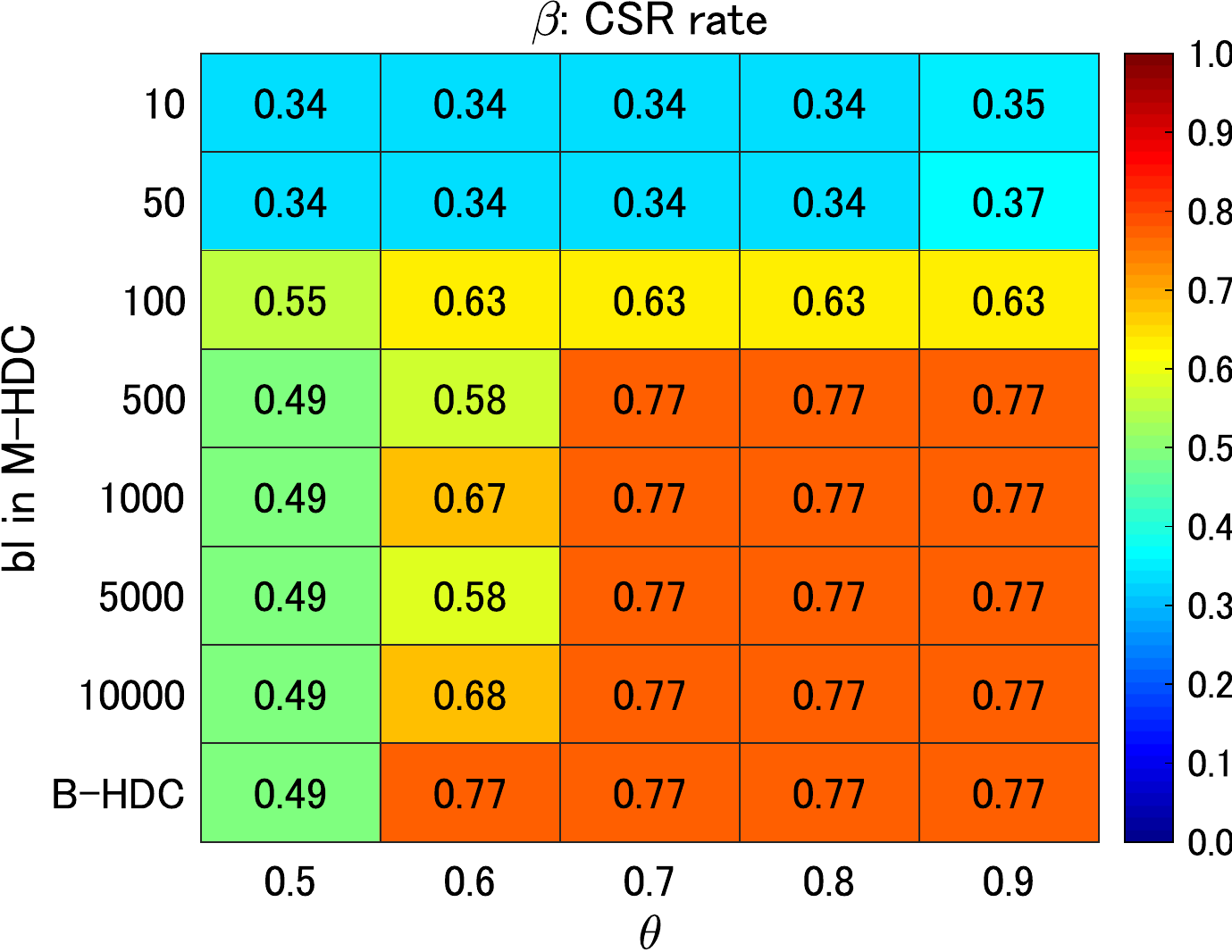}
\includegraphics[scale=0.24]{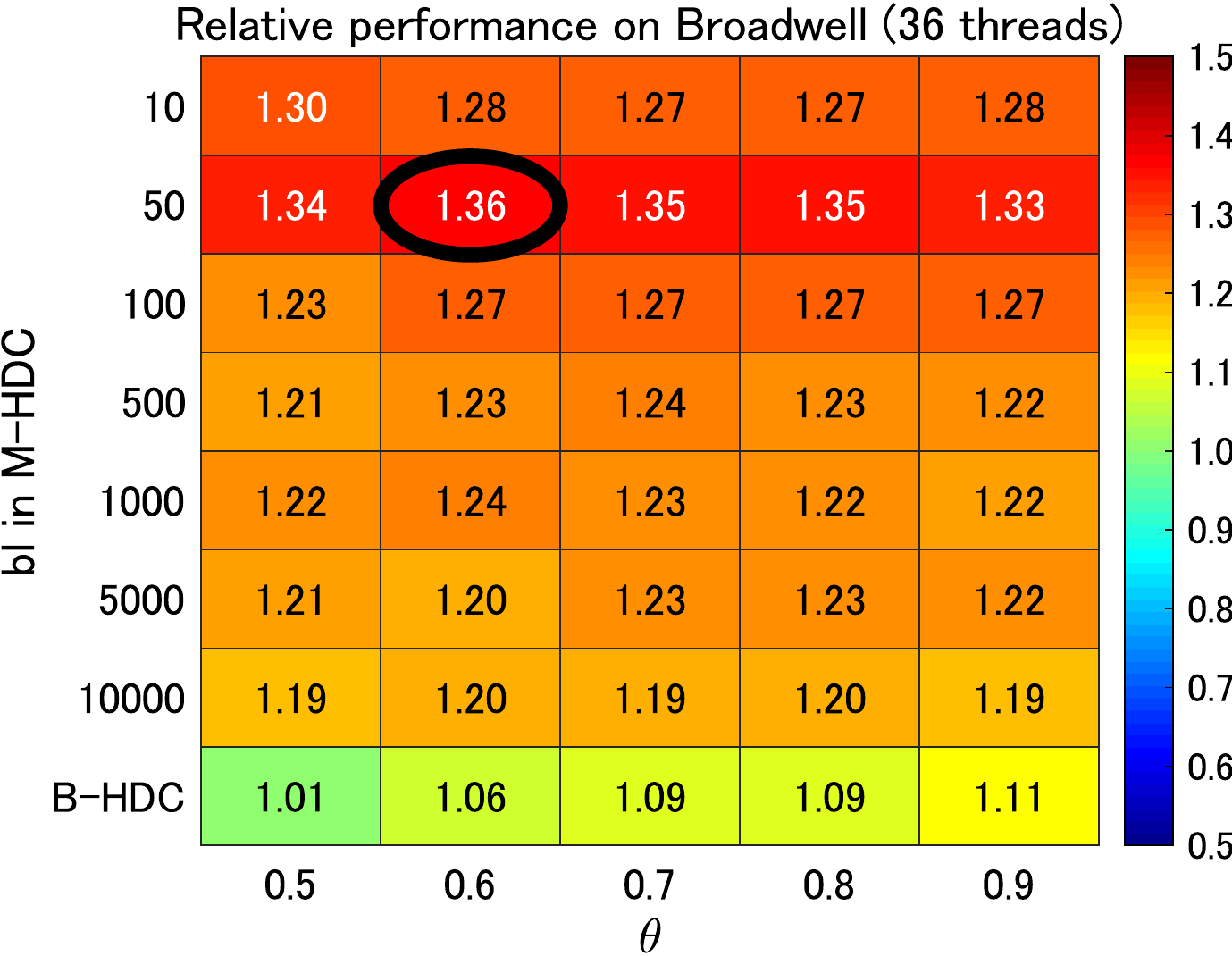}
\includegraphics[scale=0.24]{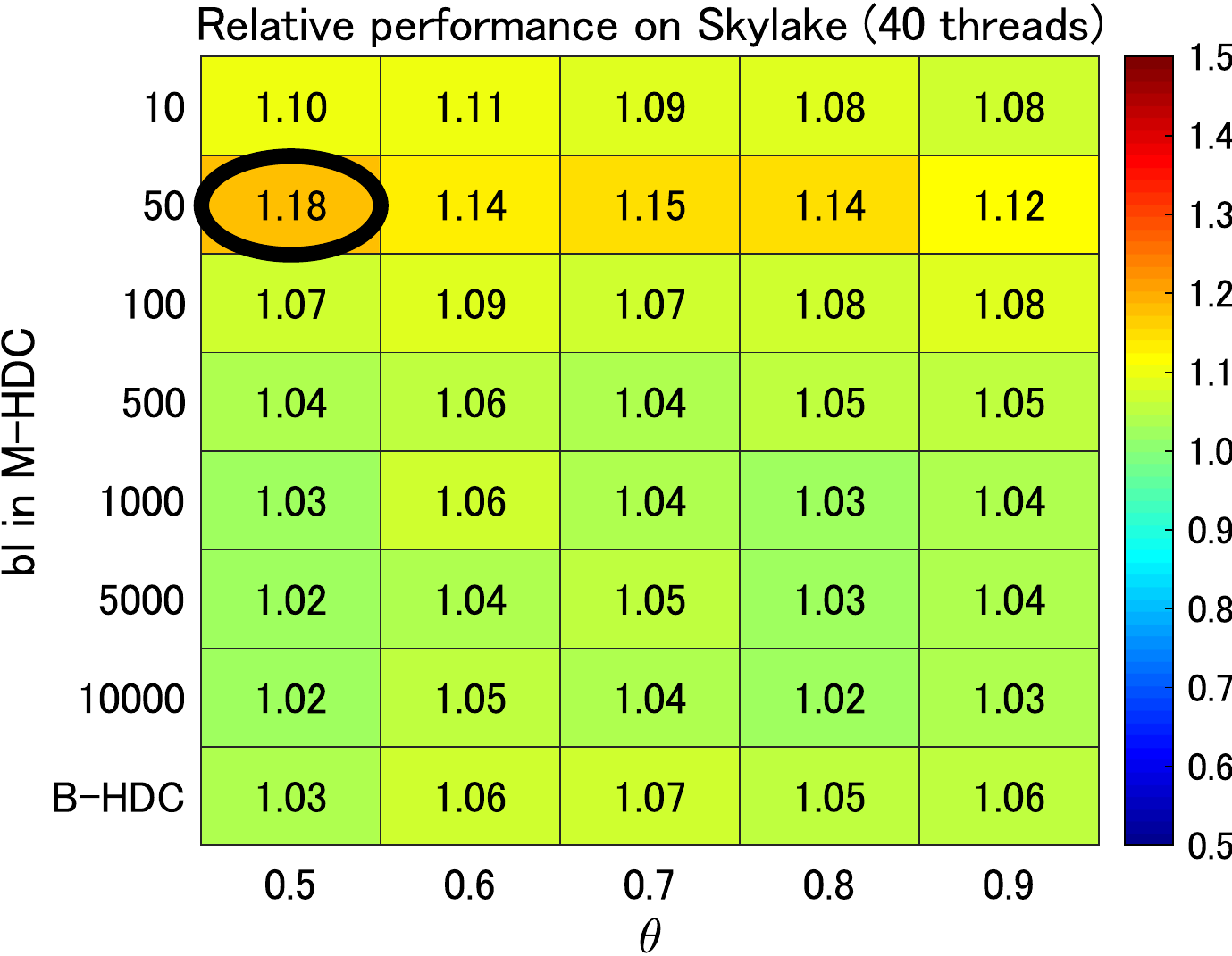}
\includegraphics[scale=0.24]{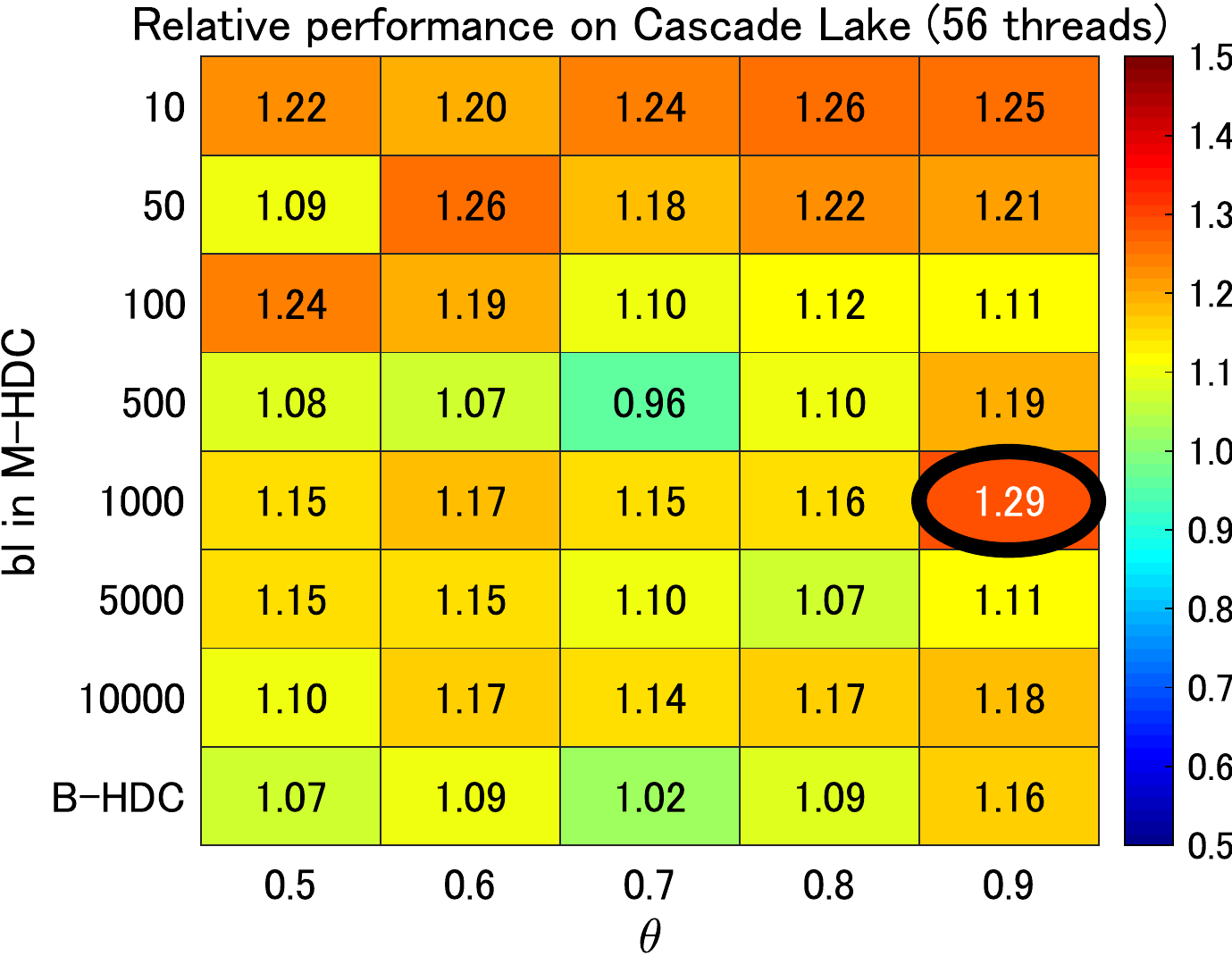}
}
\hfill
\subfloat[Matrix \#17]
{
\includegraphics[scale=0.24]{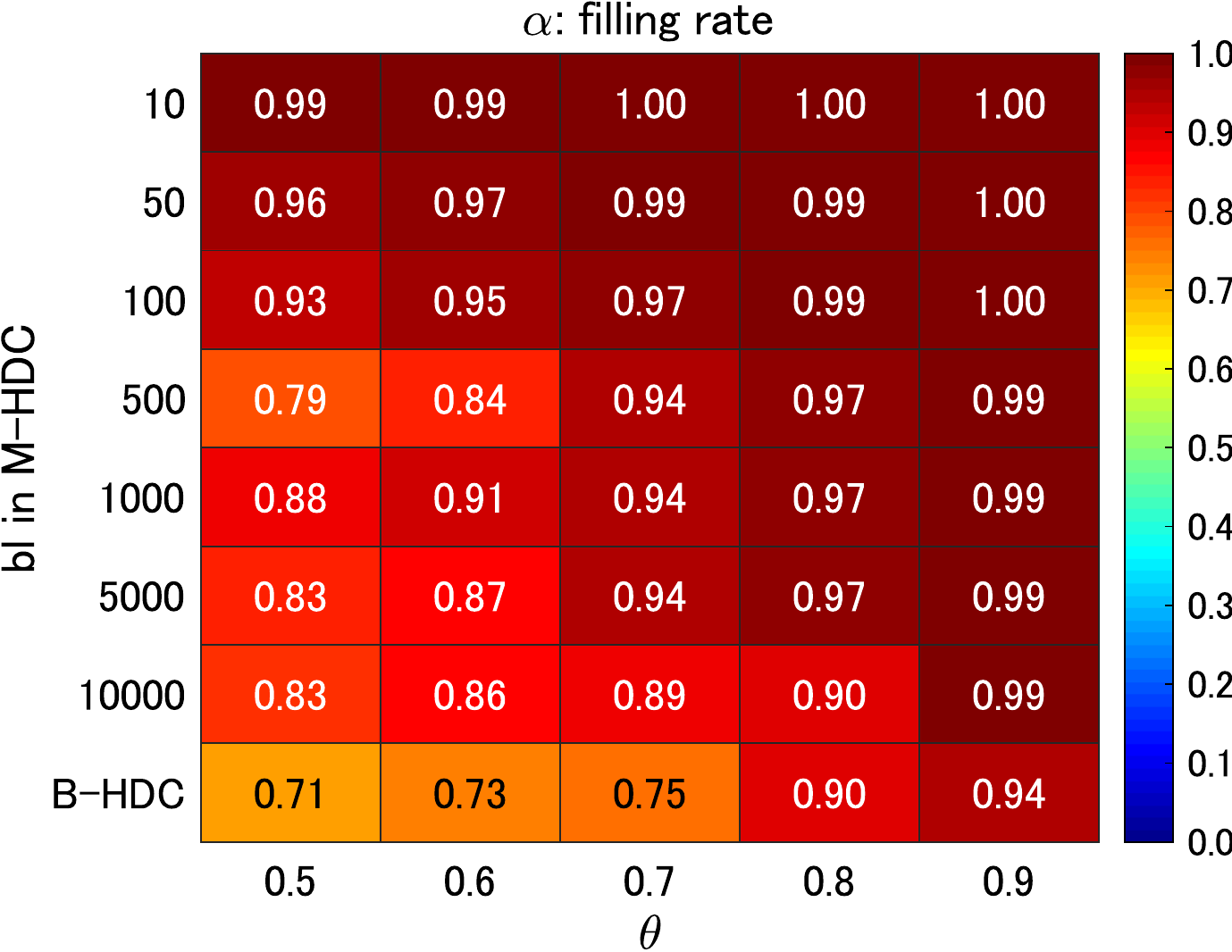}
\includegraphics[scale=0.24]{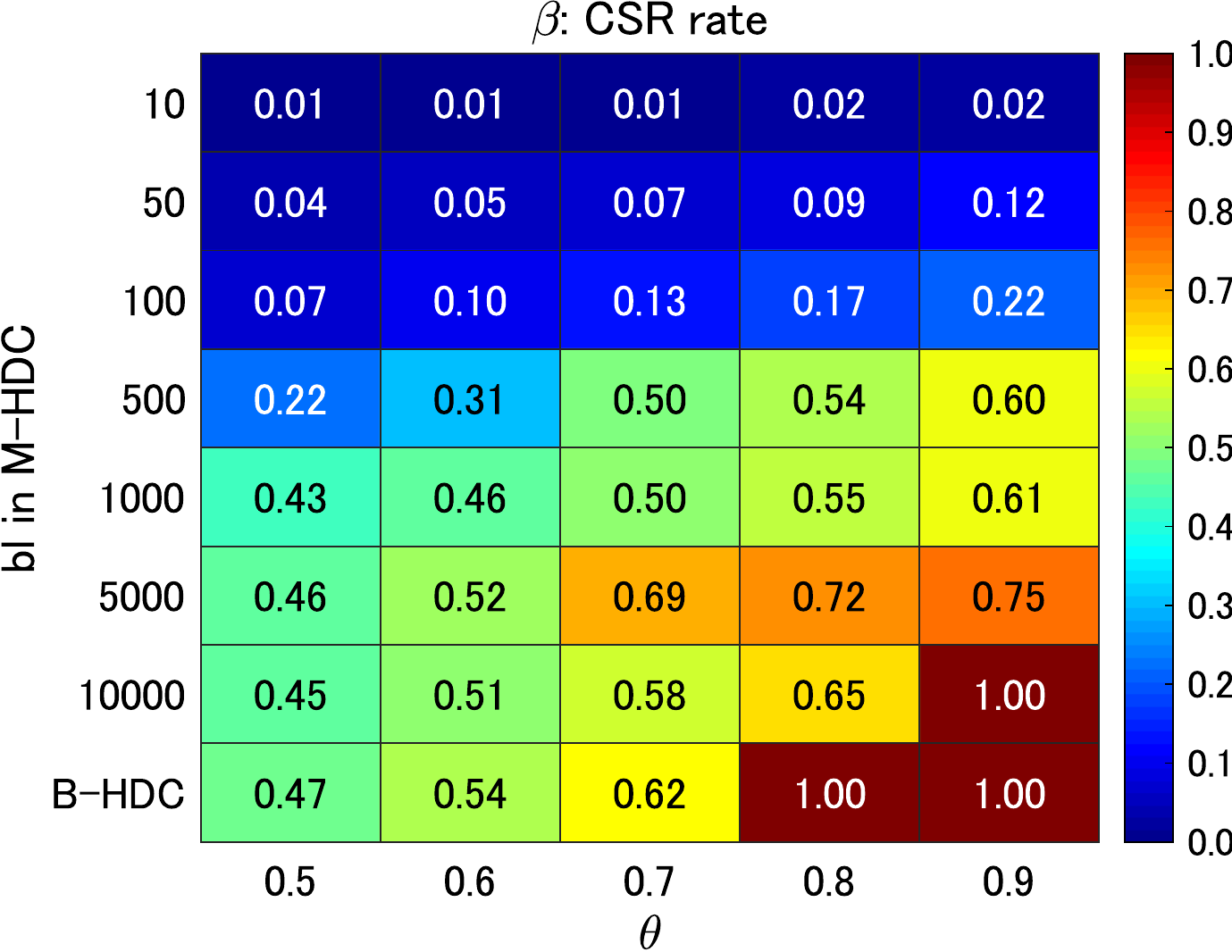}
\includegraphics[scale=0.24]{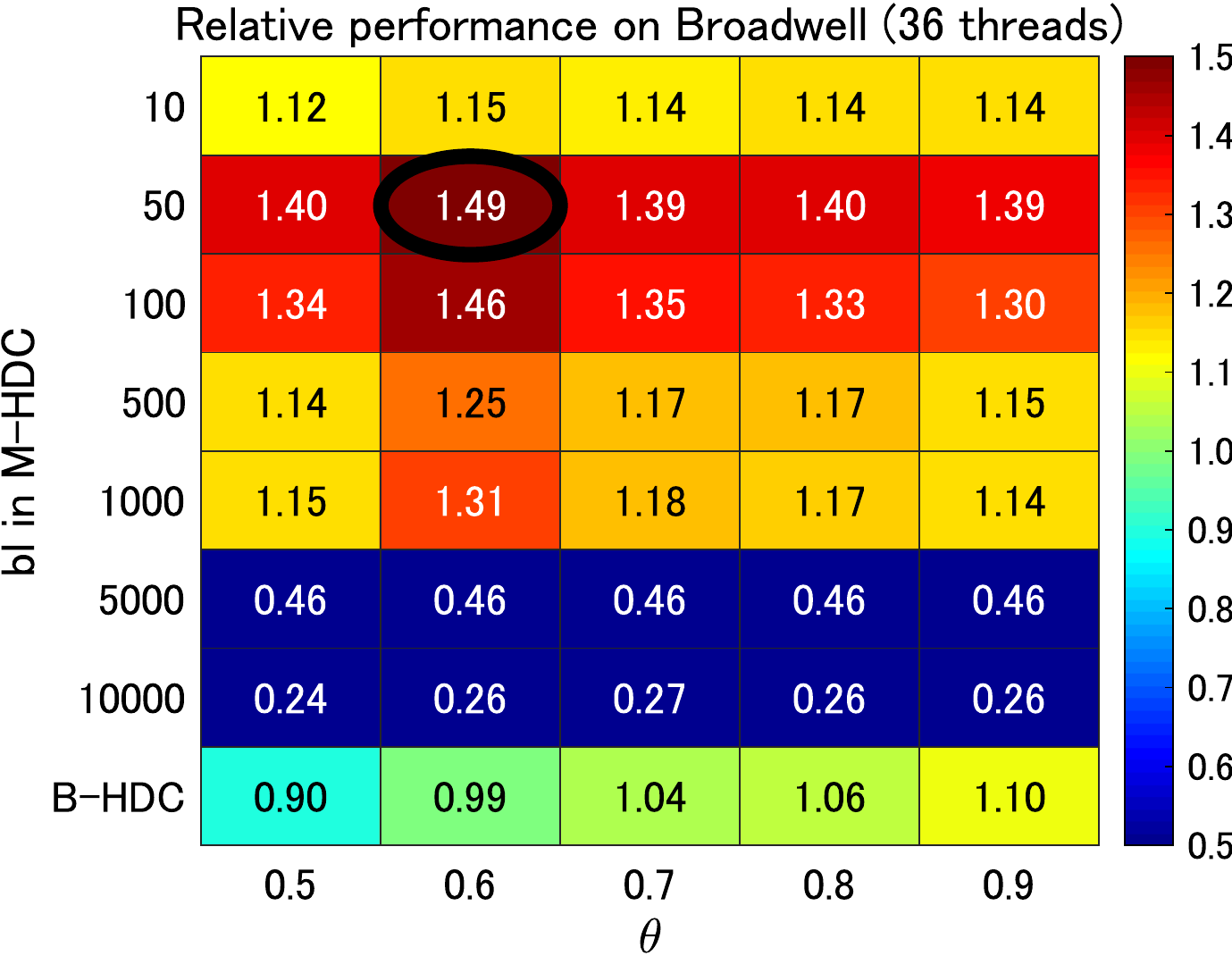}
\includegraphics[scale=0.24]{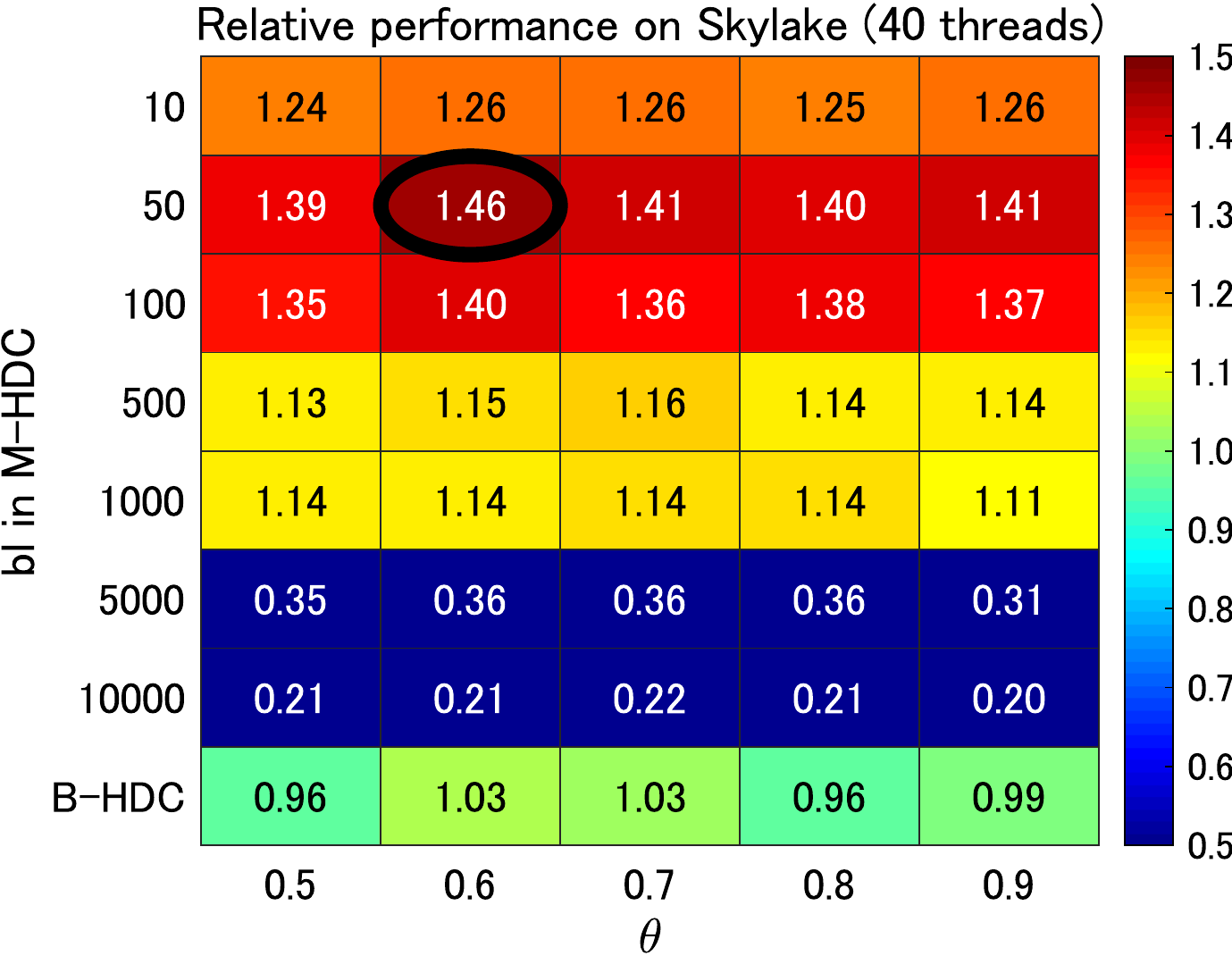}
\includegraphics[scale=0.24]{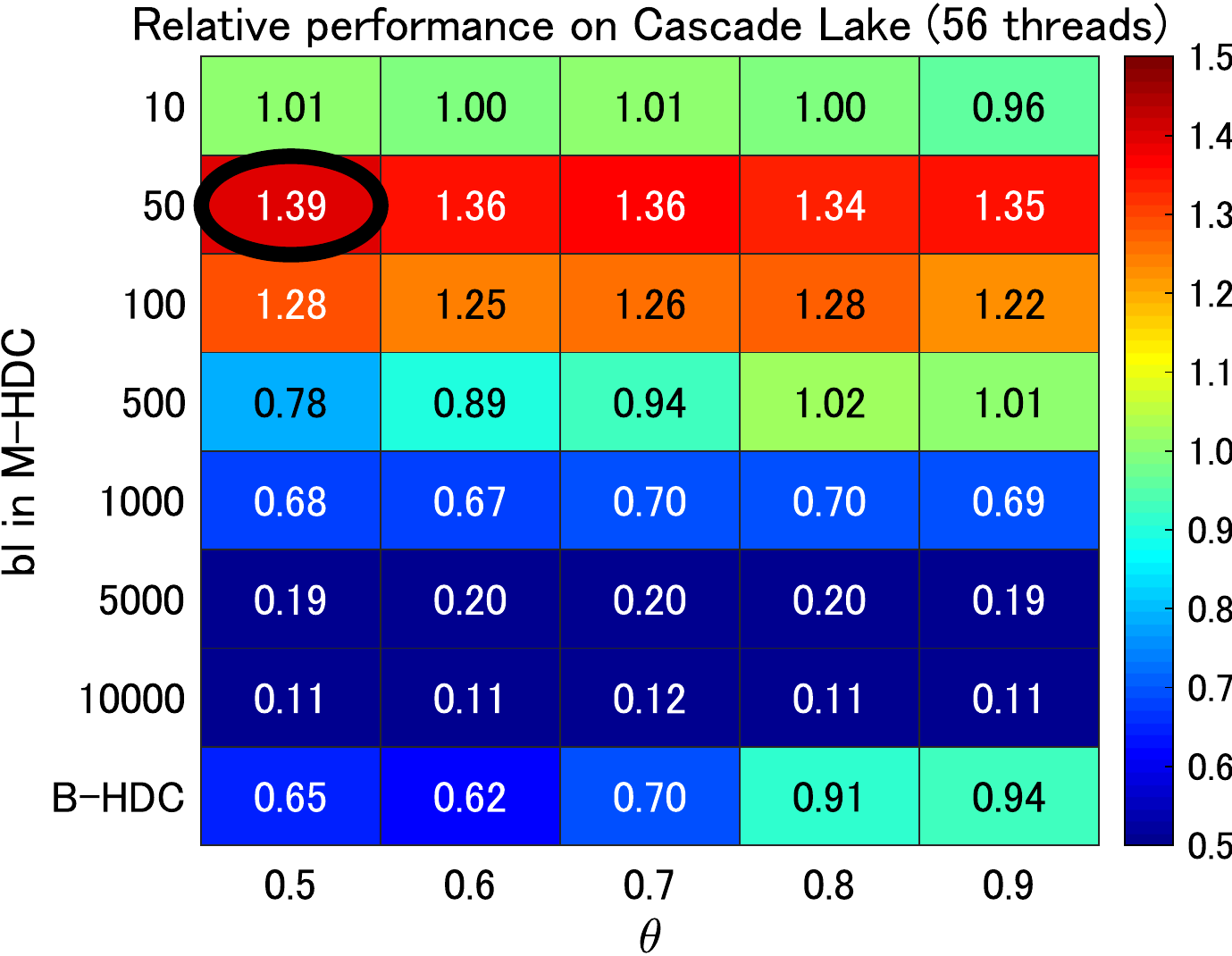}
}
\end{center}
\caption{Relationship to the parameters $bl$ and $\theta$: the left two show the filling rate ($\alpha$) and the CSR rate $(\beta$), 
and the right three show the relative performance of the M-HDC and B-HDC kernels over the CSR kernel in three environments, respectively. 
An ellipse denotes the best case in each situation. 
As the performance of the B-HDC kernel, the best case among different $bl$ candidates is presented. 
It is worth noting that $alpha$ and $\beta$ do not depend on $bl$ in the B-HDC kernel, 
and that the B-HDC kernel is essentially equivalent to the case of setting $bl = n$ in the M-HDC kernel in terms of $alpha$ and $beta$. }
\label{fig: ssmc_parameter}
\end{figure*}

\begin{figure*}[t]
\begin{center}
\subfloat[Matrix \#1]
{
\includegraphics[scale=0.25]{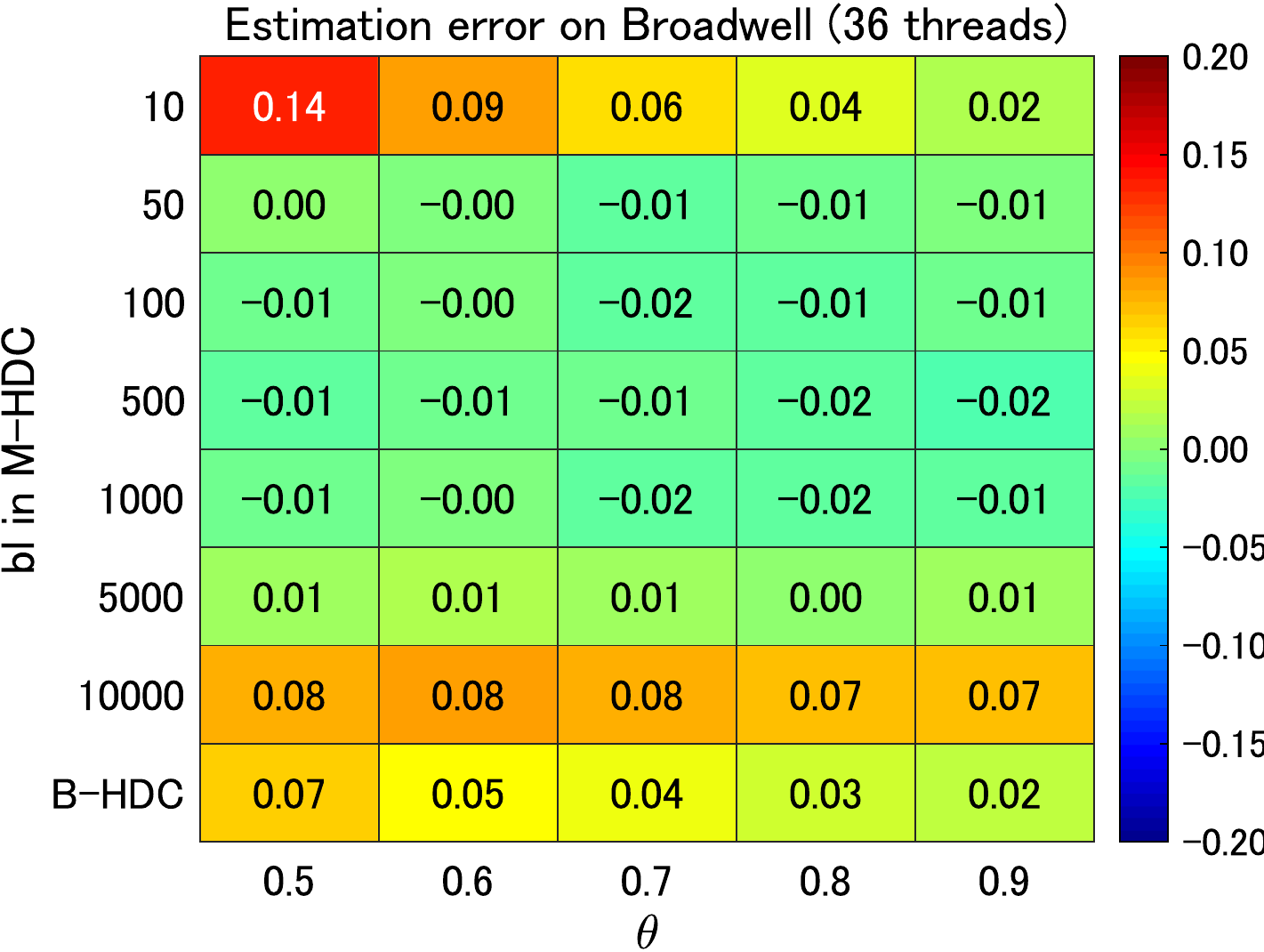}
\includegraphics[scale=0.25]{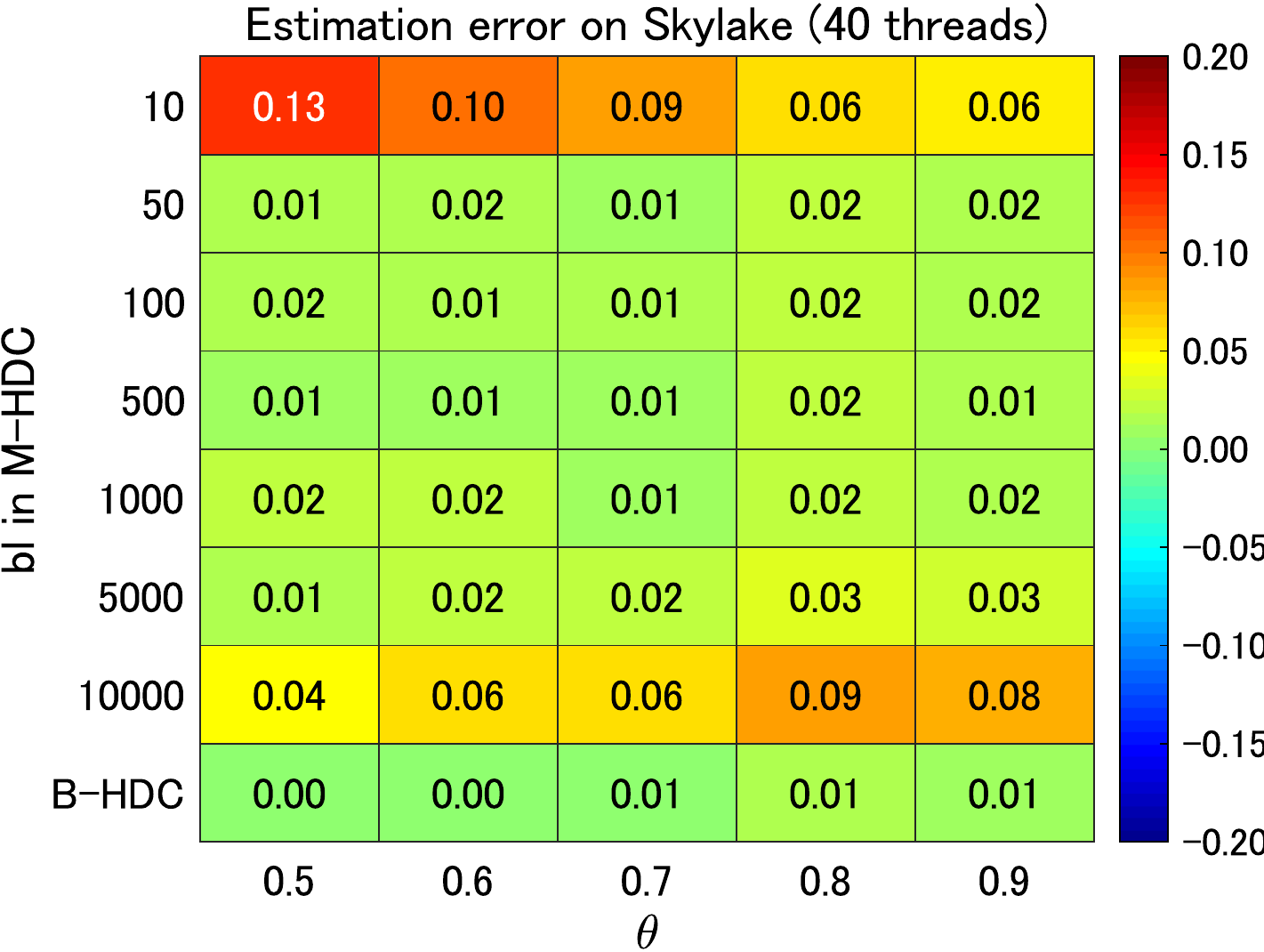}
\includegraphics[scale=0.25]{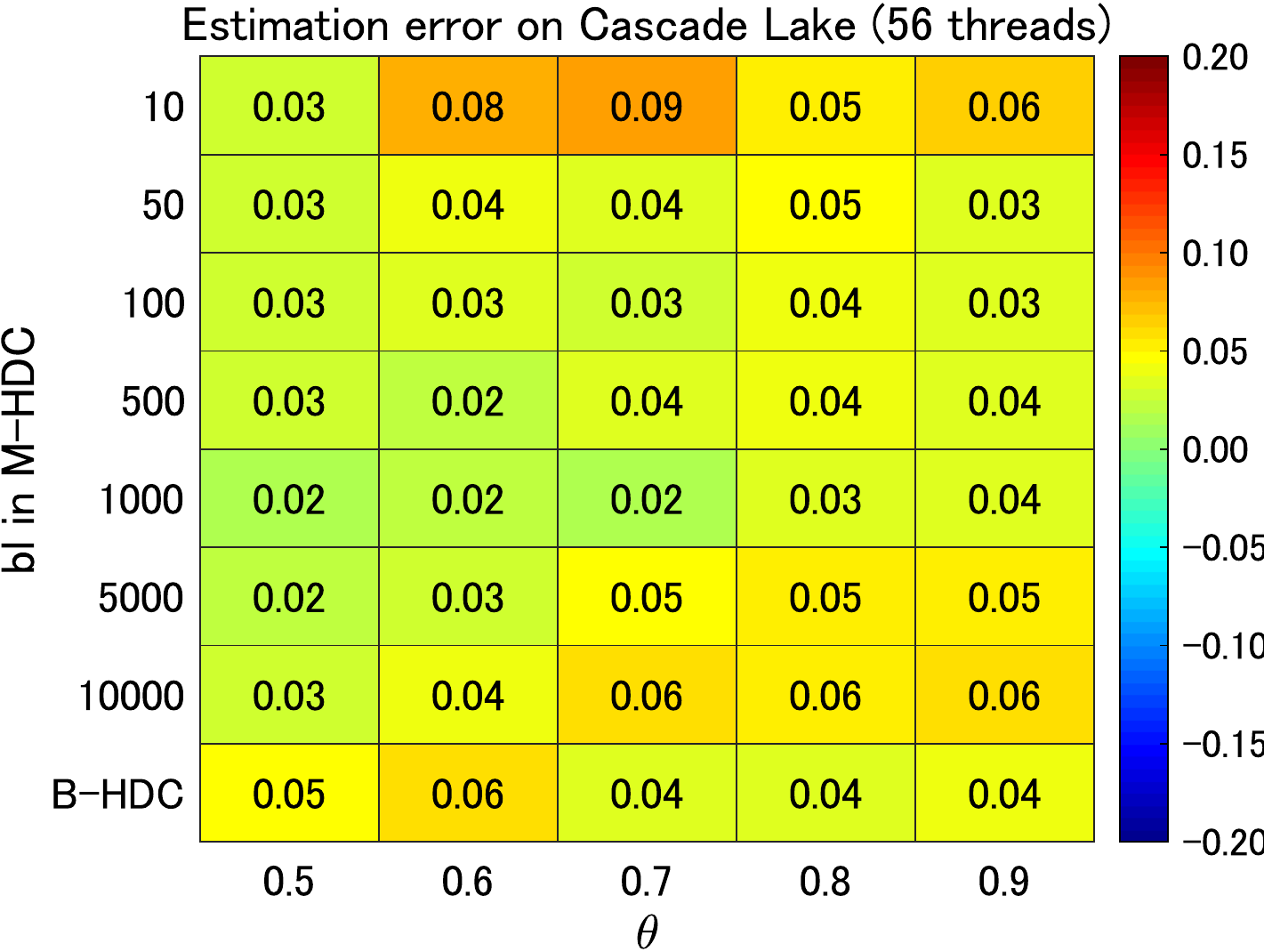}
}
\hfill
\subfloat[Matrix \#3]
{
\includegraphics[scale=0.25]{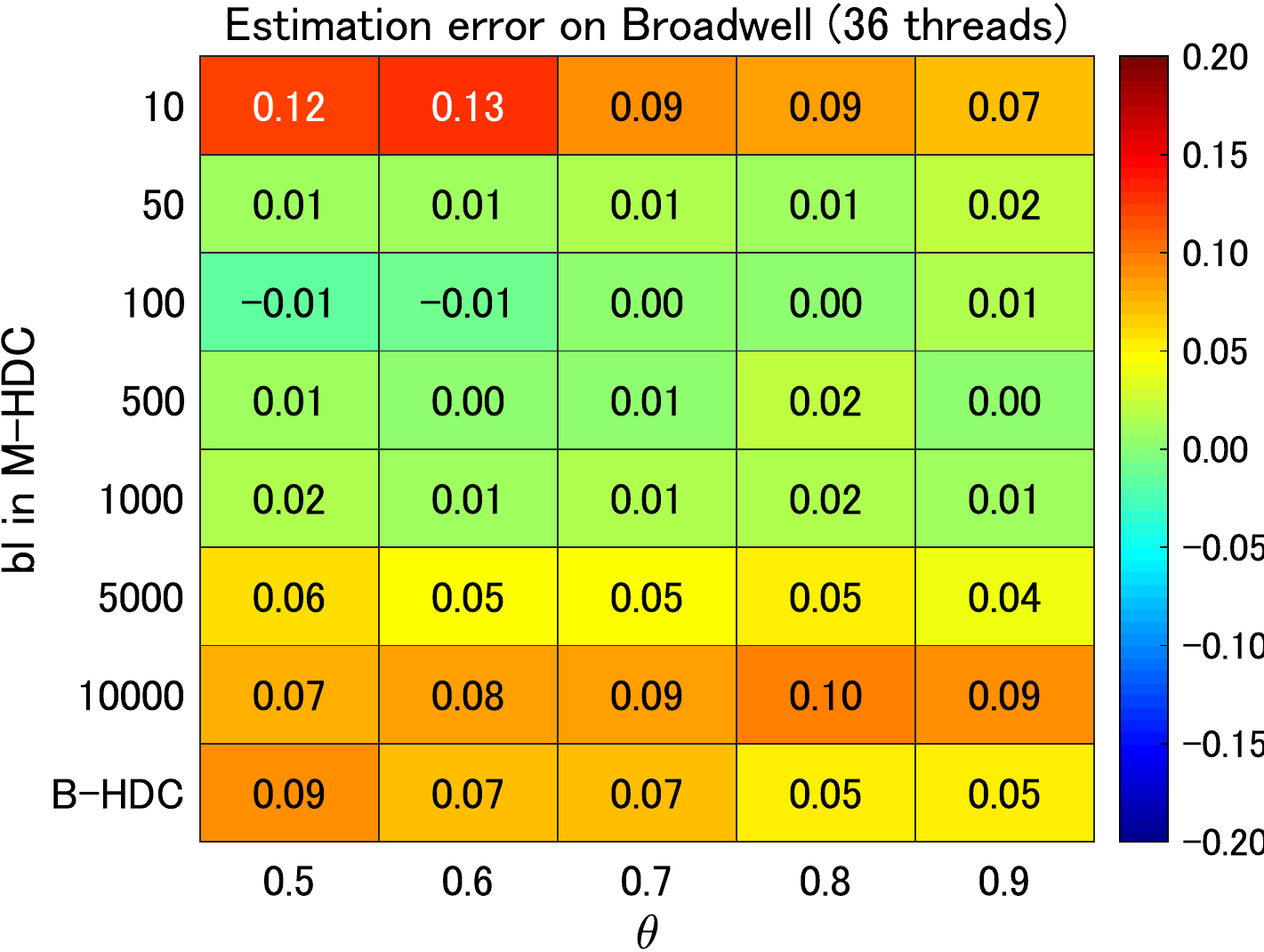}
\includegraphics[scale=0.25]{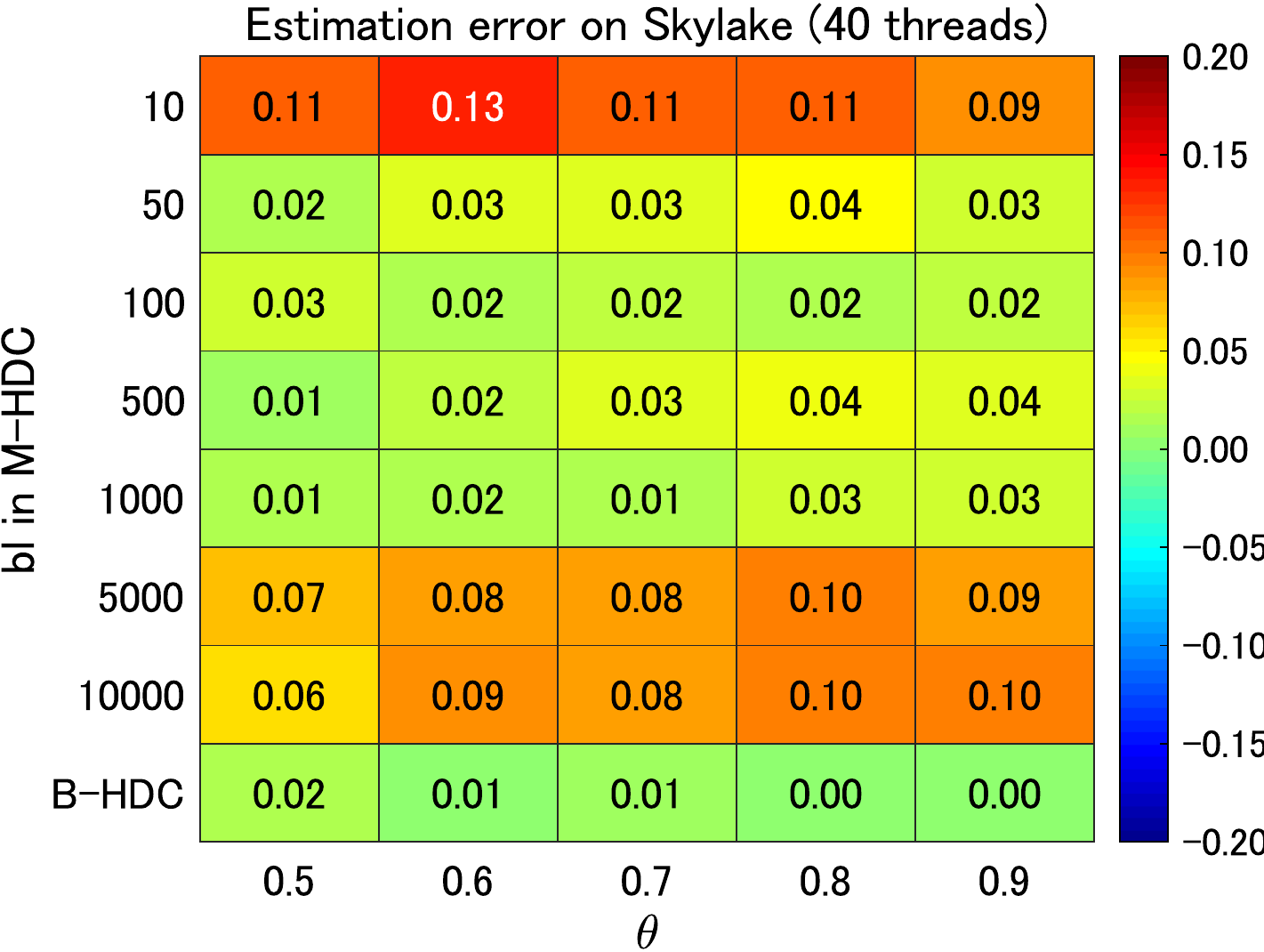}
\includegraphics[scale=0.25]{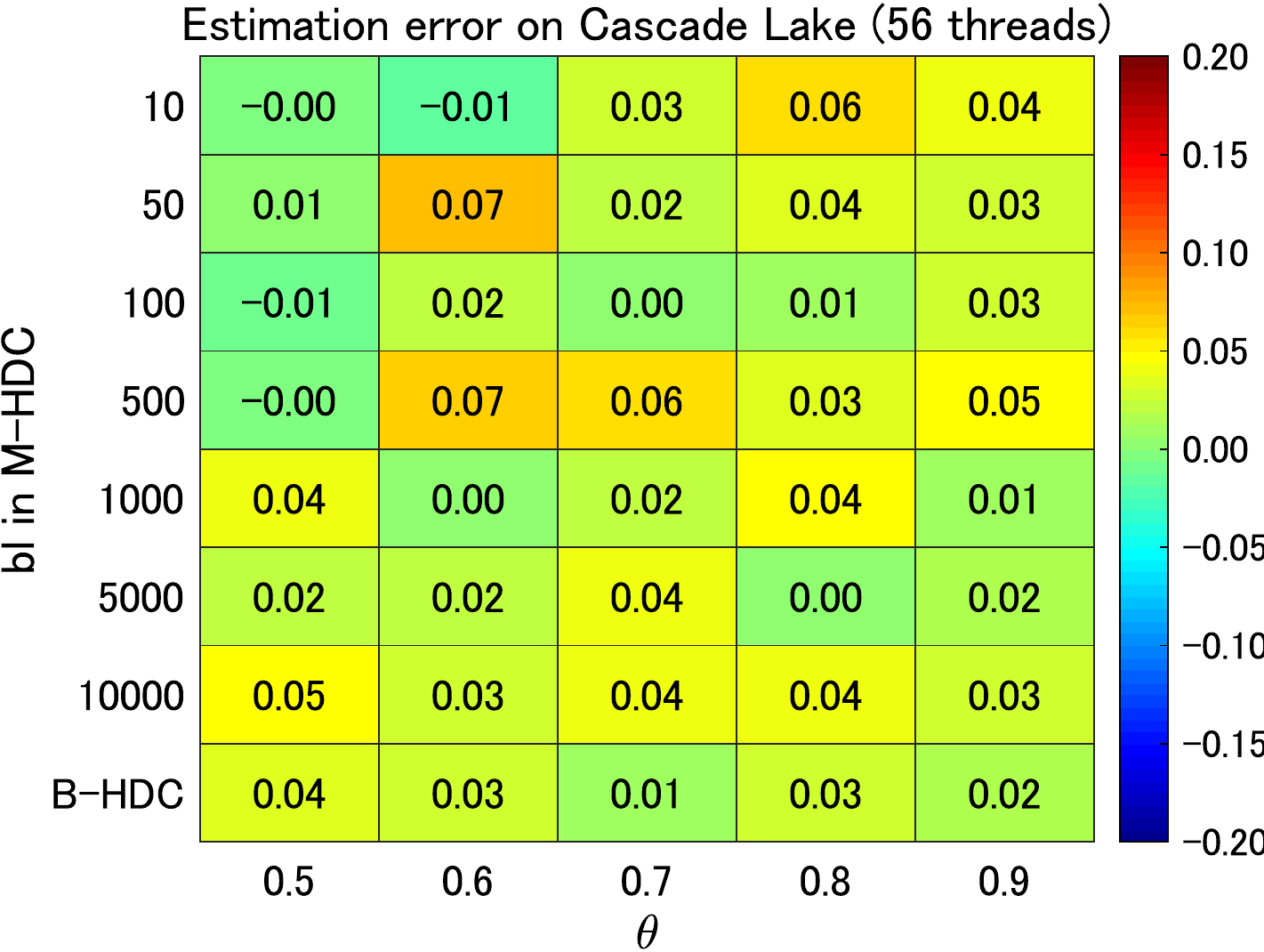}
}
\hfill
\subfloat[Matrix \#10]
{
\includegraphics[scale=0.25]{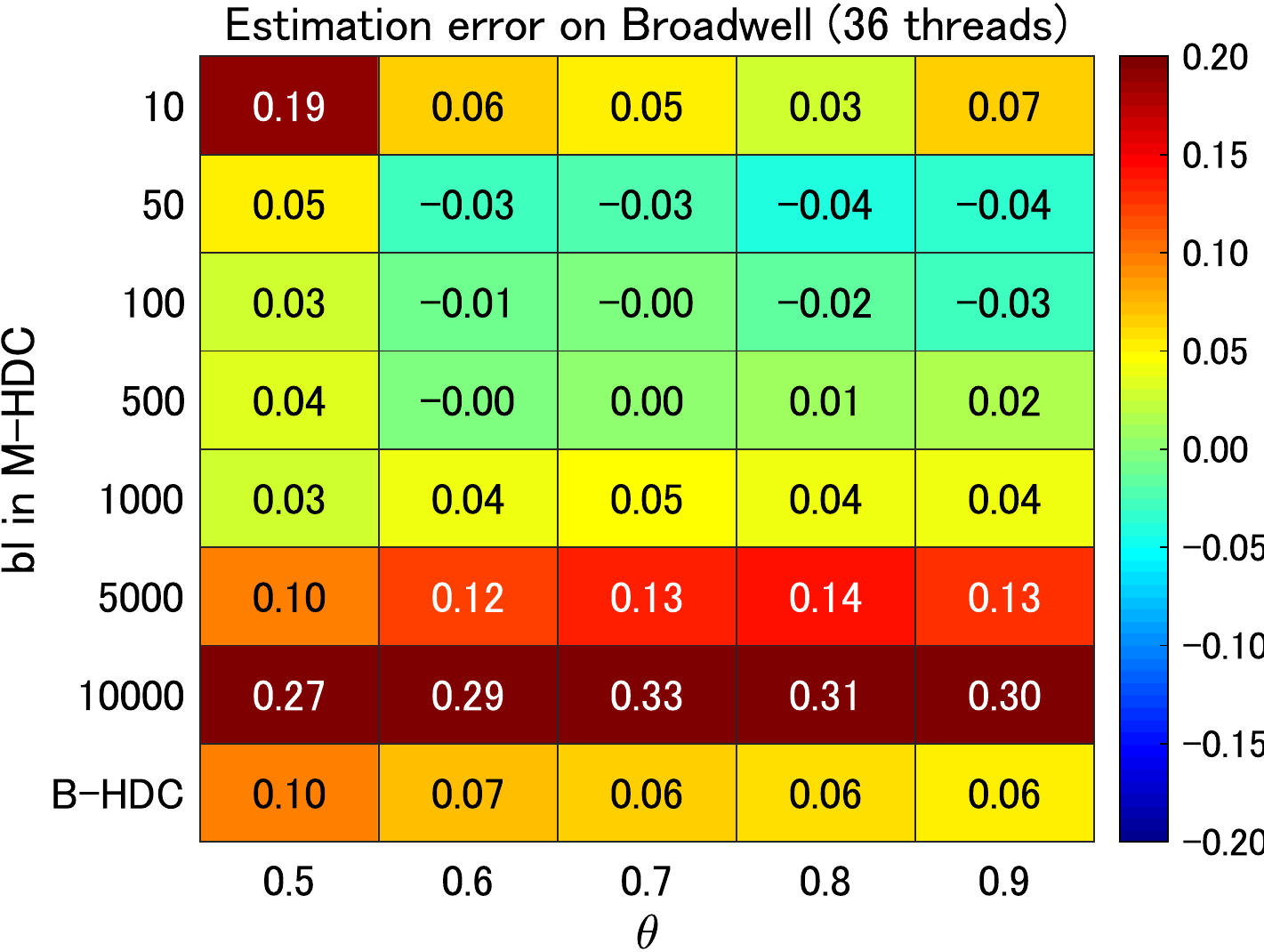}
\includegraphics[scale=0.25]{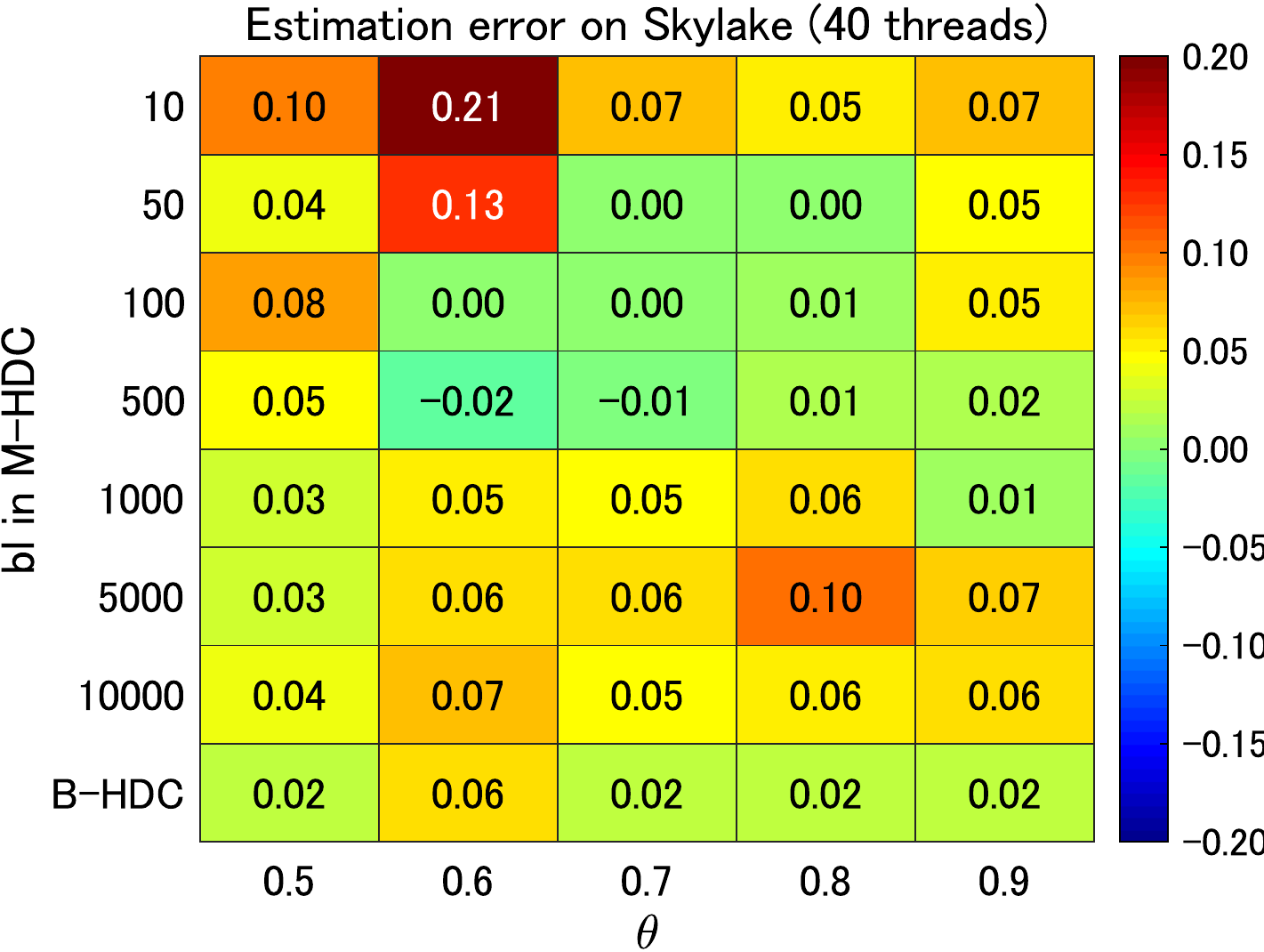}
\includegraphics[scale=0.25]{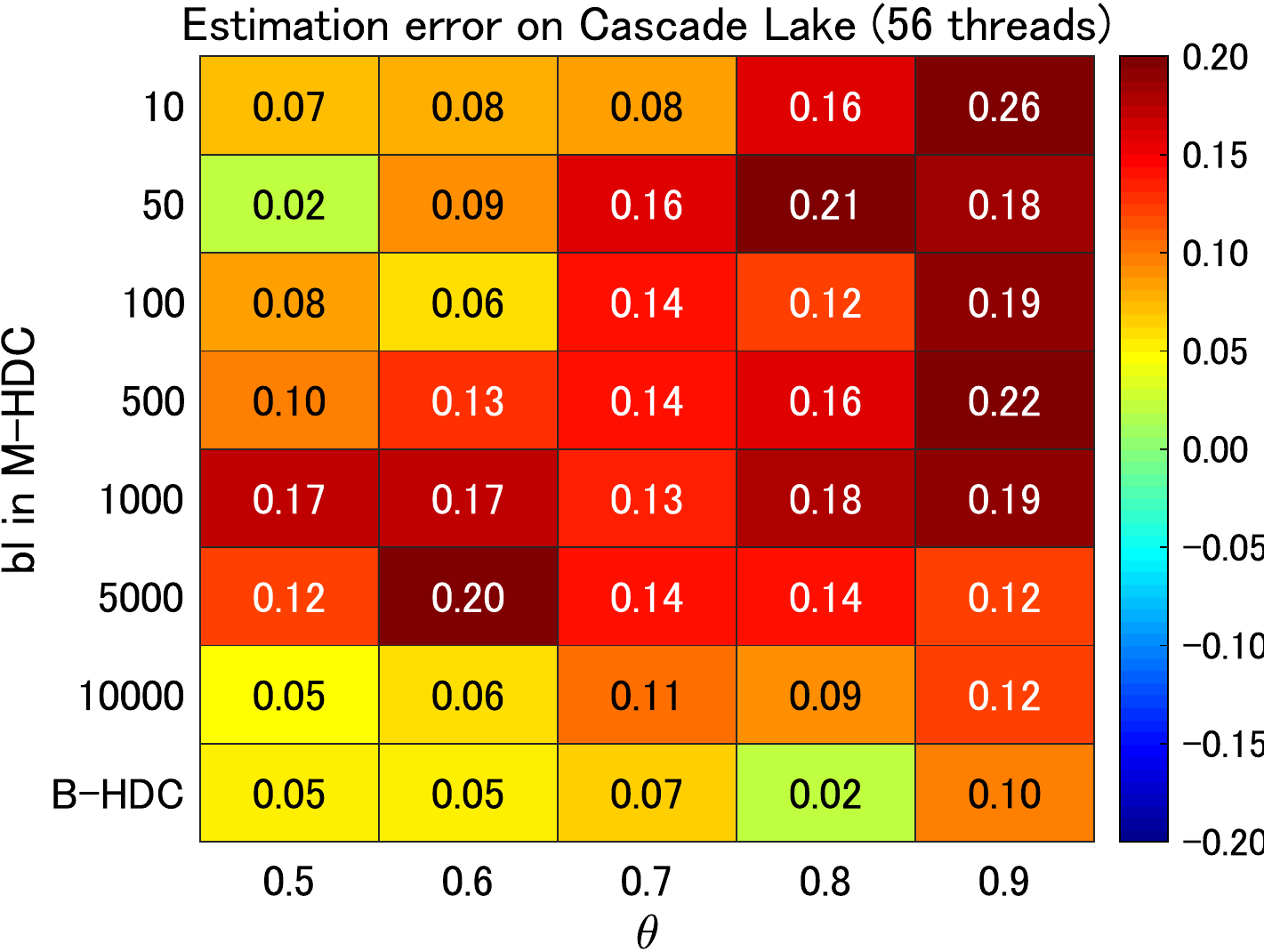}
}
\hfill
\subfloat[Matrix \#13]
{
\includegraphics[scale=0.25]{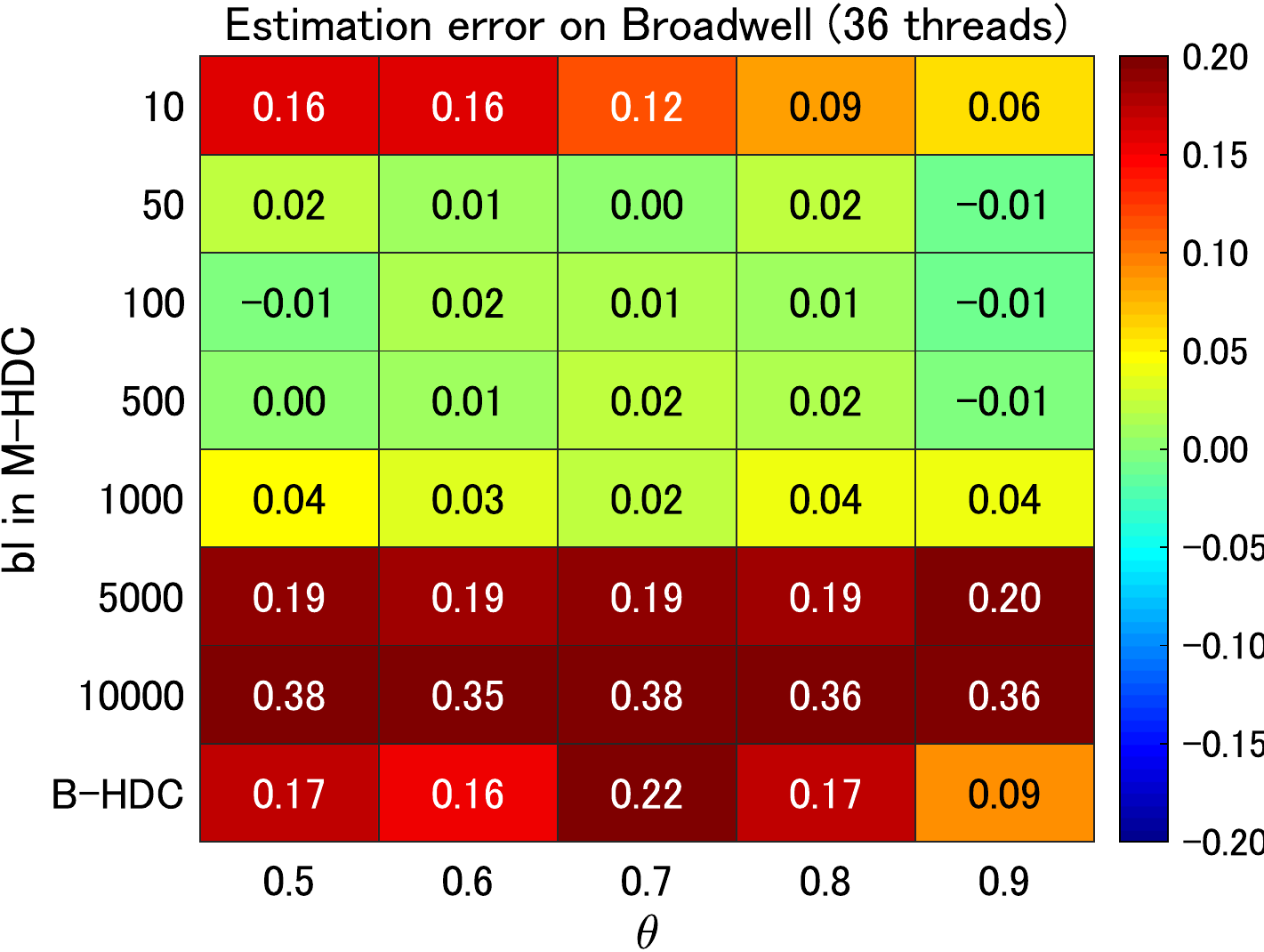}
\includegraphics[scale=0.25]{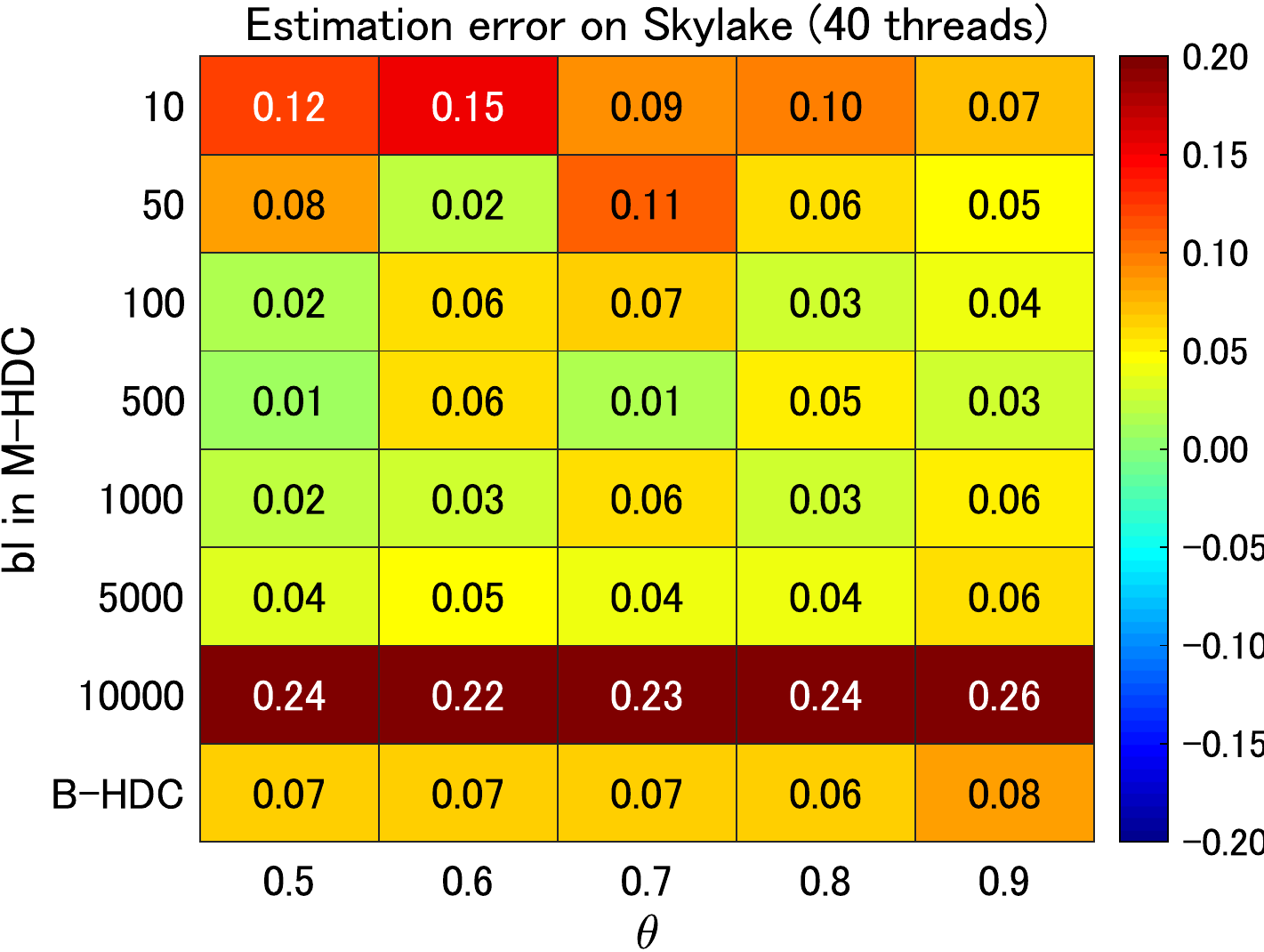}
\includegraphics[scale=0.25]{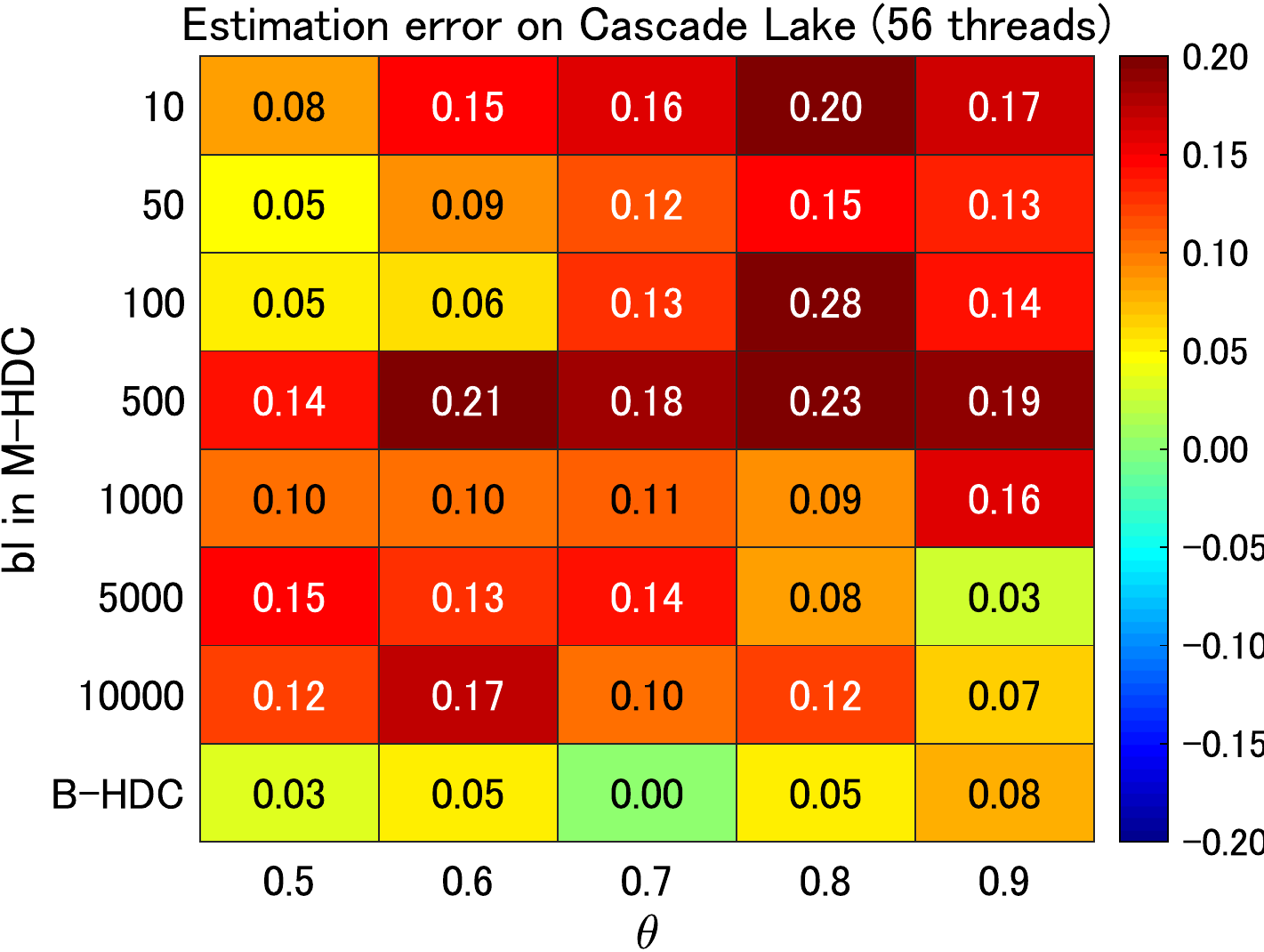}
}
\hfill
\subfloat[Matrix \#14]
{
\includegraphics[scale=0.25]{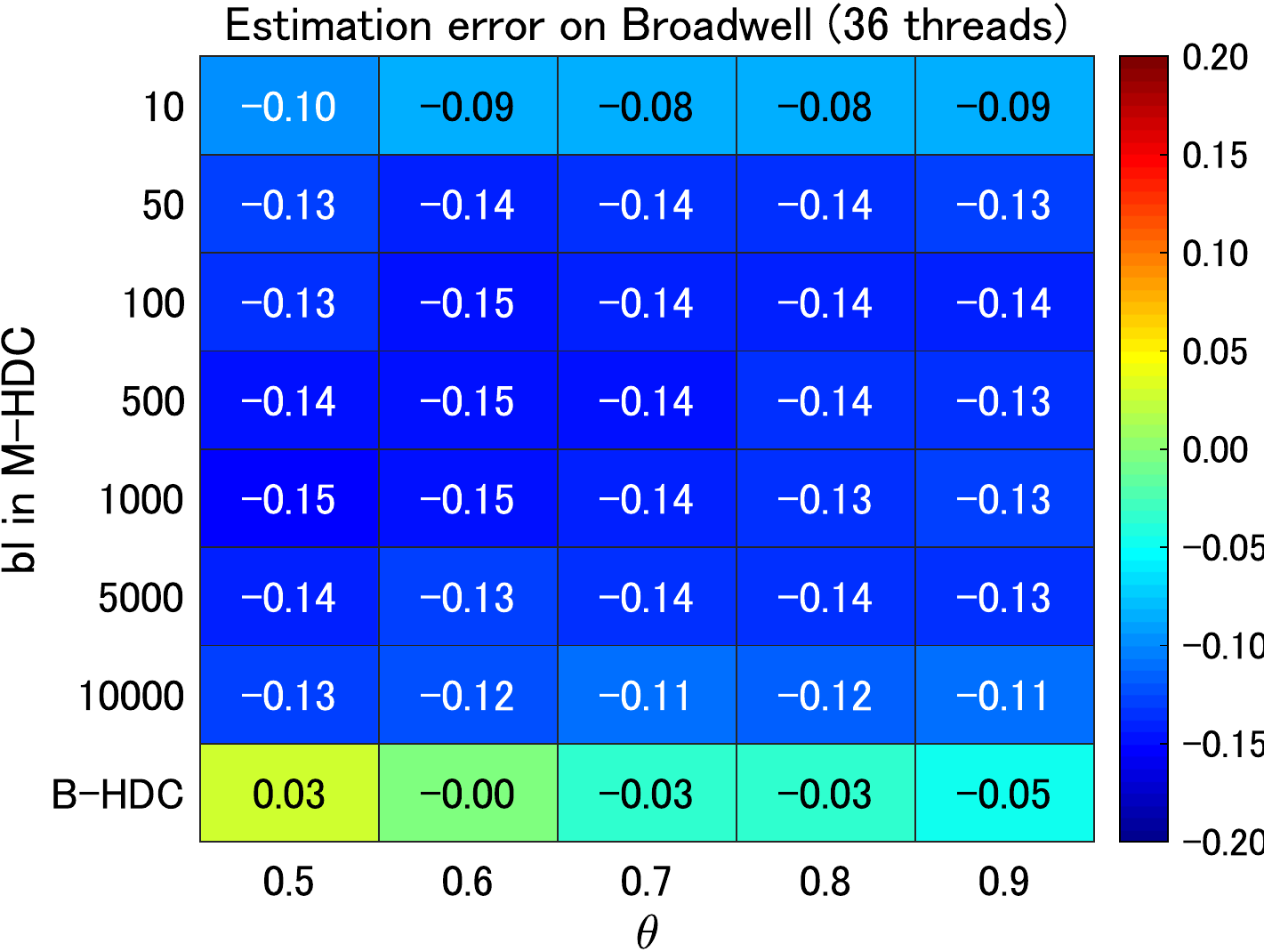}
\includegraphics[scale=0.25]{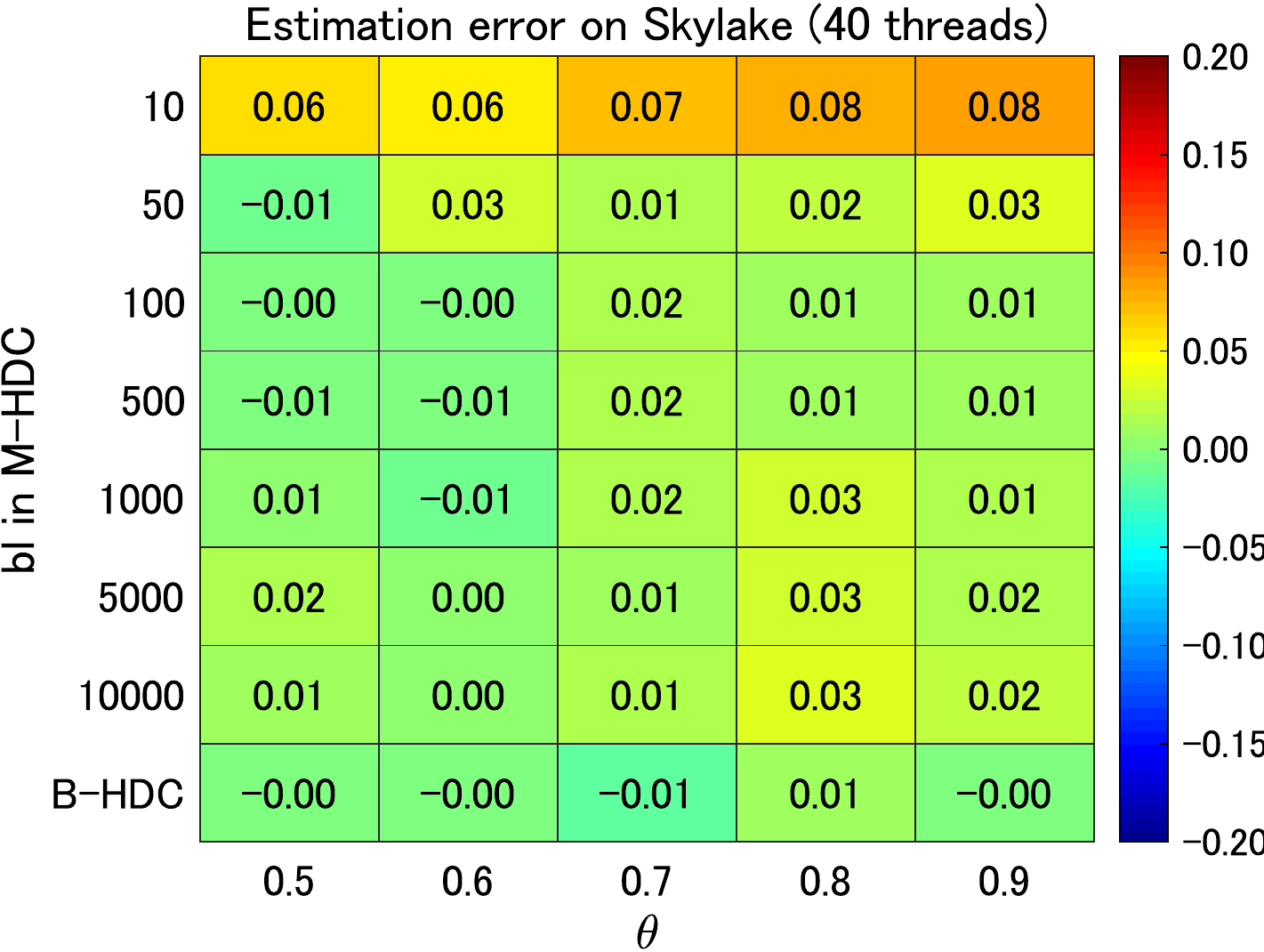}
\includegraphics[scale=0.25]{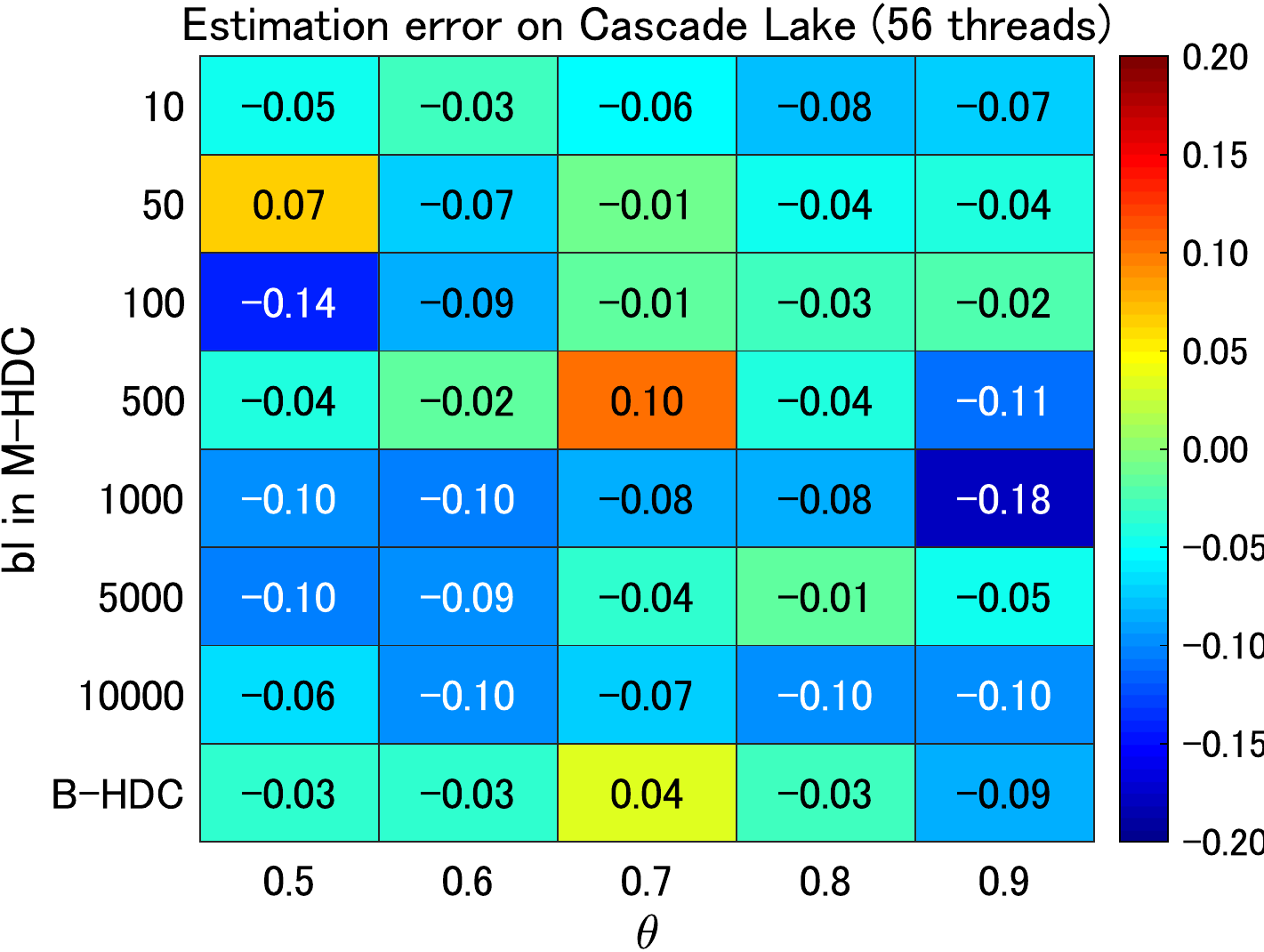}
}
\hfill
\subfloat[Matrix \#17]
{
\includegraphics[scale=0.25]{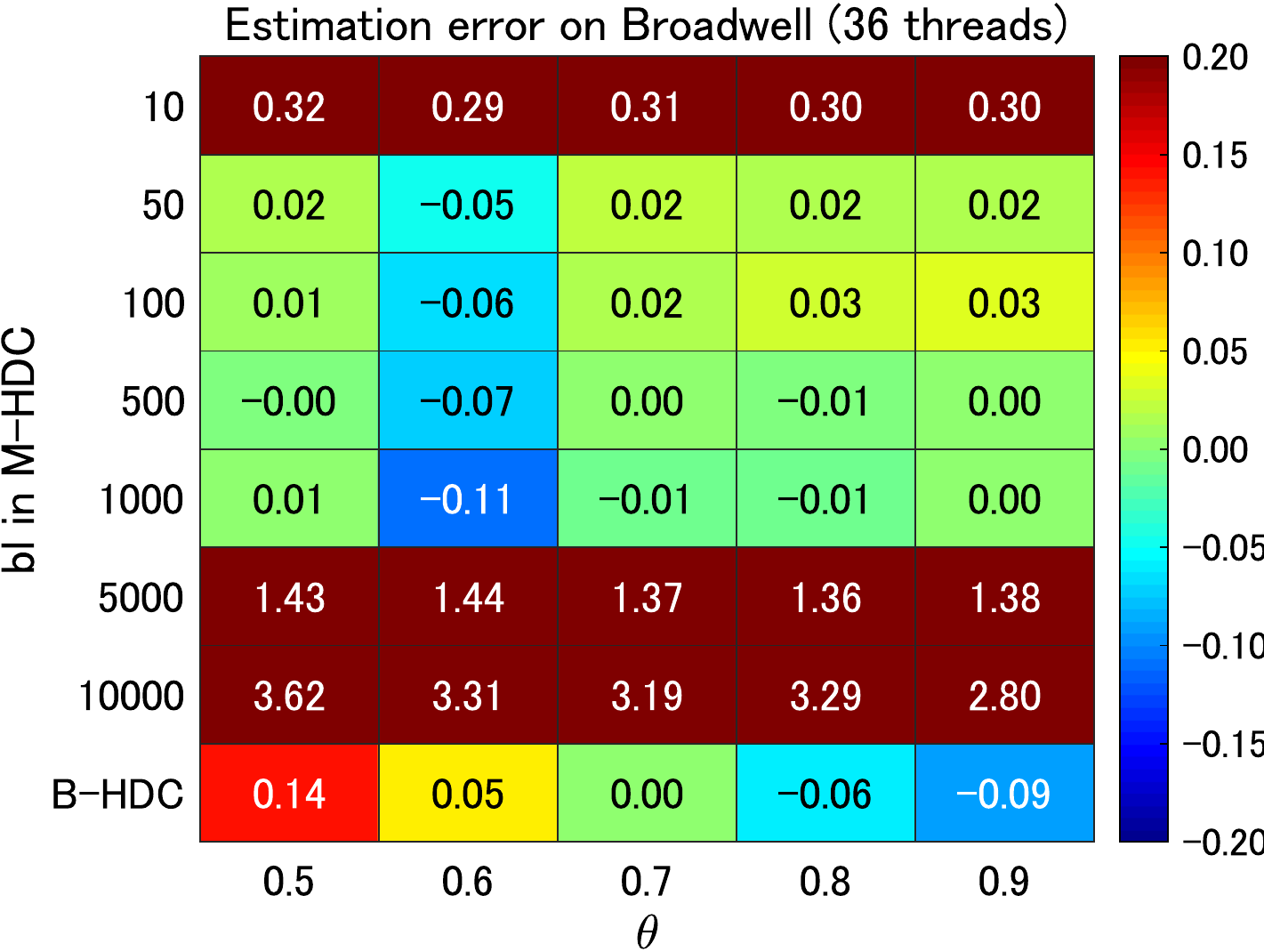}
\includegraphics[scale=0.25]{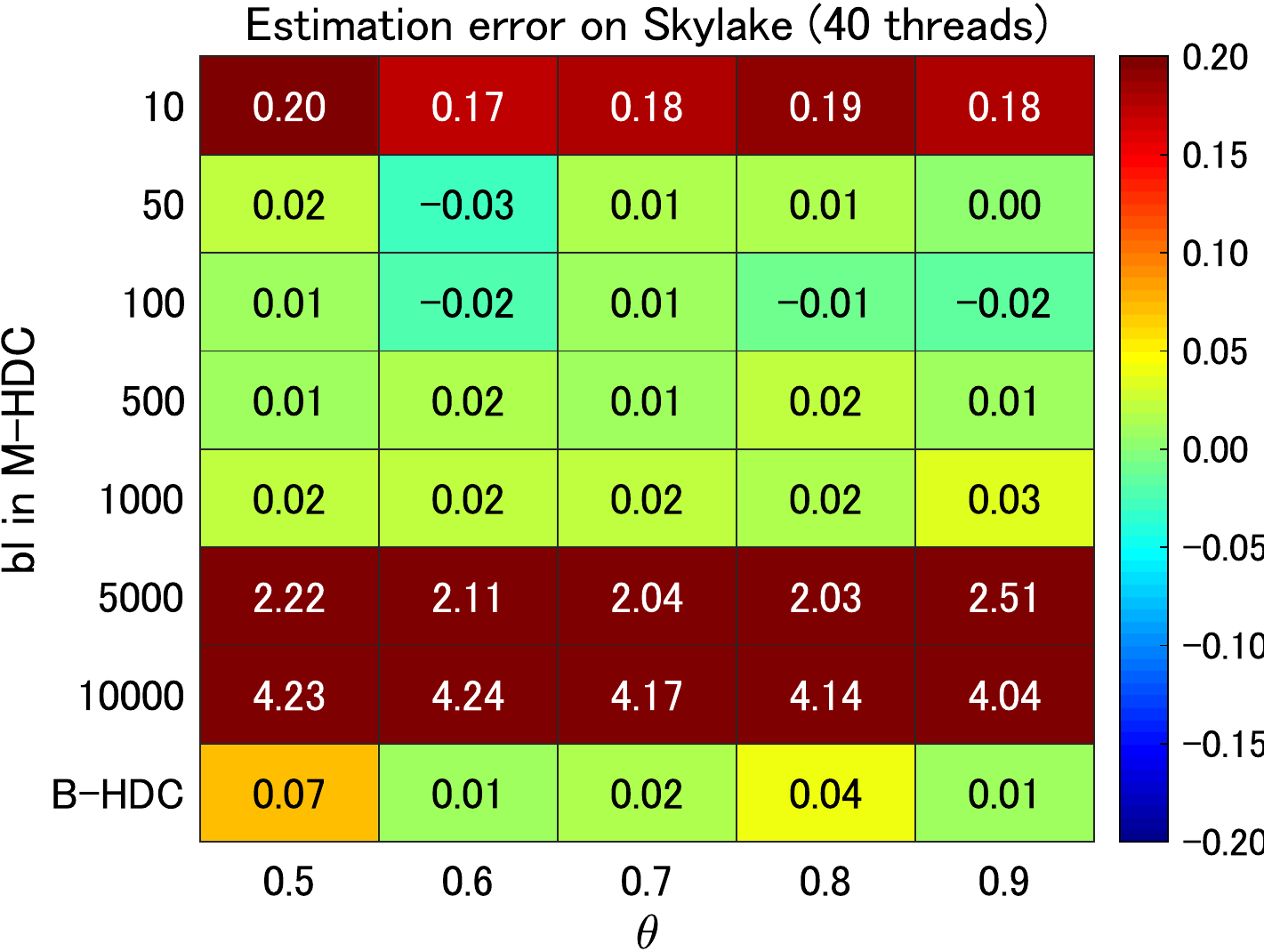}
\includegraphics[scale=0.25]{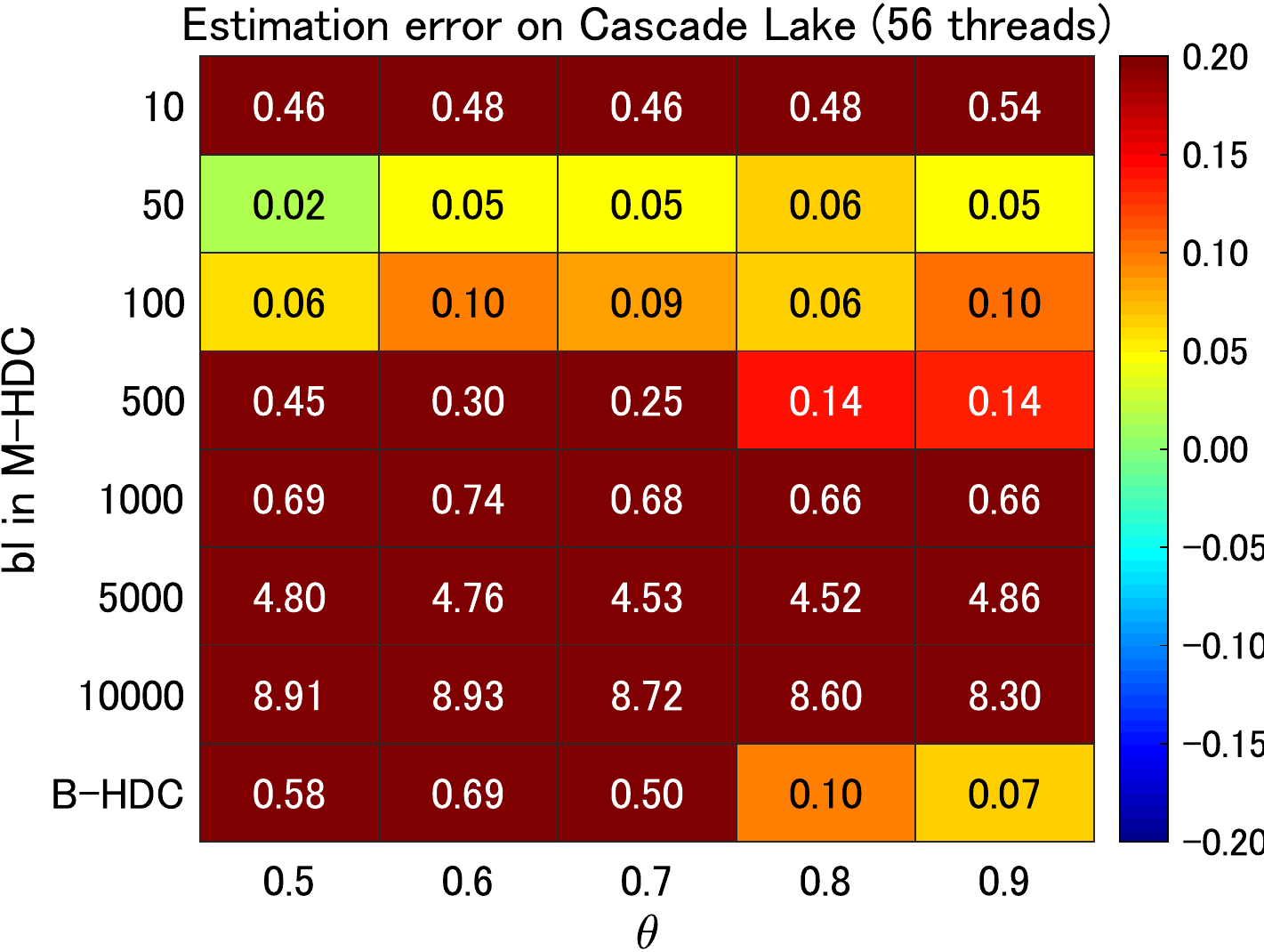}
}
\end{center}
\caption{Accuracy of the estimation by the performance models; 
let $RP$ be the relative performance of the M-HDC (or B-HDC) kernel over the CSR kernel, 
$RP_\textrm{est}$ and $RP_\textrm{exe}$ denote the estimated and obtained values, respectively, 
and $(RP_\textrm{est} - RP_\textrm{exe}) /RP_\textrm{ext}$ is plotted in each graph.}
\label{fig:ssmc_error}
\end{figure*}

\begin{figure*}[t]
\begin{center}
\includegraphics[scale=0.35]{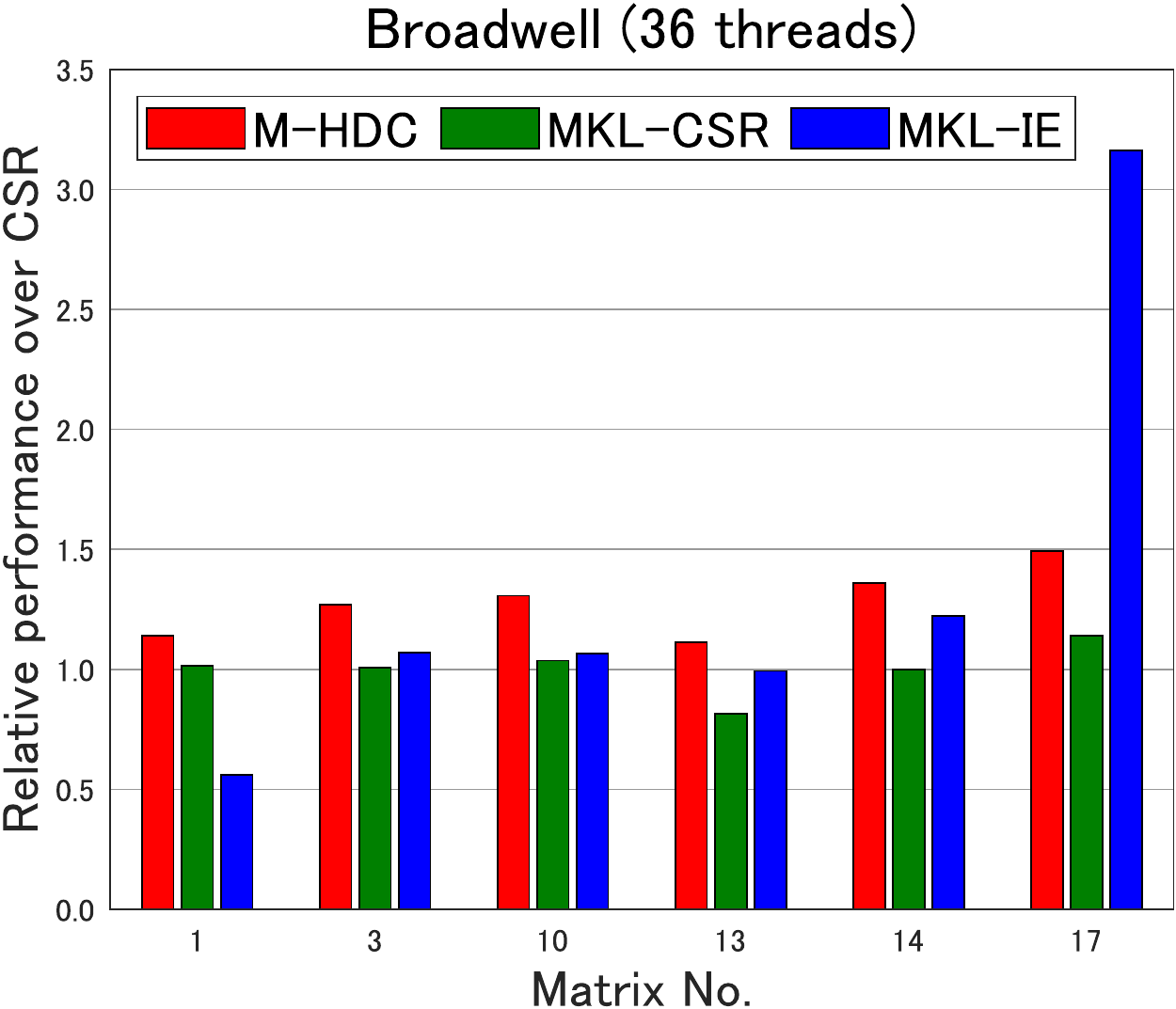}
\includegraphics[scale=0.35]{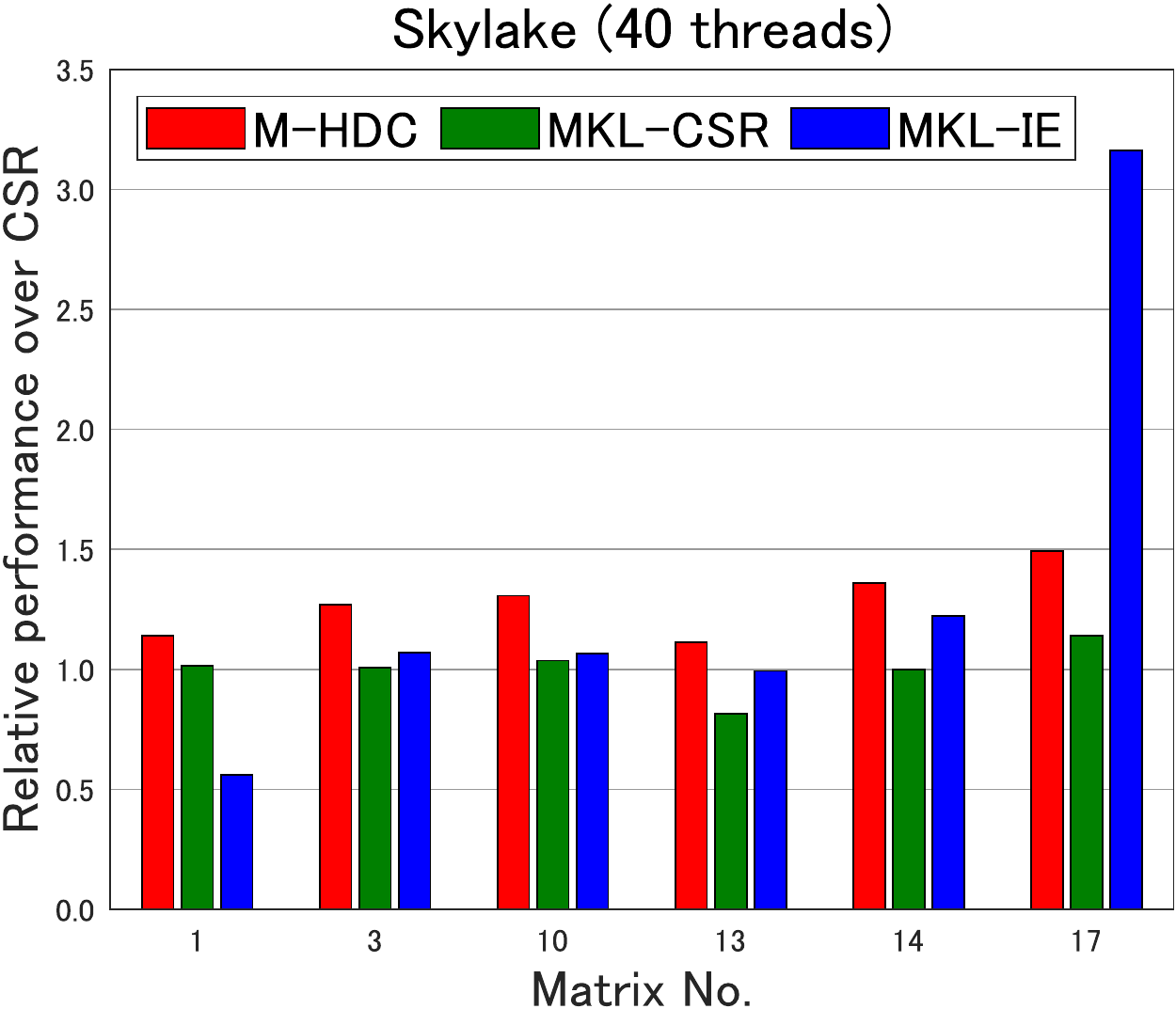}
\includegraphics[scale=0.35]{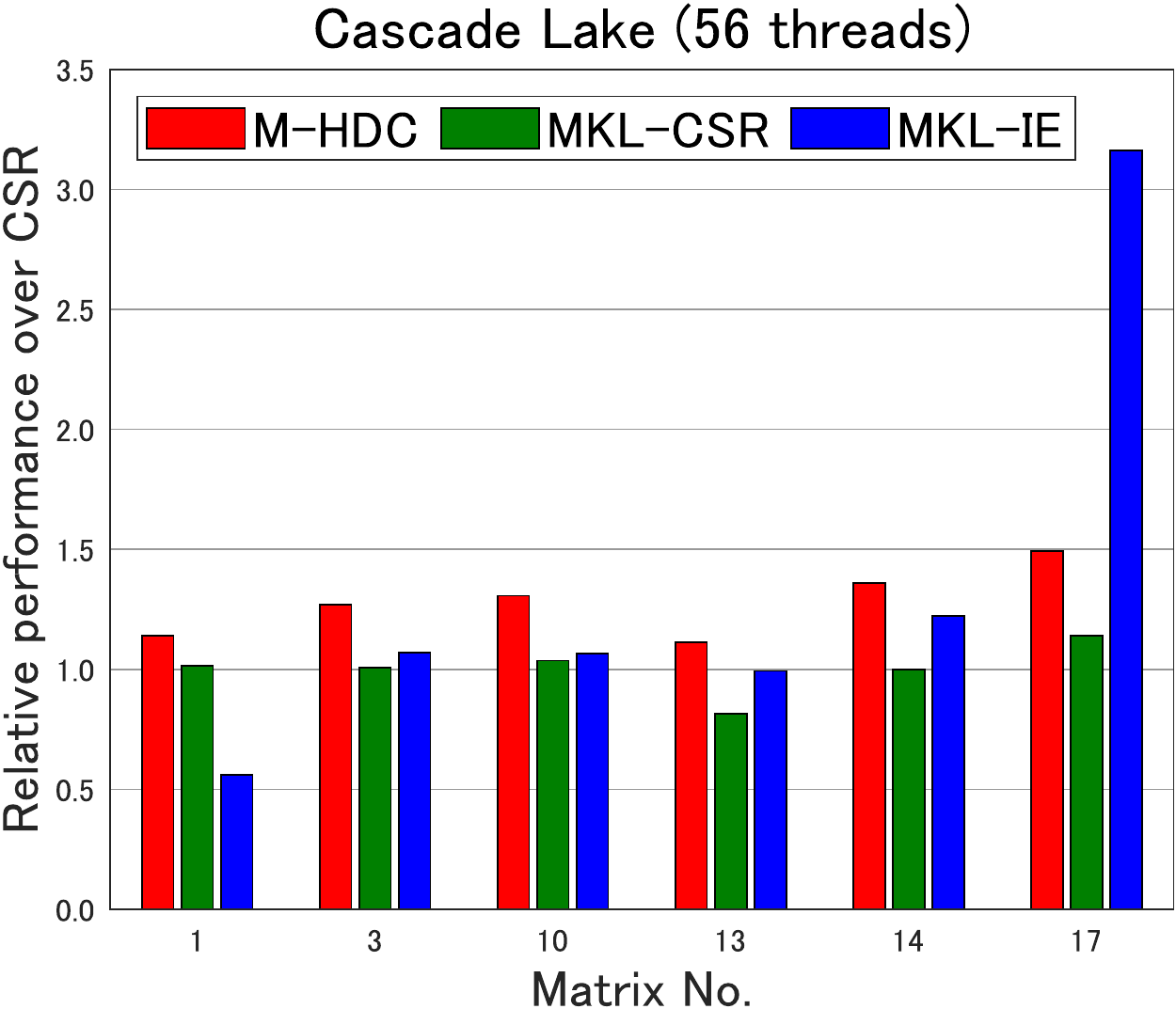}
\end{center}
\caption{Relative performance of the M-HDC kernel, the MKL CSR routine, and the MKL IE routine over the (in-house) CSR kernel; for the M-HDC kernel, the performance by the best setting of $bl$ and $\theta$ is plotted.}
\label{fig:ssmc_mkl}
\end{figure*}

\subsubsection{Overall results}
As the baseline, we give the performance (Flop/s) of the CSR kernel in each environment (Figure~\ref{fig:ssmc_csr_gflops}). 
The results shown in Figure~\ref{fig:ssmc_csr_gflops}, both the performance itself and the correlation of performance among three environments, 
seem to be consistent with the the theoretical memory performance (Table~\ref{tbl:spec}) and the effective memory performance (Figure~\ref{fig:results_stream}). 
\par
Then, we present the relative performance (i.e. speedup) of the HDC, B-HDC, and M-HDC kernels over the CSR kernel in Figure~\ref{fig:compare_ssmc}. 
For discussing the results in Figure~\ref{fig:compare_ssmc}, we also give Figure~\ref{fig:ssmc_csr_ratio}, 
which illustrates the CSR rate in each kernel when $\theta = 0.5$. 
From Figures~\ref{fig:compare_ssmc} and \ref{fig:ssmc_csr_ratio}, we have the following observations:
\begin{itemize}
\item Matrix \#12: almost all the nonzero elements are stored in the DIA format. 
This is why the performance of the B-HDC and M-HHC kernels for this matrix stands out among all the test matrices. 
This result most clearly shows the effectiveness of the cache blocking technique; the B-HDC kernel outperforms the HDC kernel, 
especially on Broadwell and Cascade Lake. 
\item Matrices \#5, \#6, \#11, and \#19: more than the 90\% nonzero elements are still stored in the CSR format in both the HDC (B-HDC) and M-HDC kernels. 
Thus, in terms of the whole performance of SpMV for these matrices, there are almost no differences among three kernels. 
\item Matrices \#1, \#3, \#10, \#13, \#14, and \#17: there is a sufficient gap of the CSR rate between the HDC and M-HDC kernels, 
which means that the M-HDC kernel can pick up partial diagonal structures but the HDC and B-HDC kernels cannot. 
For these matrices, we can find a significant performance improvement of the M-HDC kernel over the B-HDC kernel.
\item Other matrices: since the CSR rate of the HDC and M-HDC kernels are not small, 
the performance differences among three kernels are not noteworthy, 
however we can find the tendency that the B-HDC kernel is better than the HDC kernel, 
and that the M-HDC kernel is superior to the B-HDC kernel.
\end{itemize}
\par
From these experimental results, we confirm the effectiveness of the cache blocking technique employed in the B-HDC kernel and the potential of the M-HDC kernel; 
among 20 test matrices derived from practical applications, we can find matrices that the M-HDC kernel advantageously works for.

\subsubsection{Detailed analysis for representative matrices}
Now, we report detailed analyses for matrices~\#1, \#3, \#10, \#13, \#14, and \#17, which well represent the characteristics of the M-HDC kernel. 
\par
First, in Figure~\ref{fig: ssmc_parameter}, we give the relationship of the filling rate ($\alpha$), the CSR rate ($\beta$), and the performance to the parameters ($bl$ and $\theta$), together with the best parameter setting in each situation (matrix and environment). 
From Figure~\ref{fig: ssmc_parameter}, we have the following observations: 
\begin{itemize}
\item The filling rate: $\alpha \ge \theta$ is assured in theory, and high $\alpha$ is obtained in practice even when $\theta$ is not large, 
e.g. when $\theta = 0.6$, the obtained $\alpha$ is roughly higher than 0.8 in almost all the cases. 
We can also find that $bl$ has little effect on $\alpha$. 
\item The CSR rate: we can find a clear impact of $bl$ and/or $\theta$ on $\beta$; i.e. both in \#3, \#10, and \#13, mainly $bl$ in \#14 and \#17, and mainly $\theta$ in \#1. 
The tendency that smaller $bl$ and/or $\theta$ bring lower $\beta$ can be observed, which is easily expected. 
\item Performance: the impact of $bl$ and/or $\theta$ on the obtained performance can be clearly found, and its main factor seems to be caused by $\beta$. 
In the matrix \#17, the performance is significantly low when $bl = 5000$ and $10000$. 
This is reasonable because its $n$ is less than $40000$ and only partial threads work using such $bl$.  
We can find that $bl = 50$ or $100$ basically provide (near) the best performance in every environment, although there are a few exceptions. 
Typically $bl = 10$ brings lower $\beta$ than $bl = 50$, 
however its performance is not the best due to the disadvantage of increasing the memory access of the offset information in the M-HDC kernel. 
\end{itemize}
\par
Next, we confirm the accuracy of the estimation by the performance models explained in Section~\ref{sec:model}, 
e.g. the examples shown in Figure~\ref{fig:upper_bound_estimate}.
Based on Equation~\ref{eq:b_hdc_over_csr} with the assumption that $v_{\vec{x}} = 1$, 
by substituting the obtained $\alpha$ and $\beta$, which are shown in Figure~\ref{fig: ssmc_parameter}, and $c = \nnz / n$, 
we have an estimation of the relative performance of the M-HDC (or B-HDC) kernel over the CSR kernel, denoted $RP_\textrm{est}$. 
Then, we compare it with the obtained results, denoted $RP_\textrm{exe}$. 
We calculate the relative error $RE$ as 
\begin{equation*}
	RE := \frac{RP_\textrm{est} - RP_\textrm{exe}}{RP_\textrm{exe}}, 
\end{equation*}
and plot them in Figure~\ref{fig:ssmc_error}.
From Figure~\ref{fig:ssmc_error}, we have the following observations: 
\begin{itemize}
\item When $10 \le bl \le 1000$, the accurate estimations are obtained in many cases, i.e. basically $|RE| \le 0.05$ excepting a few cases. 
Moreover, basically $RE \le 0$, and this fact is consistent with our remark that a kind of upper bound is obtained by the models, 
as mentioned in Section~\ref{sec:model}. 
\item We can find the degradation of the accuracy for $bl = 10$ and $5000$ (or $10000$); the obtained performance is lower than the estimation. 
As its reasons, we guess the increase of accessing cost to the offset information when $bl = 10$ and 
the inefficient work of the cache blocking technique when $bl$ is too large. 
\item The significant large error in Matrix~\#17 with large $bl$ is due to the inefficient use of threads, 
as already mentioned in the observation on the Figure~\ref{fig: ssmc_parameter}. 
\item A tendency that $RE$ becomes large is observed on Cascade Lake, however we currently have not insights for its reasons. 
\end{itemize}
\par
What is important in practical situations is how to set the parameter $bl$ and $\theta$ appropriately. 
For this matter, based on the observations on Figures~\ref{fig: ssmc_parameter} and \ref{fig:ssmc_error}, we suggest a policy as follows: 
\begin{itemize}
\item $bl$ should be selected from $50 \lesssim bl \lesssim 500$. 
At least, we should avoid too small or too large $bl$. 
\item $\theta$ has much less impact on the performance than $bl$, and simply setting $\theta = 0.6$ will be acceptable. 
\item The performance depends mainly on the CSR rate, and calculating it in advance is thus informative. 
Moreover, for the above range of $bl$, the presented models, i.e. Equation~\ref{eq:b_hdc_over_csr}, are expected to provide accurate estimations, 
and we can roughly know the obtained performance in advance. 
\item Theoretically, $RP$ does not depend on environments, and an optimal setting of $bl$ and $\theta$ do not change. 
This indicates that we can reuse the knowledge of selecting parameters across different environments. 
\item If $n$ of a target matrix is not sufficiently large compared with the number of threads, 
we have to pay additional attention to fully using the threads.  
\end{itemize}

\subsubsection{Performance comparison with the MKL routines}
Finally, to make the evaluation of the obtained performance more clear, we compare them with routines provided in the Intel MKL library. 
For matrices~\#1, \#3, \#10, \#13, \#14, and \#17, we measured the performance of two MKL routines: 
the traditional CSR based routine ({\tt mkl\_dcsrgemv}) and the Inspector-Executor (IE) routine ({\tt mkl\_sparse\_d\_mv})~\citep{2020_fedorov}. 
In the use of the IE routine, 
we initially stored a matrix in the 4-array CSR format ({\tt mkl\_sparse\_d\_create\_csr}), 
then set the information of how many times SpMV repeats via {\tt mkl\_sparse\_set\_mv\_hint}, 
and called the optimization function ({\tt mkl\_sparse\_optimize}), 
before the execution of SpMV. 
\par
In Figure~\ref{fig:ssmc_mkl}, we present the relative performance of the M-HDC kernel (with the best setting of $bl$ and $\theta$), 
the MKL CSR routine, and the MKL IE routine over the (our in-house) CSR kernel in three environments. 
From Figure~\ref{fig:ssmc_mkl}, we have the following observations: 
\begin{itemize}
\item The relative performance of the MKL CSR routine is basically around $1.0$, which validates the obtained performance of our CSR kernel. 
This also means that almost no room of performance optimization for CSR-based SpMV kernels even by a hardware vendor, i.e Intel. 
\item Excepting matrix~\#17, the M-HDC kernel outperforms the MKL IE routine. 
This fact strengthens the potential of the M-HDC kernel. 
\item For matrix~\#17, the MKL IE routine shows the remarkable performance, however we currently have no insights on this result 
due to little open information on the technology in the MKL IE routine. 
\end{itemize}

\subsection{Summary of the experiments}
We give a brief summary of the experiments as follows:
\begin{itemize}
\item Through the experiments with stencil matrices, the effectiveness of using the DIA format with the cache blocking technique (i.e. the B-DIA kernel) was confirmed, and it was totally different in the in-cache and out-of-cache cases; 
in the in-cache case, the B-DIA kernel provides the better use of the SIMD vectorization, 
and in the out-of-cache case, it reduces the amount of the memory access cost. 
\item The experiments using 20 test matrices selected from the SuiteSparse Matrix Collection demonstrated that the cache blocking technique (i.e. the B-HDC kernel) is effective for some matrices appearing in practical applications. 
Moreover, it was fond that the M-HDC kernel more efficiently works for some matrices than the B-HDC kernel. 
\item Through the detailed analyses for the representative matrices, we figured out the impact of the parameter $bl$ and $\theta$ in the M-HDC kernel on the performance and confirmed the accuracy of the estimation by the performance models. 
We also presented the policy of how to chose the parameters in practical. 
\item Comparison with the MKL routines, especially the MKL IE routine, also proved the potential of the M-HDC kernel. 
\end{itemize}
\par
Since we compared the B-HDC and M-HDC kernels with the CSR kernel, it is a natural question to ask the comparison with SpMV kernels using ELLPACK type formats. 
As mentioned above, in the case that a matrix is sufficiently small, i.e. in the in-cache case, 
the efficiency of the SIMD vectorization is one of vital factors as our experimental results, namely Figure~\ref{fig:flops_diag}, show. 
For this case, comparison with kernels using ELLPACK type formats is important, which remains as our future work. 
In contrast to the in-cache case, when a matrix is sufficiently large, i.e. in the out-of-cache case, 
we can presume that the efficiency of the SIMD vectorization rarely affects the SpMV performance, 
which is reason we compared with only the CSR kernel in our experiments. 
Actually, in the papers by \citet{2014_kreutzer} and \citet{2020_almasri}, 
we can find that almost no performance improvement by ELLPACK type kernels over the CSR kernel was obtained 
for sufficiently large matrices on standard multi-core CPUs. 
Here, it is worth noting that this tendency differs on many-core CPUs such as Intel Xeon Phi; 
the effectiveness of SpMV kernels using ELLPACK type formats was reported by \citet{2014_kreutzer}, \citet{2020_alappat}, and \citet{2021_nakajima}.

\section{Conclusion}
\label{sec:conclusion}
In this research, we considered accelerating the SpMV computation on standard CPUs by exploiting diagonally structured sparsity patterns, 
which often appear in practical applications. 
We focused the HDC format, which combines the DIA and CSR formats, and recalled introducing cache blocking techniques into the SpVM kernel using the HDC format. 
Based on the observation of the cache blocked HDC kernel, we presented a modified version of the HDC format, which we call the M-HDC format, 
so as to more efficiently pick up partial diagonal structures. 
We carried out theoretical analyses based on performance models and theoretically provided the expected performance improvement by the B-HDC and M-HDC kernels 
over the CSR and HDC kernels. 
\par
Then, we conduced comprehensive experiments on the state-of-the-art multi-core CPUs. 
From the experiments using typical matrices, namely stencil matrices, we clarified the detailed performance characteristics of each kernel; 
it was shown that the impact of the SIMD vectorization on the performance of SpMV totally changes between the in-cache and out-of-cache cases. 
It is confirmed that the amount of data transferred from/to the main memory is much more crucial than the SIMD vectorization when a matrix is sufficiently large. 
By the experiments using practical matrices, it was established that there are matrices whose SpMV can be accelerated by exploiting the partial diagonal structures, 
which is realized by using the M-HDC format. 
The obtained experimental results were consistent with the theoretical results based on the performance models. 
This fact supports our approach to exploiting partial diagonal structures by the M-HDC format for accelerating SpMV. 
Through the present paper, we demonstrated the effectiveness of exploiting partial diagonal structures by the M-HDC format 
as a promising approach to accelerating SpMV on CPUs for a certain kind of practical sparse matrices. 
\par
In addition to the potential of the SpMV kernel using the M-HDC format,  an important insight has been also provided. 
As shown in the experimental results, namely Figure~\ref{fig:flops_diag}, 
the impact of the SIMD vectorization on the SpMV performance is clearly different between the in-cache and out-of-cache cases. 
Although many approaches have been recently proposed for the efficient SIMD vectorization on modern CPUs, 
we should pay attention to the appropriate problem setting to efficiently use them. 
For substantially large matrices, they may provide little speedup over the CSR kernel on standard CPUs.  
\par
A remained important task is involving the presented approach into application programs or numerical libraries. 
In the case of using application programs, it is better to store a matrix in the M-HDC format directly without converting from other formats. 
In this case, an efficient implementation to find out partial diagonal structures in a targeted problem such as a discretized model is needed.
In the case of introducing into numerical libraries, a matrix is expected to be given in a well-used format such as the CSR format, 
and the cost for converting to the M-HDC format is one of vital issues. 
In our research, we did not consider this cost and employed a naive way of finding partial diagonal structures, 
and developing a way that will be accepted in the practical use of numerical libraries is crucial. 
It will be also important to appropriately determining whether the M-HDC format should be used or not for a given matrix.

\bibliographystyle{unsrtnat}
\bibliography{fukaya.bib}

\section*{Acknowledgment}
This work is supported by JSPS KAKENHI Grant Nos.~JP18H03249 and JP19H04122, and JST PRESTO Grant No.~JPMJPR20M8.

\end{document}